\documentclass[journal=mamobx,manuscript=article]{achemso}
\setkeys{acs}{articletitle=true}

\SectionNumbersOn

\usepackage[version=3]{mhchem} 
\usepackage[T1]{fontenc} 

\usepackage{color}
\usepackage{gensymb}
\usepackage{graphicx}

\author{Wen-Sheng Xu}
\email{wsxu@ciac.ac.cn}
\affiliation{State Key Laboratory of Polymer Physics and Chemistry, Changchun Institute of Applied Chemistry, Chinese Academy of Sciences, Changchun 130022, P. R. China}

\author{Jack F. Douglas}
\email{jack.douglas@nist.gov}
\affiliation{Materials Science and Engineering Division, National Institute of Standards and Technology, Gaithersburg, Maryland 20899, United States}

\author{Zhao-Yan Sun}
\email{zysun@ciac.ac.cn}
\affiliation{State Key Laboratory of Polymer Physics and Chemistry, Changchun Institute of Applied Chemistry, Chinese Academy of Sciences, Changchun 130022, P. R. China}

\title{Polymer Glass Formation: Role of Activation Free Energy, Configurational Entropy, and Collective Motion}

\abbreviations{IR,NMR,UV}

\begin{document}

\begin{tocentry}

 \centering
 \includegraphics[height=3.4cm]{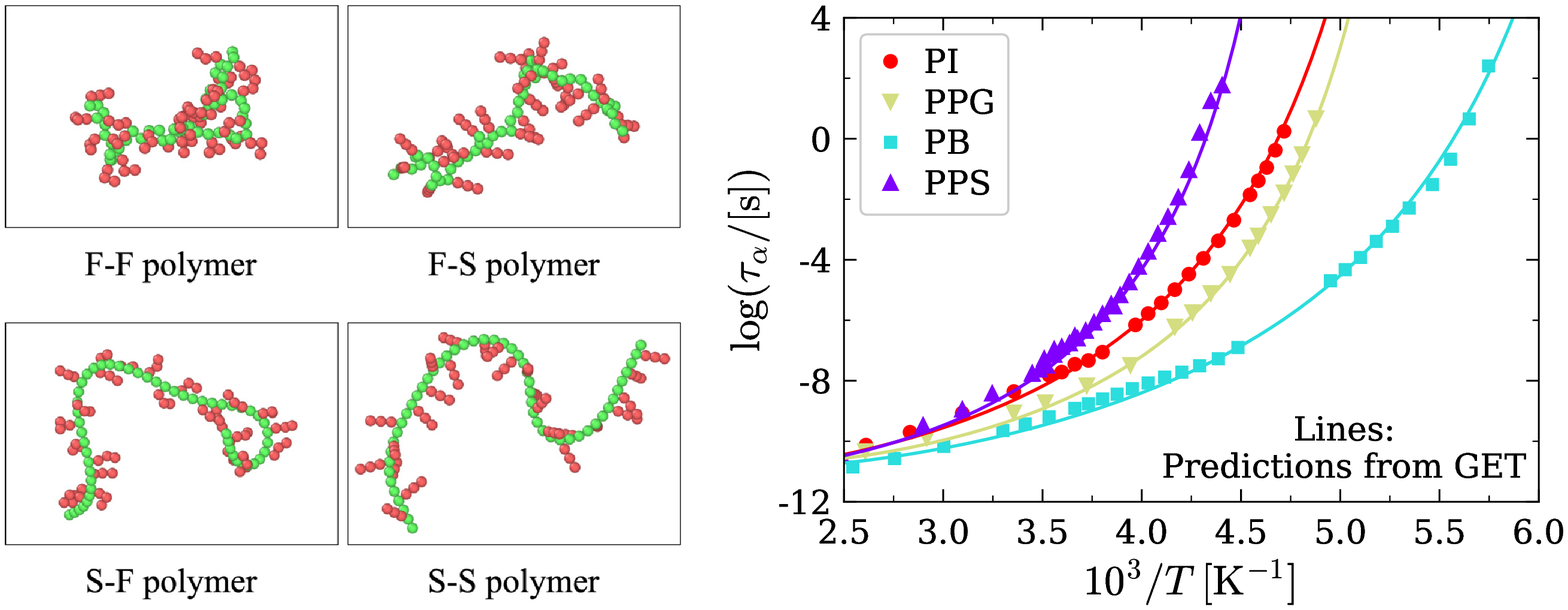}


\end{tocentry}

\newpage

\begin{abstract}
We provide a perspective on polymer glass formation, with an emphasis on models in which the fluid entropy and collective particle motion dominate the theoretical description and data analysis. The entropy theory of glass formation has its origins in experimental observations relating to correlations between the fluid entropy and liquid dynamics going back nearly a century ago, and it has entered a new phase in recent years. We first discuss the dynamics of liquids in the high temperature Arrhenius regime, where transition state theory is formally applicable. We then summarize the evolution of the entropy theory from a qualitative framework for organizing and interpreting temperature-dependent viscosity data by Kauzmann to the formulation of a hypothetical `ideal thermodynamic glass transition' by Gibbs and DiMarzio, followed by seminal measurements linking entropy and relaxation by Bestul and Chang and the Adam-Gibbs (AG) model of glass formation rationalizing the observations of Bestul and Chang. These developments laid the groundwork for the generalized entropy theory (GET), which merges an improved lattice model of polymer thermodynamics accounting for molecular structural details and enabling the analytic calculation of the configurational entropy with the AG model, giving rise to a highly predictive model of the segmental structural relaxation time of polymeric glass-forming liquids. The development of the GET has occurred in parallel with the string model of glass formation in which concrete realizations of the cooperatively rearranging regions are identified and quantified for a wide range of polymeric and other glass-forming materials. The string model has shown that many of the assumptions of AG are well supported by simulations, while others are certainly not, giving rise to an entropy theory of glass formation that is largely in accord with the GET. As the GET and string models continue to be refined, these models progressively grow into a more unified framework, and this Perspective reviews the present status of development of this promising approach to the dynamics of polymeric glass-forming liquids.
\end{abstract}

\newpage

\section{Introduction}

Glasses have been central to fabrication technologies since the dawn of civilization, and these materials are becoming increasingly important in modern technologies. Glasses are not only encountered as structural materials in our environment, such as window panes, vessels, and plastic polymeric materials, but they are also used in high technology applications as materials for non-volatile electronic memory,~\cite{2007_NatMater_6_824} organic light emitting diodes and organic electronics,~\cite{2005_JMC_15_75, 2007_CR_107_923} commercial aircraft~\cite{2005_MSEA_412_171}, molecular separations in manufacturing and water desalination and purification,~\cite{2008_JAPS_107_1039, 2010_PC_1_63, 2013_Science_339_303, 2016_NM_15_760, 2018_JPCL_8_2003, 2019_MacroLett_8_1022} etc. The physics of glass formation is also highly relevant to biological substances whose properties and functions are dictated by their complex molecular structure and intermolecular interactions.~\cite{2011_PNAS_108_4714, 2012_NJP_14_115012, 2013_JRSI_10_20130726, 2015_NatPhys_11_1074}

Polymeric materials have a general propensity to form glasses at low temperatures ($T$) and exhibit glassy dynamics over a large $T$ range, even if some of them ultimately form semicrystalline materials composed of a mixture of crystalline and noncrystalline regions in their solid state. Since polymers exhibit many of the same features in glass formation as other glass-forming (GF) materials while at the same time the many complications associated with crystallization that often arise in GF liquids composed of atomic mixtures~\cite{2019_PRX_9_031016} can be avoided, these materials have been widely used to investigate the fundamental nature of glass formation for decades. On the practical side, polymer glasses have the advantage of being light, ductile, easy to process, and, of course, relatively inexpensive, giving rise to numerous applications in daily life and emerging technologies. We are truly living in the `Polymer Age'.~\cite{2009_PTRSB_364_1973} From our own standpoint, the rich chemistry of polymers allows for an exploration of the large molecular parameter space of possible types of glass formation, the same attribute that makes these materials so attractive for applications in materials science. As a complement to our work emphasizing the entropy theory of glass formation,~\cite{1958_JCP_28_373, 1965_JCP_43_139, 2008_ACP_137_125, 2014_JCP_140_204509} and simulations for purposes of validation and refinement of the present theoretical perspective on the dynamics of GF liquids, we mention some excellent summaries of the experimental observations on polymer glass formation.~\cite{2005_RPP_68_1405, 2010_Mac_43_7875, 2014_JPCM_26_153101, 2017_RPP_80_036602, 2017_Mac_50_6333, Chapter1_PolymerGlasses, Chapter2_PolymerGlasses, Chapte5_PolymerGlasses, Chapter6_PolymerGlasses, Chapter7_PolymerGlasses, Chapter9_PolymerGlasses, Chapter10_PolymerGlasses, Chapter12_PolymerGlasses}

Experimental and computational studies over the past decades have established the rich phenomenology of GF liquids with an emphasis on the approach to the glass transition temperature $T_{\mathrm{g}}$ from above where the fluid is still in equilibrium and where strong changes in the fluid dynamics still occur.~\cite{1996_JPC_100_13200, 2001_Nature_410_259, 2011_RMP_83_587, 2012_JCP_137_080901, 2013_JCP_138_12A301, 2013_ARCMP_4_263} The most basic phenomenology is that the dynamic properties of GF liquids, such as shear viscosity $\eta$ and structural relaxation time $\tau_{\alpha}$, display a dramatic $T$ dependence upon cooling toward $T_{\mathrm{g}}$. For example, the shear viscosity of a GF material can alter by over $15$ orders of magnitude over a relatively narrow $T$ range. Angell~\cite{1991_JNCS_131_13, 1995_Science_267_1924} has introduced the concept of `fragility' to quantify the strength of the $T$ dependence of the dynamics of GF liquids. The dynamics of `strong' liquids, such as $\text{SiO}_{2}$, has a nearly Arrhenius $T$ dependence, and in contrast, `fragile' liquids, such as \textit{o}-terphenyl, exhibit dynamic properties whose $T$ dependence is highly non-Arrhenius. For polymers, $T_{\mathrm{g}}$ can be tuned roughly from $150$ K to $500$ K, and fragility varies by nearly an order of magnitude from approximately $25$ to over $200$.~\cite{1993_Mac_26_6824, 2008_Mac_41_7232, 2012_Mac_45_8430}

While models and theories are being constantly introduced for understanding the origin and nature of the slowing down of the dynamics of GF liquids, some ideas remain invariant for qualitatively understanding the basic phenomenology of glass formation. It has been generally appreciated since the works of Simon~\cite{1931_ZAAC_203_219} and Kauzmann~\cite{1948_CR_43_219} that the rapid increase in $\eta$ and $\tau_{\alpha}$ upon cooling is accompanied by a drop in the fluid entropy. Much of the early interesting literature attempting to define the glass transition and understand the role of the fluid entropy in the glass transition is difficult to access, but Schmelzer and Tropin~\cite{2018_Entropy_20_103} have provided an informative review of these early studies of glass formation, along with a discussion of how these early studies relate to more recent works. Kauzmann discussed the fate of equilibrium cooled liquids based on the excess entropy, $S_{\text{exc}} \equiv S_{\text{liq}} - S_{\text{xtal}}$, where $S_{\text{liq}}$ and $S_{\text{xtal}}$ are the entropies of a material in the liquid and crystalline states, respectively. The drops in $S_{\text{exc}}$ upon lowering $T$ are so rapid that, if $S_{\text{exc}}$ in the $T$ regime above $T_{\mathrm{g}}$ is formally extrapolated to much lower $T$ where the material can no longer remain in equilibrium, $S_{\text{exc}}$ would vanish at a finite characteristic temperature called the `Kauzmann temperature' $T_K$, below which $S_{\text{exc}}$ would formally become negative. The inferred vanishing of $S_{\text{exc}}$ is termed the `Kauzmann paradox' or the `entropy crisis', but this situation may instead represent the result of an unwarranted extrapolation, and thus have no physical meaning. Nonetheless, the reality of the drop in the entropy upon lowering $T$ is an unequivocal feature of GF liquids, and the general idea that there is some sort of `entropy crisis' associated with intrinsic glass formation remains a prevalent model of the physical nature of glass formation, an approach that we develop in our own work. In more recent works, Martinez and Angell~\cite{2001_Nature_410_663} have reviewed the parallelism between the $T$ dependences of the viscosity and fluid entropy in diverse GF liquids, and Angell~\cite{1997_JRNIST_102_171} has emphasized the role of the fluid entropy in understanding the fragility of GF liquids.

Another more recently identified feature that is often invoked to explain the dynamics of GF liquids is the occurrence of dynamic heterogeneity or collective motion of particles.~\cite{1996_JPC_100_13200, 2001_Nature_410_259, 2011_RMP_83_587, 2012_JCP_137_080901, 2013_JCP_138_12A301, 2013_ARCMP_4_263} A significant body of evidence has indicated the presence of transient clusters of particles with excessively high or low mobilities relative to simple Brownian particles. In many cases, the specific nature of dynamic heterogeneity has often been rather vague, and there are currently many uncertainties about how the various types of dynamic heterogeneity that one can imagine or identify might be related to the dynamics of GF liquids. This list seems to be essentially endless and ever growing, and we believe that the time has come to cull this proliferation by investigating which of these heterogeneity measures have a direct relation to the overall dynamics of the fluids. Interestingly, both mobile and immobile particles are found to exhibit the property of forming `fractal' dynamic polymeric clusters, regardless of the molecular bond connectivity of individual molecules, and the temperature and structural characteristics of these types of heterogeneity conform to well-known phenomenologies of self-assembling systems.~\cite{1999_JCP_111_7116, 2006_JCP_125_144907, 2008_JCP_128_224901, 2014_JCP_140_204509} This finding gives us hope for an overall generalized theoretical treatment of GF materials. It must be acknowledged at the outset that the theory of glass formation is still in a model building stage, however. The entropy theory is just one of a number of promising theoretical frameworks for glass formation. For instance, the nonlinear Langevin equation (NLE) theory~\cite{2004_JCP_121_1984, 2004_JCP_121_2001, 2010_ARCMP_1_277} and the more recent elastically collective nonlinear Langevin (ECNLE) equation theory~\cite{2014_JCP_140_194506, 2014_JCP_140_194507, 2015_Mac_48_1901, 2016_Mac_49_9655} of Schweizer and coworkers emphasize the central role of the long wavelength limit of the static structure factor, $S(0)$, in determining the dynamics of GF liquids. The free volume model of the dynamics of GF liquids also remains popular in the polymer science community.~\cite{2013_SoftMatter_9_3173, 2016_Mac_49_3987} Schweizer and coworkers~\cite{2009_JPCM_21_503101} have previously discussed many of these models, so we do not reproduce such a discussion in this Perspective.

Here, we focus on models of glass formation where the entropy and collective particle exchange motion are taken to be of primary importance in the description of the dynamics of GF liquids, as postulated by Adam and Gibbs (AG).~\cite{1965_JCP_43_139} The present perspective on the nature of polymer glass formation thus emphasizes the consequence of changes in the fluid entropy on the dynamics of cooled liquids in the low $T$ regime where the $T$ dependences of relaxation and diffusion are observed to be non-Arrhenius. We strictly avoid the term `supercooled' since it is clear that not all GF liquids, and particularly not all polymeric liquids, can crystallize, even in principle.~\cite{1981_ANYAS_371_1, 2013_NatPhys_9_554, 2016_SciRep_6_36963} It is then possible that the glass state of some materials at low $T$ is the true thermodynamic state rather than just a metastable thermodynamic condition. 

We first discuss transition state theory (TST)~\cite{Book_Eyring, 1940_JACS_62_3113, 1941_CR_28_301} describing the dynamics of liquids in the high $T$ Arrhenius regime, which, in our view, should be a natural starting point for any theory of glass formation based on the idea of thermally activated transport. We then summarize the historical development of the entropy theory of glass formation which has been `under construction' for almost a century, a development that started from observations of correlations between the fluid entropy and heats of vaporization and dynamics in the 1930s, to the organization and interpretation of these data by Kauzmann~\cite{1948_CR_43_219} in 1948, and then the formulation of the Gibbs-DiMarzio (GD) theory~\cite{1958_JCP_28_373} in 1958 based on the then newly developed theory of polymer thermodynamics formulated by Flory~\cite{1956_PRSLA_234_60} in his classical work aimed originally at understanding the crystallization and equation of state properties of polymer materials. A precise fundamental link between dynamics and thermodynamics was first discovered by Bestul and Chang~\cite{1964_JCP_40_3731} in 1964 by comparing data for relaxation and excess entropy obtained from specific heat measurements of molecular and polymer GF liquids, which, in turn, led to the introduction of the AG model~\cite{1965_JCP_43_139} in 1965 to rationalize these observations. The AG model later gave rise to the current prevailing conceptual framework and language in which the dynamics of GF liquids is characterized by fragility~\cite{1991_JNCS_131_13, 1995_Science_267_1924} and other metrics related to the $T$ variation of the configurational entropy $S_c$ associated with a complex energy landscape that gives rise to the complex dynamics of cooled liquids. These developments led to the generalized entropy theory (GET)~\cite{2008_ACP_137_125} and the string model,~\cite{2014_JCP_140_204509} which are built on concepts and methods developed in a vast number of previous experimental studies, so these newer works are just part of a long historical development for a description of liquid dynamics to increased precision and validation. For completeness, we also provide a brief overview of other models of liquid dynamics emphasizing the fluid entropy, including the random first-order transition (RFOT)~\cite{1987_PRB_36_8552, 1989_PRA_40_1045, 2004_JCP_121_7347, 2007_ARPC_58_235} and the Rosenfeld and excess-entropy scaling approaches~\cite{1977_PRA_15_2545, 1999_JPCM_11_5415, 2018_JCP_149_210901} to relaxation and diffusion in fluids.

\section{\label{Sec_TST}Transition State Theory}

When cooling upon the approach to $T_{\mathrm{g}}$, the $T$ dependence of the rapidly growing relaxation time of GF liquids is described in most experimental studies by the Vogel-Fulcher-Tammann (VFT) equation,~\cite{1921_PZ_22_645, 1925_JACS_8_339, 1926_ZAAC_156_245}
\begin{equation}
	\label{Eq_VFT}
	\tau_{\alpha} = \tau_0\exp\left(\frac{D T_{\mathrm{VFT}}}{T - T_{\mathrm{VFT}}}\right),
\end{equation}
where $\tau_0$ is a prefactor, $D$ is a `fragility parameter' quantifying the strength of the $T$ dependence of $\tau_{\alpha}$, and $T_{\mathrm{VFT}}$ is the temperature where $\tau_{\alpha}$ extrapolates to infinity. Although the VFT equation is perhaps the best-known phenomenological relationship in GF liquids, there have been attempts to challenge the supremacy of this equation.~\cite{1995_PA_219_27, 1996_PRE_53_751, 2009_PNAS_106_19780, 2009_JPCB_113_5563, 2010_JPCB_114_17113, 2012_PRE_86_041507, 2013_JCP_139_084504, 2015_Mac_48_3005, 2015_PRE_92_062304, 2020_JCP_153_124507} Nonetheless, most researchers persist in fitting their relaxation time, viscosity, and diffusion data to this simple equation because of its relative simplicity and apparent effectiveness in fitting data for a vast range of materials. It should be mentioned, however, that even some of the earliest quantitative works on relaxation in polymer materials recognized that the VFT equation, and the mathematically equivalent and heavily utilized expression of Williams, Landel and Ferry (WLF),~\cite{1955_JACS_77_3701} are limited to a $T$ range between $T_{\mathrm{g}}$ and a temperature about $100$ K higher than $T_{\mathrm{g}}$, which corresponds to a temperature comparable to the crossover temperature $T_c$ in our discussion of the GET model in Section~\ref{Sec_GET}. It is notable that the VFT relation may be \textit{derived} from the GET for a prescribed $T$ range between $T_c$ and $T_{\mathrm{g}}$, along with an analog of this equation when glass formation is achieved by increasing pressure ($P$) at fixed $T$. Moreover, the VFT parameters $D$ and $T_{\mathrm{VFT}}$, or their WLF equivalent,~\cite{2015_JCP_142_014905} may be calculated from the GET so these are not just adjustable empirical parameters in the GET. The GET also specifies an alternative functional form for the $T$ regime above $T_c$ but below the onset temperature $T_A$, where relaxation and diffusion become non-Arrhenius, again with all model parameters fixed in terms of molecular parameters describing the thermodynamics of polymer fluids.

At sufficiently high $T$, a large body of evidence indicates that the structural relaxation time $\tau_{\alpha}$ of GF liquids generally exhibits an Arrhenius $T$ dependence,
\begin{eqnarray}
	\label{Eq_Arrhenius}
	\tau_{\alpha} = A_0 \exp \left( \frac{\Delta H_0} { k_BT} \right),
\end{eqnarray}
where $A_0$ is a prefactor, $k_B$ is Boltzmann's constant, and $\Delta H_0$ is the activation enthalpy. The prefactor $A_0$ in eq~\ref{Eq_Arrhenius} involves a vibrational attempt frequency $\tau_0$ multiplied by a factor involving the activation entropy $\Delta S_0$, $\exp(-\Delta S_0/k_B)$. We append an `$o$' subscript to the activation energetic parameters to denote that these quantities are determined in the high $T$ Arrhenius regime where relaxation times are relatively short and the fluid is relatively `dynamically homogeneous'.

\begin{figure*}[htb!]
	\centering
	\includegraphics[angle=0,width=0.475\textwidth]{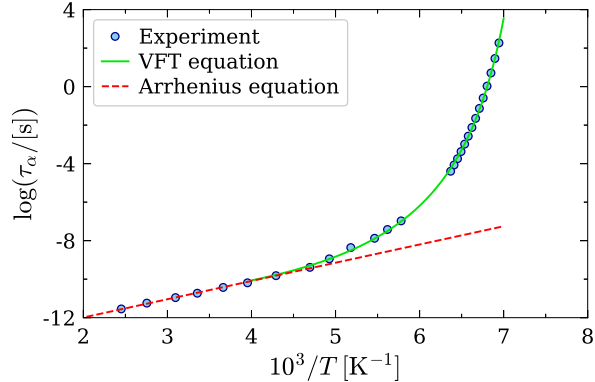}
	\caption{\label{Fig_Arrhenius}Illustration of the temperature ($T$) dependence of the dynamics of glass-forming liquids. The figure displays the logarithm of the segmental structural relaxation time, $\log \tau_{\alpha}$, versus inverse temperature $10^3/T$ for poly(dimethylsiloxane) with the molar mass of $21600$ g/mol at ambient pressure ($P$). Experimental data were provided to us by the authors of ref~\citenum{2015_Mac_48_3005}. The solid line is a fit to the Vogel-Fulcher-Tammann (VFT) equation (eq~\ref{Eq_VFT}). The dashed line is a fit to the Arrhenius equation (eq~\ref{Eq_Arrhenius}) with the fitted parameters, $A_0 = 1.3 \times 10^{-14}$ s and $\Delta H_0/k_B = 2181$ K. Despite the dramatic increase of $\tau_{\alpha}$ with lowering $T$ upon approaching the glass transition temperature $T_{\mathrm{g}}$, an Arrhenius dynamics applies instead at sufficiently high $T$.}
\end{figure*}

To highlight the presence of Arrhenius dynamics at high $T$, we show the segmental structural relaxation time $\tau_{\alpha}$ of poly(dimethylsiloxane) versus $10^3/T$ at ambient pressure measured by the R{\"o}ssler group~\cite{2015_Mac_48_3005} in Figure~\ref{Fig_Arrhenius}. Measurements of dynamics in the Arrhenius regime are generally challenging experimentally due to the thermal degradation of polymers, but the field-cycling $^1$H nuclear magnetic resonance relaxometry has recently changed this situation, at least for some polymers.~\cite{2012_PRE_86_041507, 2013_JCP_139_084504, 2015_Mac_48_3005} It is thus well established that the fluid dynamics is observed to be homogeneous in the high $T$ regime where the Arrhenius equation formally holds, and the dynamic heterogeneity sets in below the onset temperature $T_A$ for non-Arrhenius relaxation where the dynamics deviates from the Arrhenius equation.

While the physical mechanisms for explaining the dynamics of fluids must be different in the high $T$ Arrhenius and low $T$ non-Arrhenius regimes, it is essential to build an experimentally validated theoretical foundation for understanding and ultimately controlling the dynamics of cooled liquids, where collective many-body effects are prevalent, based on a thorough understanding of the dynamics in the high $T$ Arrhenius regime. In our view, this is a natural starting point for any theory of glass formation based on the idea of thermally activated transport. Molecular dynamics (MD) modeling becomes a particularly attractive tool at present because of the relative small magnitude of structural relaxation and relatively short equilibration times in this regime. It should then be possible to make relatively rapid progress in understanding the relation between molecular structure and other molecular parameters and factors controlling structural relaxation and diffusion in the high $T$ Arrhenius regime. We expect that this information should be useful for understanding the dynamics of polymers, even though many polymer materials would physically degrade before reaching such high $T$ at which the Arrhenius equation describes the dynamics of the materials. These observations again serve to underscore the fact that we need to be concerned about more than just the glass transition temperature $T_{\mathrm{g}}$, since $T_A$ is often more than twice $T_{\mathrm{g}}$~\cite{2008_ACP_137_125} so that glass formation encompasses the entire $T$ regime in which polymer materials are normally utilized and processed and thermally stable to chemical decomposition. This large $T$ range above $T_{\mathrm{g}}$ thus has fundamental importance for processing and field use applications of polymers, and for understanding the nature of glass formation, regardless of whether the polymer materials crystallize or not at temperatures lower than $T_A$.

Historically, TST of Eyring and others~\cite{Book_Eyring, 1940_JACS_62_3113, 1941_CR_28_301} has provided significant advances for understanding the dynamics of liquids in the Arrhenius regime. If viscous flow is viewed as a chemical reaction in which the elementary process is the passing of a single molecule from one equilibrium position to anther over a potential barrier, viscosity, plasticity, and diffusion of fluids can all be treated as examples of absolute reaction rates within TST.~\cite{1936_JCP_4_283, 1937_JCP_5_726} The central quantity in TST is the activation free energy $\Delta G_0$, 
\begin{equation}
	\label{Eq_DeltaGo}
	\Delta G_0 = \Delta H_0 - T \Delta S_0,
\end{equation}
which contains \textit{both} enthalpic and entropic contributions. The TST-based approach has proven to be a reliable general theoretical framework for describing the dynamic properties of condensed fluids at elevated $T$,~\cite{1983_JPC_87_2664} even though the actual development of Eyring's absolute rate theory~\cite{1940_JACS_62_3113, 1941_CR_28_301, 1937_JCP_5_726} for the dynamics of real liquids is based on the rather idealized view of the dynamics of real fluids that enables transforming the problem of explaining gas phase dynamics into a model of the dynamics of condensed liquids in terms of assumptions more acceptable for gases than for pure fluids. A rigorous TST has been developed for idealized crystalline materials,~\cite{1957_JPCS_3_121} in which the particles interact with purely harmonic interactions. Viewing a liquid as a highly defective crystal has provided an alternative model of the Arrhenius dynamics of liquids.~\cite{1992_PRL_68_974} This alternative view of structural relaxation and diffusion in condensed materials emphasizes the collective nature of transport in the condensed state, but there is still no satisfying analytic theory of dynamics, even in the Arrhenius regime where TST is expected to apply. Numerical implementations of TST in ordered condensed materials have shown that the TST framework still applies even though it currently does not allow any direct analytic treatment.~\cite{2001_PRB_64_075418, 1995_PRL_75_469, 1998_PRB_58_12667} We thus adopt TST as a working framework for understanding the dynamics of cooled liquids. Notably, the AG theory~\cite{1965_JCP_43_139} implicitly adopts the TST framework in its formulation, and thus, this assumption extends to the GET,~\cite{2008_ACP_137_125} which combines the AG relation with a thermodynamic theory that enables calculating the configurational entropy of polymer fluids. While the TST-based approach to glass formation does not require the many simplifying assumptions of Eyring,~\cite{Book_Eyring, 1940_JACS_62_3113, 1941_CR_28_301} its practical implementation requires MD simulations~\cite{2015_JCP_143_144905} to estimate the activation free energy parameters. Thus, the treatment of the dynamics of condensed fluids remains largely semi-empirical and requires simulation studies or experimental measurements to determine parameters characterizing the free energy of activation for fluids with very different chemical natures. Over the years, correlations between the energetic parameters of activation and thermodynamic properties have been observed, such as a strong correlation between $\Delta H_0$ and the heat of vaporization,~\cite{Book_Eyring, 1938_JAP_9_252} and changes in $\Delta H_0$ and $\Delta S_0$ have been found to occur in a parallel fashion when changing molecular mass,~\cite{1946_JCP_14_591, 1961_Nature_191_1292, 1943_TFS_39_48, 1959_PPS_73_153, 1940_JACS_62_3113, 2015_JCP_143_144905} film thickness,~\cite{2014_NatCommun_5_4163, 2015_JCP_142_234907} or additive concentration,~\cite{1995_PRE_51_1791, 2006_PRE_74_031501, 2008_JPCB_112_15980, 2012_SoftMatter_8_2983, 2013_SoftMatter_9_241} a phenomenon termed the `entropy-enthalpy compensation' (EEC) effect.~\cite{2001_CR_101_673, 2008_JPCB_112_15980, 1992_PRB_46_12244, 1986_JPC_19_5655} These correlations provide some sign posts indicating what properties control the basic energetic parameters, $\Delta H_0$ and $\Delta S_0$, which are attractive targets for future simulation and theoretical modeling efforts.

We have observed the EEC effect to be a significant factor in simulations of polymer systems~\cite{2014_NatCommun_5_4163, 2015_JCP_142_234907, 2015_PNAS_112_2966, 2017_SoftMatter_13_1190, 2016_MacroLett_5_1375, 2017_Mac_50_2585, 2020_Mac_53_4796, 2020_Mac_53_9678} and experiments of polymer nanocomposites and thin films.~\cite{2015_PNAS_112_2966, 2017_JCP_147_154902, 2018_NanoLett_18_7441} Notably, the magnitude of the EEC effect on relaxation and diffusion can be rather significant, even meriting the term `astounding'. For example, the prefactor of eq~\ref{Eq_Arrhenius} can change by a factor of about $10^{47}$ as the activation energy grows upon approaching the melting temperature from below. Extremely large changes in the prefactor are also encountered through this mechanism in the diffusion coefficient of metallic glasses; e.g., see Figure 13 in ref~\citenum{2003_RMP_75_237}. Calculations based on TST have provided insight into these astronomical changes of the prefactor in the context of surface dislocation nucleation of metallic materials.~\cite{2011_PNAS_108_5174} Numerical TST calculations have also given some insight into the EEC phenomenon in the interfacial dynamics of crystalline materials.~\cite{2001_PRB_64_075418, 1995_PRL_75_469, 1998_PRB_58_12667} An implication of these results is that the neglect of the activation entropy $\Delta S_0$, which is a basic assumption in the classical AG theory~\cite{1965_JCP_43_139} and the more recent GET,~\cite{2008_ACP_137_125} can lead to serious errors in the TST description of the dynamics of GF liquids. This is perhaps the most serious assumption in the GET~\cite{2008_ACP_137_125} that we must address in the future to improve this model. This point is further discussed in Section~\ref{Sec_GET}.

\section{Classical Entropy Theories of Glass Formation}

In this section, we introduce the classical entropy theories of glass formation, namely, the theories of GD~\cite{1958_JCP_28_373} and AG.~\cite{1965_JCP_43_139} We present main ideas behind these theories and discuss their strengths and weaknesses. The current section augments a previous discussion given in ref~\citenum{2008_ACP_137_125}, so our discussion is relatively brief.

\subsection{Gibbs-DiMarzio Theory}

In 1958, GD~\cite{1958_JCP_28_373} developed a systematic statistical mechanical theory of polymer glass formation based on the lattice model of semiflexible polymers that had been developed earlier by Flory~\cite{1956_PRSLA_234_60} for the description of polymer crystallization. Since the entropy and other thermodynamic properties are purely configurational in the lattice model, the term `configurational entropy' has been widely adopted in the literature. The GD theory provided the first theoretical underpinning for how estimates of the fluid entropy might be utilized to make predictions of relevance for understanding the dynamics of GF liquids by identifying an `ideal glass transition' with a thermodynamic event, the vanishing of the configurational entropy $S_c$. This result was a direct theoretical `echo' of the Kauzmann's concept of an entropy catastrophe possibly underlying glass formation.~\cite{1948_CR_43_219} In addition to providing a conceptually clear picture of the origin of glass formation upon cooling at some point where the system has `no place to go' dynamically and thus becomes arrested, the GD theory allowed for quantitative predictions for the glass transition temperature $T_{\mathrm{g}}$, which was heuristically identified by GD as being a non-equilibrium analog of the true ideal glass transition temperature $T_0$ that could not be reached because of the extreme slowing of the dynamics upon approaching this temperature, but which `tracks' $T_0$. DiMarizo and Yang~\cite{1997_JRNIST_102_135} have reviewed the many successes of the GD theory, which is commonly invoked in experimental studies to the present time.

In a more recent work,~\cite{1997_JRNIST_102_135} DiMarzio revised his interpretation of the model prediction of a vanishing of the configuraional entropy in light of information that later became available. In particular, DiMarzio~\cite{1997_JRNIST_102_135} suggested that $S_c$ became `critically small' at the glass transition, rather than actually vanishing, so that the spirit of the GD theory is preserved. Corsi and Gudrati~\cite{2003_PRE_68_031502} have reassessed the initial negative view of the GD theory by formulating a theory of semiflexible polymers on Husimi lattices to allow calculations without approximation and their work has provided an expanded view of the original work by GD~\cite{1958_JCP_28_373} that confirms the essential idea of an entropy crisis in this type of thermodynamic model. We further discuss the technical revision of the GD picture and its ramifications for glass formation based on the GET~\cite{2008_ACP_137_125} below in Section~\ref{Sec_GET}. 

The central quantity in the GD model is the configurational entropy $S_c$, defined as the total entropy with vibrational contributions being excluded. In physical terms, the configurational entropy is related to the number of distinct configurational states of the fluid. Unfortunately, the conceptual clarity of the notion of configurational entropy remains a question even in current studies of glass formation, posing difficulties and challenges for determining this quantity both experimentally and computationally. Berthier and coworkers~\cite{2019_JCP_150_160902} have recently reviewed the methods of estimating $S_c$ from a computational perspective. In particular, a new Monte Carlo swap algorithm has been developed to estimate $S_c$ at extremely low $T$,~\cite{2017_PRX_7_021039, 2019_JSM_064004} and these simulation results have been interpreted as suggesting the existence of a Kauzmann temperature in the particular model fluids studied.~\cite{2017_PNAS_114_11356, 2019_JCP_151_084504} Unfortunately, this novel simulation method seems to be only applicable to fluids of spherical particles having a high size polydispersity, a property that is not encountered in real atomic or molecular materials. It seems unlikely to us that the swap algorithm can be extended to polymeric and molecular GF liquids generally. There is still no proof that $S_c$ actually vanishes at a finite $T$. Personally, we are skeptical about the existence of a finite Kauzmann temperature in real molecular fluids, even though the GET itself predicts such a temperature rather generally in semiflexible polymer melts, as we discuss further in Section~\ref{Sec_GET}.

In addition, strong criticism has been often raised against fundamental tenets of the GD theory concerning the identification of a vanishing of $S_c$ with a second-order phase transition. Even more seriously, the vanishing of $S_c$ has been suggested to be an artifact of the inaccuracy of the mean-field calculations of $S_c$ for dense polymer fluids based on the lattice model. Gujrati and Goldstein~\cite{1981_JCP_74_2596} were the first to demonstrate that the GD model violated rigorous bounds on the entropy of polymer melts, clearly bringing the validity of the GD model into question, despite its impressive empirical success in identifying essential trends between the experimental $T_{\mathrm{g}}$ and the theoretical `ideal glass transition temperature' $T_0$ calculated from the lattice model. Binder and coworkers~\cite{1996_PRE_54_1535} clarified the situation somewhat based on simulations of a bond fluctuation lattice model of flexible polymers, which suggest that $S_c$ approaches a constant positive value at low $T$, but the validity of these results has remained a question because of the inherently difficulty of estimating thermodynamic properties reliably from simulations at such low $T$.

The original GD theory of glass formation in polymer liquids involves other assumptions that significantly limit the predictive capacity of this pioneering model of polymer glass formation. First, the theory is preoccupied with the general philosophical problem of locating and explaining the existence of an `ideal glass transition temperature', $T_0$, which, as we have discussed above, probably does not really exist. Unfortunately, a fluid cannot remain in equilibrium near $T_0$ because the structural relaxation time becomes astronomical in magnitude near this temperature so this issue is a bit academic. A more recent work suggests that the vanishing of $S_c$ at a finite temperature $T_0$ in the lattice model is just an artifact of the high $T$ expansion involved in the lattice model calculations for the free energy.~\cite{2008_ACP_137_125} Explicit calculations of $S_c$ for flexible polymer melts show that $S_c$ does not vanish at any finite positive $T$ when the high $T$ expansion associated with the treatment of chain stiffness is avoided.~\cite{2016_JCP_145_234509} When combined with the AG model,~\cite{1965_JCP_43_139} this finding also implies that no divergence generally exists in the AG theory and its extensions such as the GET.~\cite{2008_ACP_137_125} Second, we also emphasize that the GD model is based on a highly simplified description of polymer chains as semiflexible self-avoiding walks interacting with uniform short-range interactions and composed of structureless monomer units. This extremely coarse-grained model of polymer fluids does not allow one to consider how monomer structure and chain topology affect polymer glass formation. Moreover, there is more to glass formation than knowing $T_{\mathrm{g}}$. Large complex changes in the dynamics of GF liquids initiate at $T$ well above $T_{\mathrm{g}}$, but below $T_A$ for non-Arrhenius dynamics. Thus, we really need a theoretical framework that can predict the breadth of the transition temperature range by estimating temperatures characterizing the beginning, middle, and end of this broad transition phenomenon having both well-defined dynamic and thermodynamic signatures. Any discussion of this kind requires something like the AG model~\cite{1965_JCP_43_139} to interrelate the structural relaxation time to $S_c$ or some other thermodynamic property. The original GD theory describes glass formation from a purely thermodynamic perspective. It is the `marriage' of the powerful lattice cluster theory (LCT)~\cite{1998_ACP_103_335, 2014_JCP_141_044909} extension of the lattice model of type studied by GD to treat polymer melts having general monomer structure and the AG model~\cite{1965_JCP_43_139} formally linking the melt thermodynamics to the segmental structural relaxation time. The resulting GET~\cite{2008_ACP_137_125} is a highly predictive model of the segmental dynamics of polymeric liquids over the entire temperature and pressure regimes of glass formation. Of course, the question remains of whether this framework is true to the physics of real and simulated materials. This is why validation by simulation and measurement is so important in further developments of the entropy theory of glass formation.

\subsection{\label{Sec_AG}Adam-Gibbs Theory}

Although a large body of evidence indicates that the Arrhenius equation describes the $T$ dependence of relaxation and diffusion in liquids at elevated $T$,~\cite{Book_Eyring, 1940_JACS_62_3113, 1941_CR_28_301} this description ceases to apply as liquids are cooled to low $T$ (e.g., see Figure~\ref{Fig_Arrhenius}). The term `fragility' has been introduced to quantify the degree to which the dynamics deviates from the Arrhenius behavior.~\cite{1991_JNCS_131_13, 1995_Science_267_1924} The structural relaxation time provides an important quantity whose change in magnitude can be as large as $10^{15}$ between the high $T$ Arrhenius regime and the low $T$ non-Arrhenius regime where the material progressively acquires the rheological properties of a solid as $T_{\mathrm{g}}$ is approached. Of course, the apparently universal and practical nature of this phenomenon has naturally prompted many speculations as to its cause. 

Early experimental studies attempting to rationalize the non-Arrhenius $T$ dependence of the shear viscosity and other fluid transport properties emphasized that large changes in the dynamics correlate with corresponding changes in the fluid entropy,~\cite{1931_ZAAC_203_219, 1948_CR_43_219} and this correlation led some influential scientists to infer that the thermodynamic changes originate from the emergence of collective motion within cooled liquids.~\cite{1948_CR_43_219} Following up on an earlier suggestion of Goldstein~\cite{1963_JCP_39_3369} that some other thermodynamic property than density, such as entropy and enthalpy, might be a better `control parameter' for glass formation, Bestul and Chang~\cite{1964_JCP_40_3731} took a first step forward to transform this loose connection of qualitative ideas and observations into a quantitative relationship between the entropy and the rate of relaxation by noticing a direct relationship between the excess entropy $S_{\mathrm{exc}}$ and the segmental structural relaxation time $\tau_{\alpha}$. Note that Bestul and Chang~\cite{1964_JCP_40_3731} defined an excess entropy by taking the solid reference state to be the glass state, while some authors later chose the crystalline state as the reference state,~\cite{1998_JCP_108_9016, 2006_JCP_124_024906} in line with the definition of Kauzmann.~\cite{1948_CR_43_219} This choice is contingent on what form of the solid exists for a given material, since some materials have no available, or even perhaps existing, crystalline form. Regardless of how $S_{\mathrm{exc}}$ is defined, this quantity only provides a rather rough estimate of $S_c$, which is unfortunately not a directly observable property. It should also be noted that there is another excess entropy defined as the difference of entropies between the liquid and gas states, which we designate as $\widehat{S}_{\mathrm{ex}}$ to avoid confusion with other definitions of the excess entropy discussed above. $\widehat{S}_{\mathrm{ex}}$ has attracted significant interest in relation to the dynamics of GF liquids, and Dyre~\cite{2018_JCP_149_210901} has recently provided a comprehensive review on this topic. We discuss this alternative theory of the dynamics of liquids built around $\widehat{S}_{\mathrm{ex}}$ in Section~\ref{Sec_Excess}.

Bestul and Chang~\cite{1964_JCP_40_3731} also introduced a practical method of estimating $S_c$ from $S_{\mathrm{exc}}$ from specific heat measurements, in addition to performing such measurements at high resolution. They clearly made many notable contributions to the emergence of a quantitative entropy theory of glass formation, and, accordingly, we believe that their pioneering contribution to understanding of the dynamics of GF liquids deserves greater recognition.

The seminal observations of Bestul and Chang~\cite{1964_JCP_40_3731} were later rationalized by AG~\cite{1965_JCP_43_139} with formal arguments that remain highly influential to the present. In the AG model,~\cite{1965_JCP_43_139} the relaxation of the fluids is assumed to be described by an extension of TST~\cite{Book_Eyring, 1940_JACS_62_3113, 1941_CR_28_301} with a $T$-dependent barrier height $\Delta G$,
\begin{equation}
	\tau_{\alpha} = \tau_0 \exp\left(\frac{\Delta G}{k_BT}\right),
\end{equation}
where $ \tau_0$ is the high $T$ limit of $\tau_{\alpha}$. This is the first postulate of the AG model. The central idea expressed by this extension of TST to describe GF liquids, which was widely accepted as a general formulation of liquid dynamics at the time, is that the dynamic slowing down of cooled liquids has its origin in the presence of dynamic clusters, which AG termed the `cooperatively rearranging regions' (CRR) to embody the prevailing idea of why simple TST `breaks down' at low $T$. AG were rather vague about the form of these clusters and about how they might be identified in practice, but they postulated that these hypothetical clusters should have properties consistent with experimental observations. In particular, AG made a second postulate that the average number of particles in these CRR should be proportional to the activation free energy $\Delta G$. The same hypothesis was introduced earlier by Mott~\cite{1951_Mott}; see the paper of Nachtreib and Handler~\cite{1955_JCP_23_1187} for a discussion of this model of the $T$ dependence of the activation energy and its application to understand the strongly non-Arrhenius $T$ dependence of the measured diffusion coefficient of white phosphorous. It is notable that Nachtreib and Handler~\cite{1955_JCP_23_1187} also invoked EEC with the melting point being the compensation temperature in a heated crystalline material. Finally, AG assumed that the average number of particles in these hypothetical CRR should be inversely proportional to the fluid configurational entropy $S_c$, thereby tying the concepts of growing collective motion in cooled liquids to the observed decreasing fluid entropy, whereupon they adopted an estimate of $S_c$ introduced by Bestul and Chang~\cite{1964_JCP_40_3731} to finish their model of the dynamics of glass formation.

We may translate the AG arguments into a more formal symbolic representation. AG first posited that the activation free energy $\Delta G_0$ of TST at high $T$ should be multiplied by a factor $z$, corresponding to the \textit{number} of molecules or segments participating in the abstract CRR. AG then further assumed that the size of the CRR should scale inversely with $S_c$, $z = S_c^*/S_c$, where $S_c^*$ is the high $T$ limit of $S_c$, a quantity that is also assumed to exist by AG.~\cite{1965_JCP_43_139} In other words, AG basically \textit{postulated} that the activation free energy $\Delta G$ is a $T$-dependent quantity, 
\begin{equation}
	\label{Eq_DeltaG}
	\Delta G(T) = z(T) \Delta G_0 = [S_c^* / S_c(T)] \Delta G_0,
\end{equation}
which reduces to standard TST at high $T$ where $\Delta G(T)$ reduces to $\Delta G_0$. AG were silent about the characteristic temperature or the $T$ range at which the change to non-cooperative liquid dynamics should occur, however. This physically attractive picture of glass formation then indicates that collective motion simply renormalizes the activation energy by a factor equal to the number of particles involved in the collective particle exchange motion, $z$, which basically rationalized the empirical relation of Bestul and Chang.~\cite{1964_JCP_40_3731}

It is evident that the AG model is clearly more a series of hypotheses than a real theory, but this model has provided a highly influential general conceptual framework for understanding the dynamics of GF liquids, despite its apparent theoretical weaknesses. The AG model has entered a new phase in recent years as MD simulations enable estimates of collective motion and calculations of the thermodynamic properties of polymeric and other GF liquids, so we can now test this model quantitatively and extend the model as necessary. It is truly remarkable how much of the AG theory has survived in the more modern and validated description of GF liquids, which is a testament to the deep physical understanding that AG had of the essential nature of GF liquids. In particular, the AG theory has held up remarkably well over the last 50 years in comparison with numerous experiments~\cite{1967_JCP_47_2802, 1997_JRNIST_102_171, 1998_JCP_108_9016, 2004_JCP_120_10640} and simulations~\cite{1999_PRL_83_3214, 2000_Nature_406_169, 2001_Nature_409_164, 2001_Nature_412_514, 2001_PRE_63_041201, 2001_JCP_114_9069, 2002_PRE_65_041205, 2003_BC_105_573, 2004_PRE_69_041503, 2004_JPCM_16_L489, 2005_JPCB_109_15068, 2007_JPCM_19_256207, 2009_PRL_103_225701, 2011_JCP_135_194503, 2012_PRL_109_095705, 2013_JCP_138_12A541, 2013_EPJE_36_141, 2018_MP_116_3366} of diverse GF liquids. Efforts have also been made to place the AG model on a sounder theoretical foundation by extending TST to account for barrier crossing processes that involve many particles,~\cite{1994_Science_266_425, 2003_JCP_119_9088, 2004_JCP_121_7347, 2014_JCP_141_141102} basically formalizing the heuristic ideas introduced long before Mott and AG. The development of TST by Bonzel~\cite{1970_SS_21_45} in the context of understanding the non-Arrhenius dynamics of crystalline materials is particularly a detailed reformulation of TST to treat this type of problem in an analytic framework, and the theoretical framework developed by Freed~\cite{2014_JCP_141_141102} shares many mathematical features, but this theory is not formulated in a chemically specific way, so we view this theory as a promising work in progress rather than a complete theory. The RFOT theory~\cite{1987_PRB_36_8552, 1989_PRA_40_1045, 2004_JCP_121_7347, 2007_ARPC_58_235} also offers great potential for developing a sounder foundation for the AG model, as we discuss in Section~\ref{Sec_RFOT}.

While the AG theory~\cite{1965_JCP_43_139} represents an important step in our understanding of the dynamics of GF liquids, in the sense that it provides a rationale for a quantitative link between $\tau_{\alpha}$ and $S_c$, there are several drawbacks of the AG model that limit its predictive capacity. First, no effort was made by AG to calculate $S_c$ from theoretical considerations. Originally, AG~\cite{1965_JCP_43_139} argued for approximating $S_c$ by the excess entropy $S_{\mathrm{exc}}$ to make contact with the VFT equation~\cite{1921_PZ_22_645, 1925_JACS_8_339, 1926_ZAAC_156_245} and to rationalize the correlation of Bestul and Chang,~\cite{1964_JCP_40_3731} while Martinez and Angell~\cite{2001_Nature_410_663} emphasized the use of $S_{\mathrm{exc}}$ as a thermodynamic predictor of the dynamics of GF liquids in the spirit of the AG model. Second, the combination of the AG relation with a thermodynamic model, such as the GD model,~\cite{1958_JCP_28_373} was never made, so there have been no previous predictions regarding the influence of molecular structure and interactions on polymer glass formation, a problem of both fundamental and practical importance. Thus, the AG theory has been essentially used as a framework for understanding qualitative features of glass formation, such as the origin of the VFT equation~\cite{1921_PZ_22_645, 1925_JACS_8_339, 1926_ZAAC_156_245} and the origin of fragility,~\cite{1991_JNCS_131_13, 1995_Science_267_1924} a measure of the degree of non-Arrhenius relaxation. The AG model in its original form has limited value in relation to the design of new materials and prior predictions of the properties of materials in terms of given molecular parameters. Moreover, AG offered no quantitative picture of the structure and polydispersity of the hypothetical CRR so that the original AG theory has some serious shortcomings. Finally, as we have mentioned in Section~\ref{Sec_TST}, AG made the completely unwarranted assumption of neglecting the activation entropy $\Delta S_0$, presumably for reasons of mathematical expediency. It is known from even standard TST that $\Delta G_0$ contains an entropic contribution that can sometimes be quite large.~\cite{1946_JCP_14_591} This is just one of the things that must be fixed in any extended model of TST that incorporates the ideas of AG for the purpose of making quantitative comparisons with relaxation data obtained from real GF materials.

We again emphasize that despite the above problems in the classical theories of GD~\cite{1958_JCP_28_373} and AG,~\cite{1965_JCP_43_139} these models of glass formation have been quite influential in understanding the nature of glass formation and general trends of $T_{\mathrm{g}}$ with molecular structure. In particular, these classical theories~\cite{1958_JCP_28_373, 1965_JCP_43_139} have set the stage for the development of new models probing the origin of the slowing down of the dynamics of GF liquids from a molecular perspective. In the following, we discuss the development of the entropy theory along two lines of research. The first line of research merges the AG relation with explicit computations of the configurational entropy based on a statistical mechanical model of polymers, yielding the GET of segmental relaxation in polymeric GF liquids,~\cite{2008_ACP_137_125} which we discuss in Section~\ref{Sec_GET}. The second thrust of our work is built on the platform of MD simulations and tools drawn from liquid state theory to quantify collective motion and calculate structural relaxation time, along with other dynamical and thermodynamic properties. The identification of well-defined cooperative motion in cooled liquids that appears to conform exactly to the main hypotheses of AG~\cite{1965_JCP_43_139} and the finding that the geometrical properties of these clusters accord with equilibrium polymerization~\cite{1999_JCP_111_7116, 2006_JCP_125_144907, 2008_JCP_128_224901, 2014_JCP_140_204509} led to the development of the highly quantitative string model of glass formation,~\cite{2014_JCP_140_204509} as discussed in Section~\ref{Sec_String}. The development of these more recent works has continuously addressed the shortcomings of the classical GD~\cite{1958_JCP_28_373} and AG~\cite{1965_JCP_43_139} models. Again we underscore that like all extant theories of glass formation, the entropy theory of glass formation remains in an exploratory stage so that some `loose ends' are to be expected. As a complement to our works on the GET and the string model, we also discuss the RFOT~\cite{1987_PRB_36_8552, 1989_PRA_40_1045, 2004_JCP_121_7347, 2007_ARPC_58_235} and the Rosenfeld and excess-entropy scaling approaches~\cite{1977_PRA_15_2545, 1999_JPCM_11_5415, 2018_JCP_149_210901} to liquid dynamics since these models are likewise based on the fluid entropy, and may thus be characterized as alternative entropy theories of glass formation.

\section{\label{Sec_GET}Generalized Entropy Theory (GET) of Polymer Glass Formation}

\subsection{GET Essentials}

The GET~\cite{2008_ACP_137_125} takes advantage of the LCT~\cite{1998_ACP_103_335, 2014_JCP_141_044909} for the thermodynamics of polymer systems, which significantly extends the models of Flory~\cite{1956_PRSLA_234_60} and GD~\cite{1958_JCP_28_373} in the sense that it includes explicitly the description of monomer structure. The LCT represents individual monomers in terms of a set of united-atom groups with each occupying a single lattice site of volume $V_{\text{cell}}$ and enables thermodynamic descriptions of the basic molecular characteristics of polymers, including excluded volume interactions, chain connectivity, cohesive interaction strength, chain stiffness, and monomer molecular structure. For simplicity, we discuss the LCT for single-component polymer melts with the reminder that this theory enables the description of multi-component polymer systems,~\cite{1998_ACP_103_335} such as polymer blends. Extensions of the GET has been made for the glass formation of polymers with additives~\cite{2010_JCP_132_084504} and polymer blends~\cite{2014_JCP_140_244905}.

For a compressible polymer melt, the LCT yields an analytic expression for the Helmholtz free energy $f$ per lattice site in the following general form,~\cite{1998_ACP_103_335, 2014_JCP_141_044909}
\begin{equation}
	\label{Eq_FreeEnergy}
	\beta f = \beta f^{mf} - \sum_{i=1}^6C_i\phi^i,
\end{equation}
where $\beta = (k_BT)^{-1}$ and $\phi$ is the lattice `filling fraction' defined by the ratio of the total number of united-atom groups to the total number $N_l$ of lattice sites. The term $\beta f^{mf}$ corresponds to a contribution from the zeroth-order mean-field approximation,
\begin{eqnarray}
	\beta f^{mf} = \frac{\phi}{M} \ln\left(\frac{2\phi}{z^{L}M}\right) + \phi\left(1-\frac{1}{M}\right) + (1-\phi)\ln(1-\phi) - \phi \frac{N_{2i}}{M}\ln(z_b),
\end{eqnarray}
where $M$ is the molecular mass corresponding to the total number of united-atom groups in a single chain, $z$ represents the lattice coordination number, which is related to the spatial dimension via $z=2d$ for a $d$-dimensional hypercubic lattice, $L$ is the number of subchains, $N_{2i}$ is the number of sequential bond pairs lying along the identical subchains, $z_b = (z/2 - 1)\exp(-\beta E_b) + 1$ with $E_b$ being the bending energy parameter. The second term on the right-hand side of eq~\ref{Eq_FreeEnergy} represents corrections to the zeroth-order mean-field contribution, where the coefficients $C_i$ $(i=1, ..., 6)$ are obtained by collecting terms corresponding to a given power of $\phi$,
\begin{equation}
	C_i = C_{i, 0} + C_{i, \epsilon}(\beta\epsilon) + C_{i, \epsilon^2}(\beta\epsilon)^2.
\end{equation}
Here, $\epsilon$ is the microscopic cohesive energy parameter describing the net attractive van der Waals interactions between nearest-neighbor united-atom groups. $C_{i, 0}$, $C_{i, \epsilon}$, and $C_{i, \epsilon^2}$ are tabulated in the Appendix A of ref~\citenum{2014_JCP_141_044909} and are generally functions of $z$, $T$, $E_b$, $M$, as well as a set of geometrical indices that reflect the size, shape, and bonding patterns of monomers. The free energy enables computations of all thermodynamic properties.

\begin{figure*}[htb!]
	\centering
	\includegraphics[angle=0,width=0.475\textwidth]{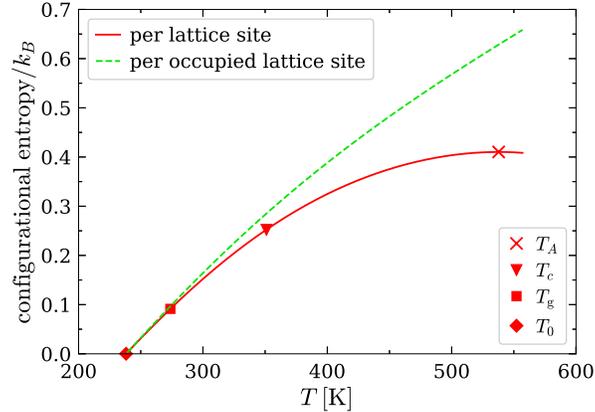}
	\caption{\label{Fig_Sc}Illustration of the $T$ dependence of the two different definitions of configurational entropy predicted by the generalized entropy theory (GET). The figure displays the configurational entropy per lattice site versus per occupied lattice site as a function of $T$ at zero pressure. The symbols cross, triangle, square, and diamond indicate the positions of the onset $T_A$, crossover $T_c$, glass transition $T_{\text{g}}$, and ideal glass transition temperatures $T_0$, respectively. The vanishing of $s_c$ for semiflexible polymer melts is probably spurious due to the high $T$ expansion associated with the treatment of chain stiffness in the lattice cluster theory. The calculations are performed for a melt of chains with the structure of polypropylene (PP), where the cell volume parameter is $V_{\text{cell}} = 2.5^3 \text{\AA}^3$, the chain length is $N_c = 8000$, the cohesive energy parameter is $\epsilon/k_B = 200$ K, and the bending energy parameter is $E_b/k_B = 600$ K. The above parameter set is utilized for the calculations considered in Figures~\ref{Fig_Tx},~\ref{Fig_Scaling},~\ref{Fig_Pressure}, and~\ref{Fig_HS}.}
\end{figure*}

Before the development of the GET,~\cite{2008_ACP_137_125} the LCT has been primarily used to explain enigmatic observations in polymer blends,~\cite{2005_APS_183_63} including the existence of an entropic contribution to the effective Flory interaction parameter $\chi$. The generalization of the LCT to consider polymer glass formation was made possible first by arriving at an analytical expression~\cite{2003_JCP_119_5730} for the configurational entropy in terms of the natural logarithm of the density of states $\Omega(U)$ in the microcanonical ensemble,~\cite{1958_JCP_28_373} 
\begin{equation}
	S_c(T) = k_B \ln\Omega(U)|_{U=U(T)},
\end{equation}
where $U(T)$ is the internal energy at the temperature $T$. The explicit form for $S_c$ derived from the LCT turns out to be quite complicated mathematically,~\cite{2003_JCP_119_5730} but a more recent work~\cite{2014_JCP_141_234903} indicates that $S_c$ approximates very closely the ordinary entropy $S$ calculated from the free energy $F = N_l f$, i.e., $S_c \approx S = -\partial F / \partial T |_{\phi}$, greatly simplifying the evaluation of $S_c$ from the LCT. While the ability to calculate $S_c$ within the LCT already enables investigating the ideal glass transition temperature $T_0$, which is the primary focus in the GD theory,~\cite{1958_JCP_28_373} the GET combines the LCT with the AG relation~\cite{1965_JCP_43_139} to explore other key aspects of polymer glass formation. This is achieved by realizing that one has to utilize the physically consistent configurational entropy when invoking the AG relation. We discuss below what we mean by `physically consistent'. For simplicity, here we perform illustrative calculations for a melt of chains with the structure of polypropylene (PP) at zero pressure, where a single bending energy $E_b$ is required for the backbone. As shown in Figure~\ref{Fig_Sc}, the configurational entropy per occupied lattice site increases progressively with increasing $T$, showing no tendency to level off at high $T$, as AG suggested.~\cite{1965_JCP_43_139} Note that AG did not consider fluids at constant pressure.~\cite{1965_JCP_43_139} On the other hand, the $T$ dependence of the configurational entropy per lattice site or the \textit{configurational entropy density} ($s_c = S_c/N_l$) does plateau at high $T$, making it a sensible candidate for use in the AG relation. This point has been elucidated in detail in ref~\citenum{2008_ACP_137_125}, and we briefly discuss some of the ramifications of the $T$ dependence of the configurational entropy density $s_c$ that serves to define the characteristic temperatures within this model of glass formation.

\begin{figure*}[htb!]
	\centering
	\includegraphics[angle=0,width=0.95\textwidth]{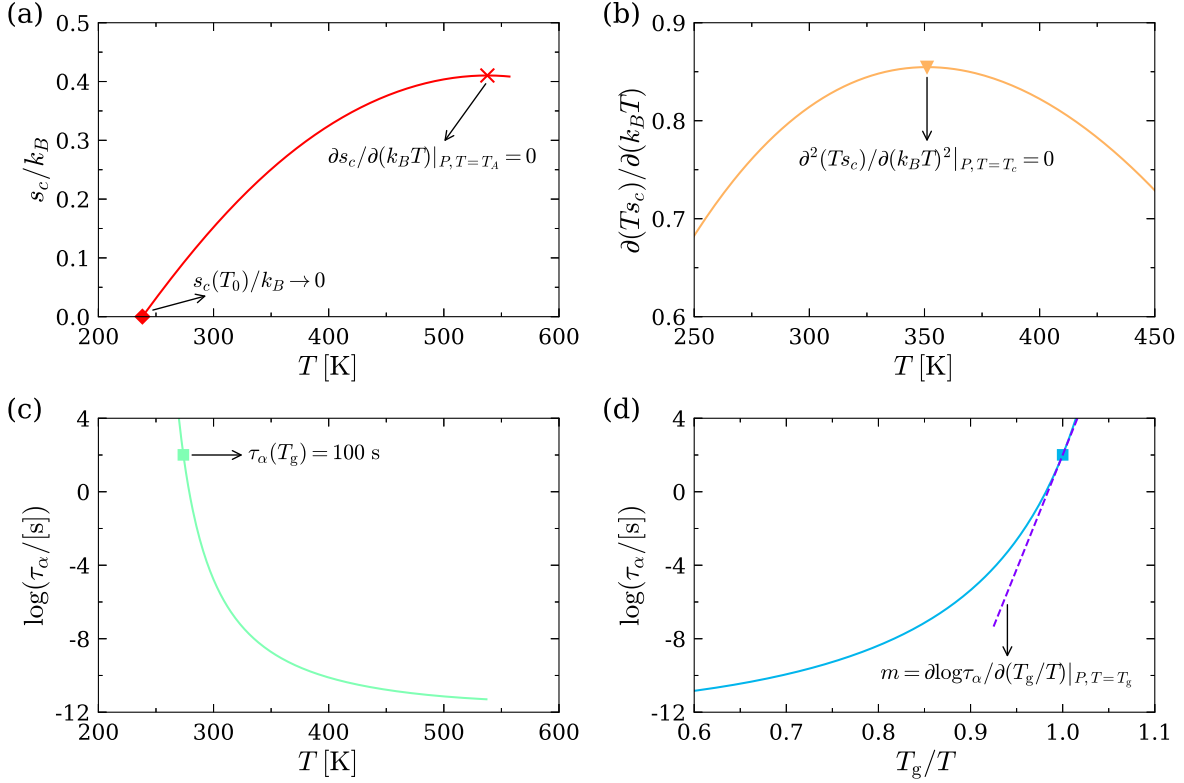}
	\caption{\label{Fig_Tx}Determination of the characteristic temperatures and fragility of glass formation in the GET. (a) Configurational entropy density $s_c/k_B$ versus $T$. $T_A$ and $T_0$ are determined from the temperatures at which $s_c$ displays a maximum and is extrapolated to zero, respectively. (b) $\partial (Ts_c)/\partial (k_BT)$ versus $T$. $T_c$ is determined from the temperature at which the $T$ dependence of $Ts_c$ displays an inflection point. (c) $\log \tau_{\alpha} $ versus $T$. $T_{\mathrm{g}}$ is determined from the temperature at which $\tau_{\alpha} = 100$ s. (d) Angell plot of $\log \tau_{\alpha}$. The fragility index $m$ is determined from its standard definition.~\cite{1991_JNCS_131_13, 1995_Science_267_1924} The calculations are performed at zero pressure and utilize the parameter set specified in the caption of Figure~\ref{Fig_Sc}.}
\end{figure*}

The GET predicts that polymer glass formation is a broad thermodynamic transition with four characteristic temperatures.~\cite{2008_ACP_137_125} Figure~\ref{Fig_Tx} summarizes the determinations of these characteristic temperatures, along with the fragility index. The first one is called the Arrhenius or onset temperature, $T_A$, which signals the onset of non-Arrhenius behavior of $\tau_{\alpha}$. In particular, $s_c$ exhibits a maximum $s_c^*$ at a temperature identified with $T_A$ (Figure~\ref{Fig_Tx}a), and the maximum $s_c^*$ is chosen to be the high $T$ limit of $s_c$ appearing in the AG relation.~\cite{1965_JCP_43_139} Following GD,~\cite{1958_JCP_28_373} the hypothetical `ideal glass transition temperature' $T_0$ is identified from the condition of $s_c(T_0) \rightarrow 0$ (Figure~\ref{Fig_Tx}a). The crossover temperature $T_c$ separates two regimes of $T$ with qualitatively different dependences of $\tau_{\alpha}$ on $T$ and is estimated from an inflection point in the $T$ dependence of $Ts_c$ (Figure~\ref{Fig_Tx}b). To determine $\tau_{\alpha}$ and thus $T_{\text{g}}$, the GET combines information on $s_c$ with the AG relation,~\cite{1965_JCP_43_139}
\begin{equation}
	\label{Eq_AG}
	\tau_{\alpha} = \tau_0 \exp\left( \frac{\Delta G_0}{k_BT} \frac{s_c^*}{s_c} \right),
\end{equation}
where $s_c^*$ and $s_c$ are inputs directly determined from the LCT, and $\tau_0$ and $\Delta G_0$ are the high $T$ limit of $\tau_{\alpha}$ and the high $T$ activation free energy of TST, respectively.

Because the LCT does not provide a method for evaluating $\tau_0$ and $\Delta G_0$, these fundamental parameters have been obtained empirically so that assumptions are necessarily invoked.~\cite{2008_ACP_137_125} The GET takes $\tau_0 = 10^{-13}$ s, following typical experimentally reported values for polymers~\cite{2003_PRE_67_031507} and arguments based on TST~\cite{1941_CR_28_301, Book_Eyring} suggest that this time scale should be on the order of an inverse molecular vibrational frequency. This time scale can be refined by calculating the decay time of the velocity autocorrelation function. In practice, the variations of $\tau_0$ with molecular parameters such as chain stiffness~\cite{2020_Mac_53_4796} have been found to be rather limited. As anticipated from TST,~\cite{1941_CR_28_301, Book_Eyring} there should be both enthalpic and entropic contributions to the activation free energy (eq~\ref{Eq_DeltaGo}). However, as $\Delta S_0$ is notoriously difficult to estimate from a theoretical viewpoint, the GET follows the original treatment by AG~\cite{1965_JCP_43_139} and assumes that the entropic contribution to the activation free energy is negligible,~\cite{2008_ACP_137_125} i.e., $\Delta S_0 = 0$. The GET also assumes based on experimental evidence~\cite{2008_ACP_137_125} that $\Delta G_0 = \Delta H_0 \approx 6 k_B T_c$. We have discussed this relation at length in previous works,~\cite{2008_ACP_137_125, 2020_Mac_53_9678} so the reader is referred to these necessarily technical discussions for details. We emphasize here that these simplifying assumptions are not required by the GET, but they allow for the prediction of polymer properties without any further information than the determination of molecular parameters (e.g., monomer structure, cohesive interaction strength, and chain stiffness) and thermodynamic parameters (e.g., $T$ and $P$) required to specify the thermodynamics of polymer melts, properties that the LCT has been established to describe with good reliability. The enthalpy $\Delta H_0$ and entropy $\Delta S_0$ of activation can be independently determined from either simulation or measurement, and in Section~\ref{Sec_GET_Challenge} we explore an extension of the GET where these parameters are determined from experiment to gain some insight into general trends in these energetic parameters. Jeong and Douglas~\cite{2015_JCP_143_144905} have studied the dependence of these energetic parameters on molecular mass in unentangled linear alkane chains by MD simulations, for which there are substantial experimental data to compare to and validate the simulation results. We expect this activity of studying the activation free energy parameters of polymers and GF liquids to expand in the future, given the fundamental importance of these energetic parameters revealed by recent simulation and experimental studies.~\cite{2014_NatCommun_5_4163, 2015_JCP_142_234907, 2015_PNAS_112_2966, 2017_SoftMatter_13_1190, 2016_MacroLett_5_1375, 2017_Mac_50_2585, 2020_Mac_53_4796, 2020_Mac_53_9678, 2017_JCP_147_154902, 2018_NanoLett_18_7441}

Once $\tau_{\alpha}$ is computed as a function of $T$, then $T_{\text{g}}$ can be determined using the common empirical definition, $\tau_{\alpha}(T_{\text{g}}) = 100$ s (Figure~\ref{Fig_Tx}c). The fragility index proposed by Angell,~\cite{1991_JNCS_131_13, 1995_Science_267_1924}
\begin{equation}
	\label{Eq_Angell}
	m = \left. \frac{\partial \log\tau_{\alpha}}{\partial (T_{\text{g}}/T)} \right|_{P, T=T_{\text{g}}},
\end{equation}
is also readily obtained. Alternatively, we can determine $m$ by fitting $\tau_{\alpha}$ in the $T$ range between $T_c$ and $T_{\text{g}}$ to the VFT equation (eq~\ref{Eq_VFT}).~\cite{1921_PZ_22_645, 1925_JACS_8_339, 1926_ZAAC_156_245} Since the VFT equation turns out to be a rather good description for the $T$ variation of $\tau_{\alpha}$ calculated from the GET in the $T$ range between $T_c$ and $T_{\text{g}}$, $T_{\mathrm{VFT}}$ is basically equivalent to $T_0$ so that we do not differentiate them. $m$ is related to $D$ simply via $m = DT_{\text{g}}T_{\mathrm{VFT}} / [(T_{\text{g}} - T_{\mathrm{VFT}}) \ln10]$.

Because the LCT~\cite{1998_ACP_103_335, 2014_JCP_141_044909} is a powerful theoretical framework for addressing the thermodynamics of a vast array of polymer systems, a number of important problems relevant to polymer glass formation can be investigated by the GET.~\cite{2008_ACP_137_125} For instance, the GET has been demonstrated to quantitatively describe the characteristic properties of poly($\alpha$-olefins),~\cite{2009_JCP_131_114905} to clarify the meaning of activation volume of polymer liquids,~\cite{2020_Mac_53_7239} to elucidate the influence of cohesive energy and chain stiffness on polymer glass formation,~\cite{2014_Mac_47_6990,2015_Mac_48_2333, 2014_JCP_141_234903} and to better understand a variety of phenomena related to polymer glass formation, such as plasticization and antiplasticization of polymer melts by molecular additives,~\cite{2010_JCP_132_084504} density-temperature scaling of the segmental dynamics of polymer melts,~\cite{2013_JCP_138_234501} two glass transitions in miscible polymer blends,~\cite{2014_JCP_140_244905} and the interpretation of the `universal' WLF parameters of polymer liquids~\cite{2015_JCP_142_014905} and the related observation that the structural relaxation time depends nearly universally on $T - T_{\mathrm{g}}$ in related families of GF liquids.~\cite{2011_PRL_106_115702} The LCT has also been extended to model polymers with specific interactions~\cite{2014_JCP_141_044909, 2015_Mac_48_2333} where cohesive interaction strengths are different between united-atom groups and telechelic polymers~\cite{2016_JCP_144_214903, 2015_JCP_143_024901, 2015_JCP_143_024902} where associative groups at the chain ends have strong interactions, enabling predictions of glass formation in these polymer systems when combined with the AG model.~\cite{1965_JCP_43_139} It is also worth mentioning that the GET has been shown to be useful for coarse-grained modeling of polymeric GF materials.~\cite{2019_SciAdv_5_eaav4683, 2020_MTS_3_234509} Thus, the GET has the unique capacity to investigate the influence of structural details and interactions on polymer glass formation and better understand a variety of important problems in polymer glass formation.

As we have shown above, a particular advantage of the GET is that the characteristic temperatures ($T_A$, $T_c$, $T_{\mathrm{g}}$, and $T_0$) of glass formation can be calculated precisely in the theory because of the analytic definitions of these fundamental characteristic temperatures of glass formation. This stands in contrast to experimental and computational studies of glass formation, where large uncertainties often arise in the estimates of these temperatures. It is also interesting to note that the ECNLE theory of Schweizer and coworkers~\cite{2014_JCP_140_194506, 2014_JCP_140_194507, 2015_Mac_48_1901} allows for predictions of the characteristic temperatures of glass formation. It would be quite interesting to compare predictions of this model with the GET to better understand the physical meaning of the characteristic temperatures of glass formation.

\subsection{\label{Sec_ScGET}Temperature Dependence of Configurational Entropy}

From the AG relation (eq~\ref{Eq_AG}), it is evident that the $T$ dependence of $s_c$, along with $\Delta G_0$, determines the dynamics of GF liquids within the GET. It is thus crucial to know how $s_c$ varies with $T$ and then better comprehend the implications of this information on glass formation. While GD~\cite{1958_JCP_28_373} focused primarily on a low $T$ regime where the configurational entropy extrapolates to zero and AG~\cite{1965_JCP_43_139} essentially made some guesses about the $T$ dependence of the configurational entropy, one advantage of the GET~\cite{2008_ACP_137_125} is that it allows us to analyze $s_c$ over the entire $T$ regime of glass formation in great detail. In particular, the GET has made detailed predictions for the variation of $s_c$ with $T$ in both the high and low $T$ regimes of glass formation. This has been discussed at length in the original review on the GET.~\cite{2008_ACP_137_125} However, our more recent calculations~\cite{2016_ACP_161_443, 2016_JCP_145_234509} based on the GET have revealed certain features in the $T$ dependence of $s_c$ that are not anticipated from the original GET. Here, we summarize these features, along with those basic aspects identified in the original GET.~\cite{2008_ACP_137_125}

\begin{figure*}[htb!]
	\centering
	\includegraphics[angle=0,width=0.975\textwidth]{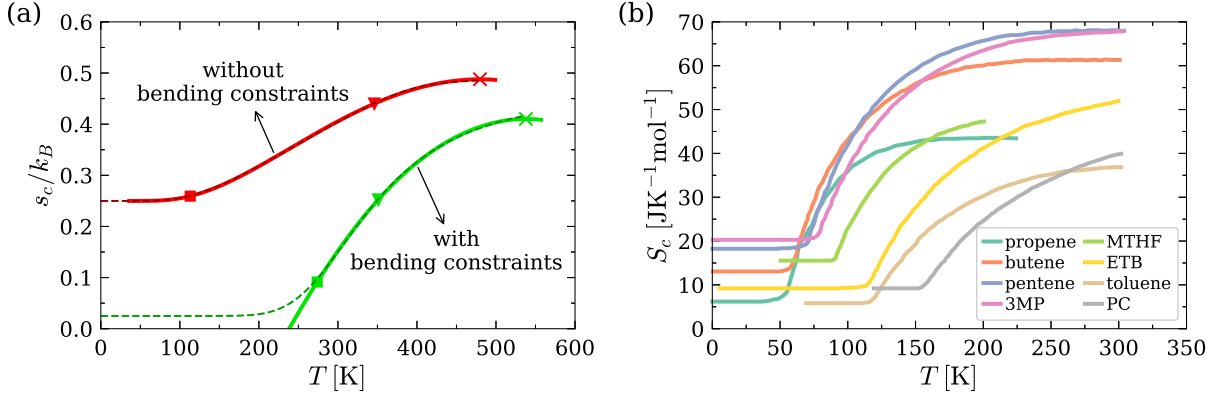}
	\caption{\label{Fig_Sr}$T$ dependence of the configurational entropy in the entire $T$ regime of glass formation. (a) $s_c/k_B$ as a function of $T$ predicted by the GET for polymer melts with and without bending constraints at zero pressure, where $E_b/k_B = 600$ K and $0$ K, respectively. Dashed lines are the descriptions based on eq~\ref{Eq_TwoState}. The symbols cross, triangle, and square indicate the positions of $T_A$, $T_c$, and $T_{\text{g}}$, respectively. The calculations are performed for a melt of chains with the structure of PP at zero pressure, where $V_{\text{cell}} = 2.5^3 \text{\AA}^3$, $N_c = 8000$, and $\epsilon/k_B = 200$ K. (b) Configurational entropy $S_c$ determined from the heat capacity measurements as a function of $T$ for various molecular liquids. 3MP, MTHF, ETB, and PC are short for 3-methylpentane, 2-methyl-tetrahedrofuran, ethylbenzene, and propylene carbonate, respectively. Panel (b) was adapted with permission from ref~\citenum{2012_PRL_109_045701}.}
\end{figure*}

We begin by noting the behavior of $s_c$ at $T$ well below $T_{\mathrm{g}}$. While the vanishing of $s_c$ at low $T$ shown in Figures~\ref{Fig_Sc} and~\ref{Fig_Tx} was found to be a typical behavior in the original GET for polymer melts with bending constraints,~\cite{2008_ACP_137_125} as in the GD model,~\cite{1958_JCP_28_373} our more recent calculations~\cite{2016_ACP_161_443, 2016_JCP_145_234509} have demonstrated that this potentially spurious behavior can be avoided in certain models. Specifically, $s_c$ was found to level off rather than vanish at low $T$ in polymer models with variable spatial dimension $d$ when $d$ is above a critical value in the vicinity of $d = 8$~\cite{2016_ACP_161_443} and in fully flexible polymer models in the physical dimension of $d = 3$.~\cite{2016_JCP_145_234509} Here, we take the fully flexible polymer model to illustrate our point, as shown in Figure~\ref{Fig_Sr}a, where we compare the $T$ dependence of $s_c$ of a polymer melt with bending constraints to that without bending constraints. As can be seen, $s_c$ for fully flexible polymer melts no longer vanishes, but instead evidently approaches a positive constant $s_r$ at low $T$, providing a different perspective for the nature of glass formation than the original GD theory.~\cite{1958_JCP_28_373} This finding is consistent with the suggestion of DiMarzio in his later work~\cite{1997_JRNIST_102_135} that $S_c$ should become `critically small' rather than vanishing at the glass transition. The explicit expression for $s_r$ from the GET has been derived in ref~\citenum{2016_JCP_145_234509} as a function of molecular parameters, such as chain length and monomer structure. Interestingly, $S_c$ has also been estimated to be effectively constant at low $T$ in the heat capacity measurements of a number of molecular liquids by Yamamuro and coworkers.~\cite{2012_PRL_109_045701, 1998_JPCB_102_1605} Although this type of observation is often interpreted to be an inherently nonequilibrium phenomenon, this result is highly attractive as a possibly equilibrium phenomenon from a viewpoint of the entropy theory, so we reproduce the excess entropy data of ref~\citenum{2012_PRL_109_045701} for an array of materials in Figure~\ref{Fig_Sr}b.

Philosophically, the presence of a positive residual configurational entropy has important ramifications for glass formation within the entropy theory. First, the Kauzmann's `entropy crisis'~\cite{1948_CR_43_219} is naturally avoided in the GET for fully flexible polymer melts. Moreover, when combined with the AG relation in eq~\ref{Eq_AG}, a finite residual configurational entropy evidently indicates that structural relaxation becomes of an Arrhenius form. In this case, $\tau_{\alpha}$ does not diverge at any finite $T$, and thus, the material is a `liquid' at low $T$ from a mathematical standpoint. However, since the activation energy $\Delta G(T)$ in the low $T$ regime below $T_{\mathrm{g}}$ is higher than that which the fluid has at high $T$ above $T_A$ and since liquids are cooled to such low $T$, the relaxation times at low $T$ are much larger than those in the high $T$ regime, and by all practical measures, the polymer melt in the low $T$ regime can be considered to be a `solid' in a rheological sense. We are thus tempted to term the low $T$ state as being a type of a `glass'. O'Connell and McKenna~\cite{1999_JCP_110_11054} and Novikov and Sokolov~\cite{2015_PRE_92_062304} have tentatively suggested that relaxation in GF materials generally approaches an Arrhenius behavior at low $T$, consistent with the prediction of the GET for fully flexible polymers. While our finding of a finite residual configurational entropy and a corresponding highly viscous `glass state' at low $T$ in flexible polymer melts provides a clear mechanism of how the `entropy crisis' might be avoided in real polymer melts and a counterexample to the claim that the AG model predicts a diverging structural relaxation time at a finite low $T$, we must admit that it does not imply that $s_c$ always exhibits a plateau at low $T$. There is also no doubt that it would be extremely difficult for materials to reach equilibrium over `reasonable' timescales at low $T$ due to the very long relaxation times, making it quite challenging to separate equilibrium from non-equilibrium contributions to the `effective' configurational entropy. This situation makes the utility of the GET of uncertain value below $T_{\mathrm{g}}$.

We can analyze the variation of $s_c$ with $T$ more quantitatively. In particular, we have shown in ref~\citenum{2016_JCP_145_234509} that $T$ dependence of $s_c$ can be reasonably described by the following empirical equation describing the generally sigmoidal variation of $s_c$,
\begin{equation}
	\label{Eq_TwoState}
	s_c = s_r + (s_c^* - s_r)[1 + \exp(a + b/T)]^c,
\end{equation}
where the constants $a$, $b$, and $c$ are obtained by numerically fitting $(s_c-s_r)/(s_c^* - s_r)$ as a function of $T$ since this normalized quantity varies from zero to unity as $T$ varies, thereby motivating the fitting functional form, $[1 + \exp(a + b/T)]^c$. The description based on eq~\ref{Eq_TwoState} is shown as a dashed line in Figure~\ref{Fig_Sr}a, where we see that this functional form provides an excellent description of $s_c(T)$ for polymer melts without bending constraints over the entire $T$ regime of glass formation. The origin of eq~\ref{Eq_TwoState}, and a similar expression utilized in our recent study of polymer coarse-graining,~\cite{2019_SciAdv_5_eaav4683} can be traced back to the string model of glass formation,~\cite{2014_JCP_140_204509} which establishes a relation between emergent elasticity and cooperative motion in polymeric GF liquids~\cite{2015_PNAS_112_2966} and a quantitative inverse relation between the extent of cooperative motion and the configurational entropy,~\cite{2013_JCP_138_12A541} as suggested by AG.~\cite{1965_JCP_43_139} A relation between the activation free energy and material stiffness is further found to qualitatively accord with the shoving model of Dyre and coworkers.~\cite{1996_PRB_53_2171} Motivated by these observations and arguments, eq~\ref{Eq_TwoState} is obtained by first invoking a simple two-state model of the shear modulus $G$ of amorphous materials,~\cite{2010_SoftMatter_6_3548} $G(T) / G(0) = 1/ [1 + \exp(a + b / T)]$, and then by utilizing the observation that the activation free energy for relaxation scales approximately as a positive power of $G$ in a model GF material.~\cite{2016_JSM_054048} While these arguments are admittedly intuitive, eq~\ref{Eq_TwoState} turns out to provide a good numerical approximation for $s_c(T)$ calculated from the full GET, whose complexity does not lend itself readily to the simple analytic forms that seem to be highly desired by experimentalists and engineers. Since the vanishing of $s_c$ in polymer melts with bending constraints is likely an artifact of the high $T$ expansion involved in the lattice model calculations for the free energy when the bending constraints are introduced, we may also invoke eq~\ref{Eq_TwoState} to describe the variation of $s_c$ with $T$ in polymer melts with bending constraints, for which eq~\ref{Eq_TwoState} is also a reasonable approximation, as evidenced in Figure~\ref{Fig_Sr}a. Note that we choose a somewhat arbitrary value of $s_r/k_B = 0.025$ for the polymer melt with bending constraints shown in Figure~\ref{Fig_Sr}a.

Given the good description of eq~\ref{Eq_TwoState} for $s_c(T)$ predicted by the GET, we can suggest the following functional form for $\tau_{\alpha}$,~\cite{2016_JCP_145_234509}
\begin{equation}
	\label{Eq_TauT}
	\tau_{\alpha} = \tau_0\exp\left\{\frac{\Delta G_0 / k_BT}{\delta s + (1 - \delta s)[1 + \exp(a + b/T) ]^c}\right\},
\end{equation}
where $\delta s \equiv s_r / s_c^*$ is the ratio of the activation energy in the high $T$ regime to that in the low $T$ regime. This approximation is expected to work in the entire $T$ regime of glass formation below $T_A$.

\begin{figure*}[htb!]
	\centering
	\includegraphics[angle=0,width=0.975\textwidth]{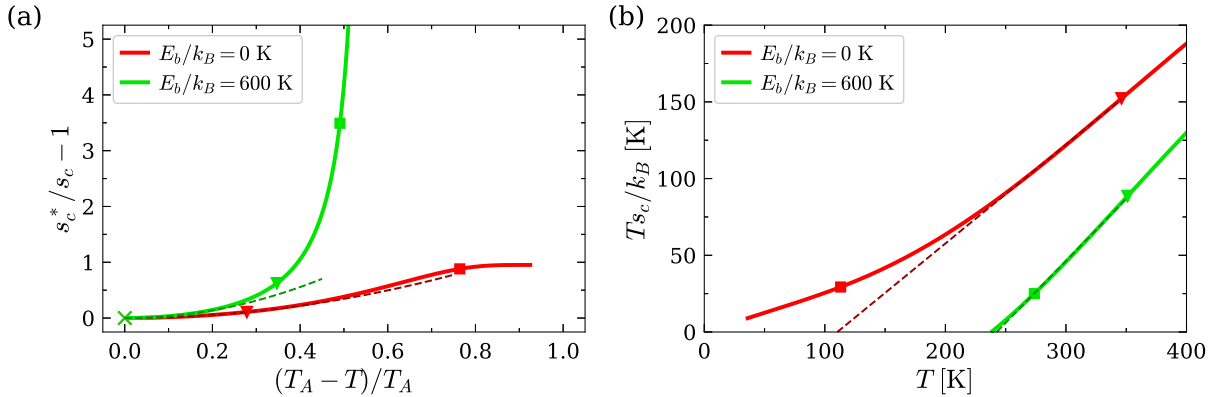}
	\caption{\label{Fig_FitSc}$T$ dependence of the configurational entropy in the high and low $T$ regimes of glass formation predicted by the GET. (a) $s_c^*/s_c - 1$ versus $(T_A - T)/T_A$ and (b) $Ts_c/k_B$ versus $T$ for polymer melts with $E_b/k_B = 0$ K and $600$ K, respectively. Dashed lines in panels (a) and (b) correspond to eqs~\ref{Eq_HighT} and~\ref{Eq_LowT}, respectively. The symbols cross, triangle, and square indicate the positions of $T_A$, $T_c$, and $T_{\text{g}}$, respectively. The calculations are performed for a melt of chains with the structure of PP at zero pressure, where $V_{\text{cell}} = 2.5^3 \text{\AA}^3$, $N_c = 8000$, and $\epsilon/k_B = 200$ K.}
\end{figure*}

While the mathematical forms given by eqs~\ref{Eq_TwoState} and~\ref{Eq_TauT} are a bit complicated in comparison to the VFT relation (eq~\ref{Eq_VFT}) and may not be utilized readily for describing experimental and simulation results, the GET predicts that the functional form for describing the $T$ variation of $s_c$ can be simplified greatly when we focus on certain \textit{restricted} regimes of glass formation. For instance, in the high $T$ regime of glass formation between $T_A$ and $T_c$, the GET predicts a simple functional form for $s_c(T)$ as,~\cite{2008_ACP_137_125}
\begin{equation}
	\label{Eq_HighT}
	s_c^*/s_c - 1 = C_s [(T_A - T)/T_A]^2,\ T_c < T_A - 100\ \mathrm{K} < T < T_A,
\end{equation}
where the quantity $C_s$ measures the steepness of the $T$ dependence of $s_c$. $C_s$ appears to provide a measure of fragility in the high $T$ regime of glass formation where the VFT equation is \textit{not} valid. Figure~\ref{Fig_FitSc}a indicates that eq~\ref{Eq_HighT} provides a good description for both polymer melts with and without bending constraints. Equation~\ref{Eq_HighT} has been confirmed by simulations of a coarse-grained polymer melt with and without antiplasticizer additives,~\cite{2007_JCP_126_234903} and this functional form has been applied in studies of metallic GF materials as a method for estimating $T_A$.~\cite{2009_PNAS_106_7735, 2016_JSM_054048, 2020_JCP_153_124508} The quantity $C_s$ should be of great interest from a simulation viewpoint since simulations are mostly restricted to a $T$ regime above $T_c$ in GF liquids because of the very long relaxation times at lower $T$, so we think that eq~\ref{Eq_HighT} deserves greater consideration in the future studies.

Kivelson and coworkers~\cite{1995_PA_219_27, 1996_PRE_53_751} have suggested a similar form as eq~\ref{Eq_HighT} based on experimental estimates of the reduced activation energy for diverse fluids, and the particular exponent was reported to be $8/3$ for their analog of eq~\ref{Eq_HighT}, a value motivated by a `frustrated-limited cluster model' of glass formation. Chandler and coworkers~\cite{2009_JPCB_113_5563, 2010_JPCB_114_17113} have also argued for the same $T$ dependence of the activation energy, as the GET predicts for eq~\ref{Eq_HighT}, which is termed the `parabolic law' involving a characteristic temperature comparable to $T_A$, along with an adjustable parameter corresponding to $C_s$, and they have made comparisons with experimental data for structural relaxation times for numerous GF liquids, where a good `fitting' was claimed. In comparison of the parabolic law proposed by Chandler and coworkers to measurements,~\cite{2015_SciRep_5_13837, 2015_PNAS_112_12020} a high $T$ Arrhenius term is added to the parabolic law to improve the comparison of the model to experiments at high $T$ where relaxation is Arrhenius, and this modification leads to an expression for $\Delta G(T)$ exactly in accord with eq~\ref{Eq_HighT}. We also mention that the parabolic functional form for the activation energy can be derived from the NLE and ECNLE theories of Schweizer and coworkers,~\cite{2004_JCP_121_1984, 2004_JCP_121_2001, 2010_ARCMP_1_277, 2014_JCP_140_194506, 2014_JCP_140_194507, 2015_Mac_48_1901, 2016_Mac_49_9655} and the details have been given in a previous review on the GET.~\cite{2008_ACP_137_125} The above comparisons seem to indicate that the GET has much in common with other models of glass formation. Note that $C_s$ and $T_A$ are derived in terms of molecular structure and interactions in the GET. We again emphasize that eq~\ref{Eq_HighT} is only applicable in the high $T$ regime above $T_c$ but below $T_A$. The GET predicts that the VFT relation applies in the low $T$ regime of glass formation below $T_c$ where the VFT parameters are likewise calculated from the theory as a function of molecular parameters, as we discuss below based on the $T$ dependence of $s_c$.

Figure~\ref{Fig_FitSc}b shows the prediction from the GET for the product $Ts_c/k_B$ as a function of $T$ for both polymer melts with and without bending constraints. Over a limited $T$ range in the low $T$ regime of glass formation between $T_c$ and $T_0$, a linear relation between $Ts_c/k_B$ and $T$ approximately holds with the slope $K_T$,~\cite{2013_JCP_138_12A548}
\begin{equation}
	\label{Eq_LowT}
	Ts_c/k_B = K_T(T/T_K - 1),
\end{equation}
where $T_K$ is the Kauzmann temperature at which $s_c$ is extrapolated to zero. Since $K_T$ bears no direct relation to the strength of the $T$ dependence of the relaxation time, we prefer to call this quantity the low $T$ fragility parameter. We see from Figure~\ref{Fig_FitSc}b that eq~\ref{Eq_LowT} holds for $T$ down to $T_0$ for the polymer melt with bending constraints. In the absence of bending constraints, however, the $T$ range where eq~\ref{Eq_LowT} is applicable seems to be limited, which clearly arises from the sigmoidal variation of $s_c$ with $T$, as discussed earlier.

\subsection{Packing Frustration and Glass Formation}

Despite the fact that the GET~\cite{2008_ACP_137_125} is essentially a thermodynamic theory of glass formation that relies on predictions of fluid dynamics based on the configurational entropy, calculations based on this model have provided \textit{molecular} insight into the physical nature of polymer glass formation and the most relevant factors governing this phenomenon. Here, we discuss how the GET is utilized to understand the fragility and $T_{\mathrm{g}}$ of GF polymers, two of the most important parameters of polymer materials from a practical viewpoint. A previous work~\cite{2009_JCP_131_114905} based on the GET investigated the influence of molecular parameters, such as cohesive interaction strength, chain stiffness, and molecular mass, on fragility quantified by the index $m$ or $1/D$ in the VFT relation (eq~\ref{Eq_VFT}), and this work indicated a remarkable regularity. Specifically, the changes in fragility arising from varying molecular parameters can be understood at fixed measures of cohesive interaction strength from how these parameters influence the \textit{packing frustration} of the molecules, as measured quantitatively by the `free volume' at $T_{\text{g}}$, i.e., $\phi_v(T_{\text{g}})$, where $\phi_v$ characterized by the extent to which the lattice is not occupied by chain segments, i.e., $\phi_v = 1 - \phi$. The concept of free volume has a long history in modeling the dynamics of cooled liquids going back to Batschinski,~\cite{1913_ZPC_84U_643} Hildebrand,~\cite{Book_Hildebrand} Doolittle,~\cite{1951_JAP_22_1031} Ferry,~\cite{Book_Ferry} and many others, and this concept remains popular in the polymer science community.~\cite{2013_SoftMatter_9_3173, 2016_Mac_49_3987} The most widely used definitions of free volume in the literature and the differences between different definitions have been recently reviewed by White and Lipson,~\cite{2016_Mac_49_3987} who have developed a `cooperative free volume' model to explain the dynamics of GF liquids.~\cite{2017_JCP_147_184503, 2018_Mac_51_4896, 2018_Mac_51_7924, 2019_MacroLett_8_41}

\begin{figure*}[htb!]
	\centering
	\includegraphics[angle=0,width=0.975\textwidth]{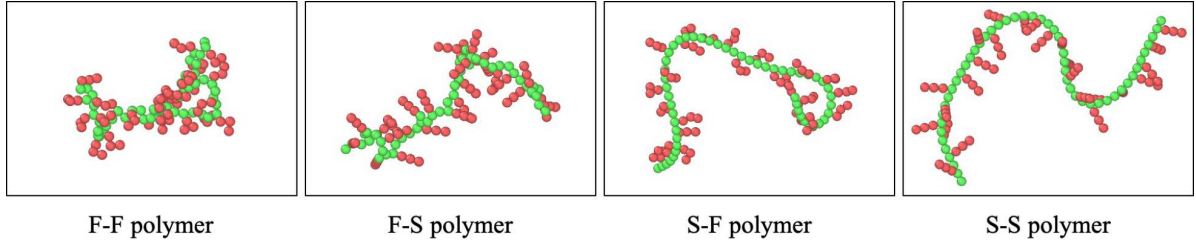}
	\caption{\label{Fig_Class}Schematic illustration of chain configurations for general classes of polymers. Polymers are categorized on the basis of the relative rigidities of the backbone and side chains, leading to flexible-flexible (F-F), flexible-stiff (F-S), stiff-flexible (S-F), and stiff-stiff (S-S) polymers.}
\end{figure*}

While the general concept of packing frustration is rather vague, the GET~\cite{2008_ACP_137_125} allows us to quantify this property of fluids in terms of how molecular parameters, such as the relative rigidities of the polymer backbone and side chains, chain length, monomer structure, and cohesive interaction strength, influence macroscopic properties such as the thermal expansion coefficient and isothermal compressibility that are highly dependent on the efficiency of molecular packing and the complex intermolecular interactions of molecular fluids. Early studies with the GET~\cite{2008_ACP_137_125} indicated that these molecular parameters can greatly influence packing efficiency, and, accordingly, polymers were classified into general classes based on the results of the model~\cite{2008_ACP_137_125} and measurements that also suggested such a classification. Specifically, this early work~\cite{2008_ACP_137_125} led to a classification of polymers into classes of polymers sharing similar packing characteristics. In particular, the flexible-flexible (F-F), flexible-stiff (F-S), stiff-flexible (S-F), and stiff-stiff (S-S) classes of polymers were defined to be chains with a flexible backbone and flexible side groups, chains with a flexible backbone and relatively rigid side branches, chains with a relatively stiff backbone and flexible side groups, and chains with both a stiff backbone and stiff side groups, respectively. The S-S class of polymers has not been encountered frequently in the past, but such polymers have recently become of interest in connection with the applications of molecular filtration in which the large free volume and $T_{\text{g}}$ of such `polymers of intrinsic microporsity' (PIMs) are highly physical attributes,~\cite{2010_PC_1_63, 2008_JAPS_107_1039, 2013_Science_339_303, 2016_NM_15_760} motivating the theoretical study of S-S polymers in more detail. For illustrative purposes, Figure~\ref{Fig_Class} displays representative chain configurations for these general classes of polymers.

Intuitively, F-F polymers can pack efficiently in space, and accordingly, these polymer systems exhibit a relatively weak $T$ dependence of $s_c$ and are relatively strong glass-formers. In comparison, polymers with either a stiff backbone or stiff side groups exhibit more packing frustration, and hence, these polymers have a relatively strong variation of $s_c$ with $T$, provided that the cohesive interaction strength is fixed, as in the case of molecules having van der Waals or other fixed types of intermolecular interactions. The GET then predicts that variable fragility is a direct outgrowth of the extent of packing frustration of the molecules composing the material,~\cite{2008_ACP_137_125} a property that can be engineered through exertion of control of molecular structure by chemical synthesis. We note that the experimental measurements of Sokolov and coworkers~\cite{2008_Mac_41_7232} have confirmed the general trends predicted by the GET of fragility with varying molecular structure.

\begin{figure*}[htb!]
	\centering
	\includegraphics[angle=0,width=0.975\textwidth]{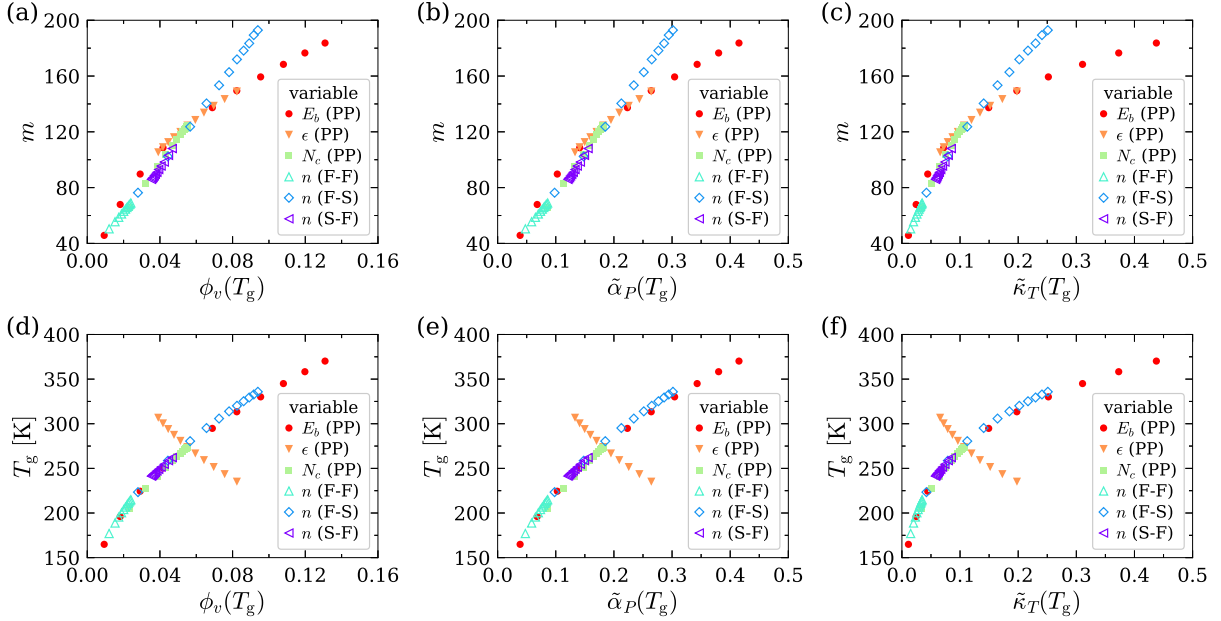}
	\caption{\label{Fig_PF}Correlations between packing frustration and the characteristic properties of glass formation predicted by the GET. (a--c) $m$ and (d--f) $T_{\text{g}}$ versus thermodynamic metrics for packing frustration for varying individual molecular parameters. The proposed metrics, $\phi_v$, $\tilde{\alpha}_P$, and $\tilde{\kappa}_T$, are the `free volume' defined in terms of lattice occupancy, reduced thermal expansion coefficient, and reduced isothermal compressibility, respectively, which are all given at $T_{\text{g}}$. In these calculations, the variables $E_b$, $\epsilon$, and $N_c$ are systematically varied for polymer melts with the structure of PP, and the variable side group length $n$ is systematically varied for F-F, F-S, and S-F polymers. Adapted with permission from ref~\citenum{2020_Mac_53_7239}.}
\end{figure*}

The above discussion naturally leads us to expect that $m$ increases with the extent of packing frustration, as quantified by the thermodynamic measures discussed above. This general trend is verified in Figure~\ref{Fig_PF}a, where $\phi_v(T_{\text{g}})$ is utilized as a candidate for a quantitative measure of packing frustration. In Figure~\ref{Fig_PF}, we have systematically varied $E_b$, $\epsilon$, and $N_c$ for polymer melts with the structure of PP and the side group length $n$ for F-F, F-S, and S-F polymers. Since S-S polymers exhibit the same trends in $m$ and $T_{\mathrm{g}}$ as F-F polymers, this class of polymers is not included here for our analysis. Reference~\citenum{2020_Mac_53_7239} provides more details regarding the calculations. We next discuss these measures of packing frustration in detail and their determination in the GET as well as their correlation with the fragility of glass formation as the structural properties of polymers are modified.

From an experimental point of view, $\phi_v(T_{\text{g}})$ is difficult to measure, and it is thus natural to compare $m$ to properties that are related to $\phi_v(T_{\text{g}})$, but more readily measured, such as the thermal expansion coefficient $\alpha_P$ and isothermal compressibility $\kappa_T$,
\begin{equation}
	\label{Eq_Thermo}
	\alpha_P = \frac{1}{V}\left. \frac{\partial V}{\partial T} \right|_P,\ \kappa_T = -\frac{1}{V} \left. \frac{\partial V}{\partial P} \right|_T.
\end{equation}
These properties are of both fundamental and practical interest and can be measured experimentally for polymers and other materials.~\cite{Book_Floudas} For equilibrium fluids, $\kappa_T$ is related to the long wavelength limit of the static structure factor, $S(0)$, via $\kappa_T = \rho k_BT/S(0)$ with $\rho$ being the number density. The density is given by $\rho = \phi/V_{\text{cell}}$ in the GET. Instead of focusing directly on $\alpha_P$ and $\kappa_T$, the GET suggests that the reduced properties be considered as measures of packing frustration,~\cite{2016_ACP_161_443, 2020_Mac_53_7239, 2020_Mac_53_9678}
\begin{equation}
	\label{Eq_ThermoReduce}
	\tilde{\alpha}_P = T\alpha_P,\ \tilde{\kappa}_T = \rho k_BT \kappa_T.
\end{equation}
The experimental studies of Simha and Boyer~\cite{1962_JCP_37_1003} have demonstrated that the reduced thermal expansion coefficient $\tilde{\alpha}_P$ can be used to estimate the dependence of $T_{\text{g}}$ on both cohesive interaction strength and chain stiffness for many polymers, thereby emphasizing the significance of $\tilde{\alpha}_P$ rather than $\alpha_P$ itself. The NLE and ECNLE theories of glass formation of Schweizer and coworkers~\cite{2004_JCP_121_1984, 2004_JCP_121_2001, 2010_ARCMP_1_277, 2014_JCP_140_194506, 2014_JCP_140_194507, 2015_Mac_48_1901, 2016_Mac_49_9655} also emphasize the central role of $S(0)$ in predicting the dynamics of GF liquids. Sanchez~\cite{2014_JPCB_118_9386} has utilized the dimensionless thermodynamics to discuss the liquid state properties, and this interesting work is recommended to the reader for further details on reduced thermodynamic properties. We have investigated the dimensionless thermodynamic properties in our simulation studies of polymer glass formation.~\cite{2016_Mac_49_8341, 2017_Mac_50_2585, 2020_Mac_53_4796, 2020_Mac_53_9678}

We show a comparison $m$ versus $\tilde{\alpha}_P$ and $\tilde{\kappa}_T$ in Figures~\ref{Fig_PF}b and~\ref{Fig_PF}c, respectively, where both quantities are given at $T_{\text{g}}$. We again see the expected trend that $m$ varies with packing frustration, as quantified by $\tilde{\alpha}_P$ and $\tilde{\kappa}_T$. We emphasize that our discussion here is based on the \textit{reduced} thermal expansion coefficient and isothermal compressibility.

We may also utilize the concept of packing frustration to understand $T_{\text{g}}$, a quantity that is naturally appreciated from the famous WLF relation~\cite{Book_Ferry, 1955_JACS_77_3701} and its universal parameters $C_1$ and $C_2$, which are often invoked in the description of the dynamics of polymer materials. The WLF relation can be derived from the VFT expression for $\tau_{\alpha}$ if the fragility parameter is taken to be simply proportional to $T_{\text{g}}$, an assumption that accounts for why fragility is not an explicit parameter in the WLF equation and for why the dynamics depends on the temperature difference $T - T_{\text{g}}$ for materials for which the WLF equation applies.~\cite{2015_JCP_142_014905} A detailed discussion of the WLF relation along with references relevant to $C_1$ and $C_2$ has been given in ref~\citenum{2015_JCP_142_014905}. Figures~\ref{Fig_PF}d--f show that $T_{\text{g}}$ strongly correlates with the metrics for packing frustration, provided that materials are compared at fixed cohesive interaction strength, which informs us that materials with inherently high packing frustration exhibit an inherently higher $T_{\text{g}}$ and fragility. This finding accords with the use of high free volume polymer materials or PIMs~\cite{2010_PC_1_63, 2008_JAPS_107_1039, 2013_Science_339_303, 2016_NM_15_760} for applications in gas separation processes, heterogeneous catalysis, and hydrogen storage as well as the socially important problem of large-scale water desalination, to name a few emerging applications of these new materials. $T_{\text{g}}$ in some commercial PIMs is so high that its estimates have previously been difficult, but recent techniques have overcome this difficulty, at least in some PIMs.~\cite{2019_MacroLett_8_1022, 2018_JPCL_8_2003} Finally, it is worth noting in Figures~\ref{Fig_PF}d--f that an inversion of the trend of $T_{\text{g}}$ with the measures of packing frustration occurs when varying $\epsilon$, a result first identified in ref~\citenum{2009_JCP_131_114905}. This more complicated behavior should be borne in mind when discussing the glass formation of polar and charged polymer fluids, which generally exhibit high cohesive interaction strength.

\subsection{Glass Formation under Pressure}

It has been well established that glass formation can often be induced by increasing $P$ at constant $T$.~\cite{Book_Floudas, 2005_RPP_68_1405, 2010_Mac_43_7875} The large alterations in the dynamics of GF liquids upon the application of pressure are evidently of significance in numerous manufacturing applications. In particular, there has been intense interest in separating out aspects of glass formation relating to the effects of attractive intermolecular interactions, temperature, and density,~\cite{2004_JCP_120_6135, 2007_JPCM_19_205117, 2005_RPP_68_1405} These efforts have naturally led to systematic studies of the dynamics of GF liquids over a wide range of $P$~\cite{2005_RPP_68_1405, 2001_JCP_114_8048} to complement the traditional studies where glass formation occurs upon cooling. The GET is naturally suitable for addressing polymer glass formation under applied pressures since calculations can be performed under different constant $P$ conditions. In this section, we utilize the GET to discuss several interesting aspects of polymer glass formation when $P$ is a variable.

\begin{figure*}[htb!]
	\centering
	\includegraphics[angle=0,width=0.475\textwidth]{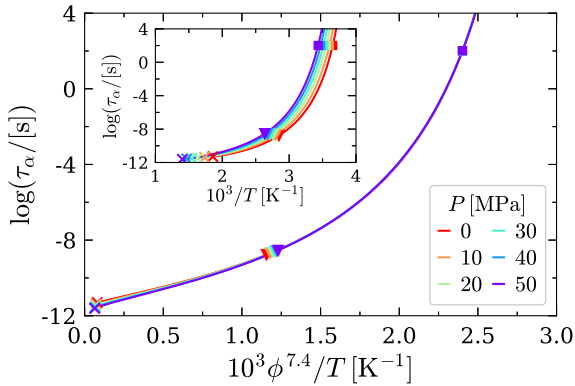}
	\caption{\label{Fig_Scaling}Density-temperature scaling of $\tau_{\alpha}$ predicted by the GET. The main plot shows $\log \tau_{\alpha}$ versus $10^3 \phi^{\gamma}/T$ with $\gamma = 7.4$ for a range of fixed $P$. The inset displays $\log \tau_{\alpha}$ versus $10^3/T$ for the same $P$. The calculations utilize the parameter set specified in the caption of Figure~\ref{Fig_Sc}.}
\end{figure*}

We begin by noting the extremely interesting density-temperature scaling phenomenon,~\cite{Book_Floudas, 2005_RPP_68_1405, 2010_Mac_43_7875} which indicates that dynamic properties such as structural relaxation time and diffusion coefficient in suitably reduced units become universal functions of $T$ times density to a power $\gamma$ characteristic of the material. Dyre and coworkers~\cite{2009_JCP_131_234504} have suggested that the existence of this type of scaling should play the role of a `filter' for acceptable models of GF liquids. In a previous study based on the GET,~\cite{2013_JCP_138_234501} we indeed found density-temperature scaling to hold for $\tau_{\alpha}$ over the entire $T$ regime of glass formation in polymer melts. The GET thus passes this test. For illustrative purposes, we show the density-temperature scaling of $\tau_{\alpha}$ for a range of fixed $P$ predicted by the GET in Figure~\ref{Fig_Scaling}, the inset of which displays $\log \tau_{\alpha}$ as a function of $10^3/T$. Experimental and computational studies indicate that the scaling exponent $\gamma$ obtained from estimates of viscosity and relaxation time lies in the broad range of $0.18$ to $8.5$.~\cite{2005_RPP_68_1405} Our previous study~\cite{2013_JCP_138_234501} indicates that the scaling exponent $\gamma$ varies in a large range when varying molecular parameters, such as chain length, chain stiffness, cohesive interaction strength, and side group length, factors that evidently influence the fragility of glass formation. Our calculations based on the GET~\cite{2013_JCP_138_234501} also confirmed the approximate inverse relation between $\gamma$ and the constant volume fragility parameter $m_V$ found experimentally by Casalini and Roland.~\cite{2007_JNCS_353_3936, 2005_PRE_72_031503}

Recently, we have made additional progress on the mysterious density-temperature scaling phenomenon based on the GET and simulations. We have also noticed that some other popular models of glass formation do not pass the test associated with density-temperature scaling. We will report on these findings in a separate paper, where we discuss the origin of density-temperature scaling and its many implications. It is notable that the GET is based on a lattice model where the polymer intersegment interactions are described by the rough equivalent of a square potential in off-lattice liquids. The scaling exponent $\gamma$ depends on molecular parameters discussed above, but it has \textit{nothing to do} with the shape of the pair potential in our calculations. Density-temperature scaling in polymer and other molecular fluids then seems to arise from the variation of the anharmonic intermolecular interactions that arise from packing frustration, and the GET provides a powerful computational framework for studying this phenomenon that fundamentally links the thermodynamics and dynamics of fluids and the relative balance of repulsive and attractive interparticle interactions governing the dynamics of real molecular fluids.

\begin{figure*}[htb!]
	\centering
	\includegraphics[angle=0,width=0.95\textwidth]{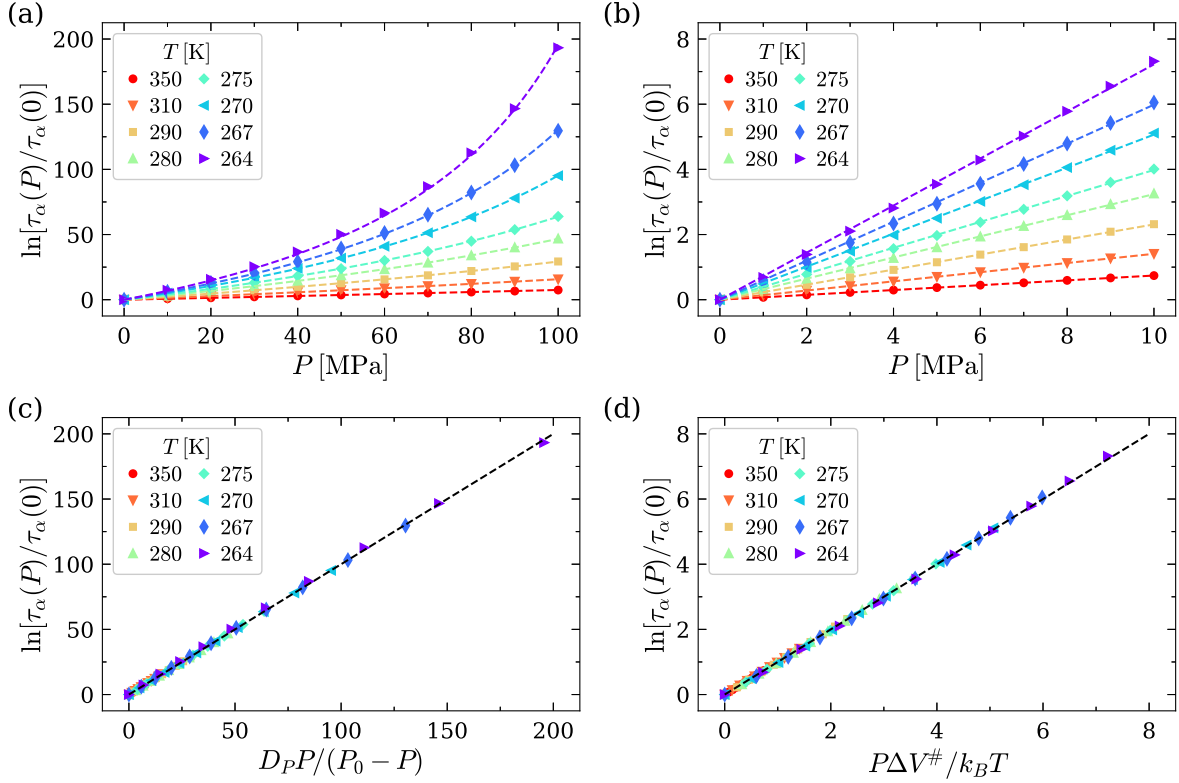}
	\caption{\label{Fig_Pressure}$P$ dependence of $\tau_{\alpha}$ at constant $T$ predicted by the GET. Panels (a) and (b) show $\ln [\tau_{\alpha}(P)/\tau_{\alpha}(0)]$ as a function of $P$ for a range of $T$ in a wide $P$ range and in the low $P$ limit, respectively. Lines in panels (a) and (b) are the descriptions based on eqs~\ref{Eq_PVFT} and~\ref{Eq_Vact}, respectively. Panels (c) and (d) show the universal reduction of the $P$ dependence of $\tau_{\alpha}$. Lines indicate $\ln [\tau_{\alpha}(P)/\tau_{\alpha}(0)] = D_P P/(P_0 - P)$ in panel (c) and $\ln [\tau_{\alpha}(P)/\tau_{\alpha}(0)] = P\Delta V^{\#}/k_BT$ in panel (d), respectively. The calculations utilize the parameter set specified in the caption of Figure~\ref{Fig_Sc}.}
\end{figure*}

The GET also allows us to quantitatively analyze the variation of $\tau_{\alpha}$ with $P$ at constant $T$. Experimental~\cite{Book_Floudas, 2005_RPP_68_1405, 2010_Mac_43_7875} and computational~\cite{2016_MacroLett_5_1375, 2017_Mac_50_2585} studies have indicated that the $P$ dependence of $\tau_{\alpha}$ generally displays a pressure analog of the VFT equation (PVFT) in non-associating GF liquids, in which $P$ and the critical pressure $P_0$, respectively, replace $T$ and the critical temperature $T_{\mathrm{VFT}}$ in the conventional VFT equation given in eq~\ref{Eq_VFT},
\begin{equation}
	\label{Eq_PVFT}
	\tau_{\alpha}(P) = \tau_{\alpha}(0) \exp \left( \frac{D_P P}{P_0 - P} \right),
\end{equation}
where $\tau_{\alpha}(0)$ is the structural relaxation time at zero pressure. The GET predicts the PVFT relation and further provides a rationale based on the variation of the configurational entropy of the fluid with $P$.~\cite{2009_JCP_131_114905} We show an illustrative result for the $P$ dependence of $\tau_{\alpha}$ calculated from the GET for a range of $T$ in a wide $P$ range in Figure~\ref{Fig_Pressure}a, along with a universal reduction of the data based on the PVFT relation in Figure~\ref{Fig_Pressure}c.

While the PVFT relation is evidently required to describe the $P$ variation of $\tau_{\alpha}$ in a large $P$ range, there is a simple linear variation of $\ln\tau_{\alpha}$ with $P$ in the low $P$ limit,
\begin{eqnarray}
	\label{Eq_Vact}
	\tau_{\alpha}(P) = \tau_{\alpha}(0) \exp \left( \frac{P\Delta V^{\#}}{k_BT} \right),
\end{eqnarray}
where $\Delta V^{\#}$ defines the activation volume. Figure~\ref{Fig_Pressure}b displays $\ln [\tau_{\alpha}(P)/\tau_{\alpha}(0)]$ as a function of $P$ in a much narrower regime of $P$ than that in Figure~\ref{Fig_Pressure}a. We see that the GET predicts the linear variation of $\ln\tau_{\alpha}$ with $P$ when $P$ is restricted to small values. Again, we show a universal reduction of the data from the GET based on eq~\ref{Eq_Vact} in Figure~\ref{Fig_Pressure}d. This analysis evidently indicates an onset pressure $P_A$ above which the variation of $\ln\tau_{\alpha}$ with $P$ deviates from the linear relation given by eq~\ref{Eq_Vact}, the analog of the Arrhenius regime of GF liquids when $T$ is instead varied at fixed $P$. We have also found that the $P$ dependence of $\tau_{\alpha}$ calculated from the GET seems to follow a power law in a restricted $P$ range, $\tau_{\alpha}(P) \sim (P_c - P)^{-\gamma_{c}}$, which $P_c$ may be defined by a crossover pressure and $\gamma_{c}$ is an effective $T$-dependent `crossover exponent'. We do not think that this exponent estimated from the GET has anything to do with the mode-coupling theory,~\cite{Book_Gotze} but it is sometimes attributed to this theory because it likewise predicts a power-law relation of this kind, albeit in a $T$ range closer to $T_A$ than $T_c$.~\cite{2008_ACP_137_125} Moreover, a `glass transition pressure' $P_{\mathrm{g}}$ can be identified for a given $T$ by the pressure at which $\tau_{\alpha} = 100$ s, and as discussed above, we may also define a pressure analog $P_0$ of the VFT temperature by a limiting pressure at which $\tau_{\alpha}$ extrapolates to infinity. Therefore, the GET allows us to estimate the different characteristic pressures ($P_A$, $P_c$, $P_{\mathrm{g}}$, and $P_0$) of glass formation when $P$ is used as a control variable for glass formation at fixed $T$. There is then an interesting `duality' between glass formation with variable $T$ at fixed $P$ and with variable $P$ at fixed $T$. This is another topic that we will investigate in the future.

Parenthetically, the activation volume $\Delta V^{\#}$ is often encountered in an industrial and materials science setting when materials are subjected to large changes in $P$ or applied steady stresses of various kinds, but the physical meaning of $\Delta V^{\#}$ is often rather unclear. In our recent work,~\cite{2020_Mac_53_6828, 2020_Mac_53_7239} we have systematically investigated $\Delta V^{\#}$ based on the GET and simulations. Our study indicates that $\Delta V^{\#}$ is related to the differential change of the activation free energy as a function of $T$, and thus bears a close relationship to the fragility of glass formation. 

\subsection{\label{Sec_GET_Challenge}Opportunities and Challenges}

The GET is just one of a number of promising models of glass formation which has the particular advantage of predicting the segmental structural relaxation time of polymeric GF liquids in terms of essential molecular parameters over the entire $T$ regime of glass formation. The model is also advantageous because it describes the equation of state and other basic thermodynamic properties of polymer melts within a mature and validated theoretical model in a consistent framework. Moreover, the GET naturally allows for the validation of thermodynamic and dynamic properties based on the string model~\cite{2014_JCP_140_204509} of glass formation in conjunction with MD simulations of coarse-grained polymer melts that cover a wide range of molecular parameters of interest in understanding the thermodynamic and dynamic properties of real polymer materials. As discussed earlier, there are assumptions in the GET relating to the precise relation between thermodynamic and dynamic properties whose validity needs to be further assessed and modified if necessary, but this type of limitation exists for all current models of GF liquids. Here, we discuss how the GET might be improved.

We first note that, while $\Delta H_0$ can be estimated from an Arrhenius fit to the dynamics of fluids at high $T$ if such information is available, it remains a difficult matter to determine and understand $\Delta S_0$. The theoretical estimation of $\Delta S_0$ is generally appreciated to be the primary source of uncertainty in TST,~\cite{1949_PR_76_1169} and inevitably, $\Delta S_0$ must be estimated from either experiment~\cite{1946_JCP_14_591} or simulation.~\cite{2015_JCP_143_144905} Theoretical attempts to calculate $\Delta S_0$ has a long history in condensed materials, which we briefly mention here. Vineyard~\cite{1957_JPCS_3_121} proposed a rigorous TST for idealized crystalline materials with ideal harmonic interactions, which has been validated for simple ordered molecular clusters.~\cite{2014_PSS_56_1239} Unfortunately, anharmonic effects, as found characteristically in cooled liquids, heated crystals, and the interfacial dynamics of crystals for $T > 2T_m/3$ with $T_m$ being the melting temperature,~\cite{2015_JCP_142_084704} cause the Vineyard theory to break down. A general discussion of TST for crystalline materials has been provided by Rice,~\cite{1958_PR_112_804} who emphasized the relation between the Arrhenius prefactor and the stable and saddle point frequencies. This work is particularly interesting from a philosophical viewpoint in the sense that it explains why one must consider a free energy of activation, $\Delta G_0$, in condensed materials rather than just an activation energy, $\Delta H_0$. Collective motion associated with many-body effects is thus essential for barrier crossing events associated with relaxation and diffusion in condensed materials.

While the first principles analytic estimation of $\Delta S_0$ for fluids from TST has remained challenging, numerical implementations of TST have recently enabled precise estimations of $\Delta S_0$ for model condensed materials.~\cite{1995_PRL_75_469, 2001_PRB_64_075418, 2011_PNAS_108_5174} A large body of semi-empirical works have combined careful measurements with physical arguments to explain $\Delta S_0$ in terms of a presumed connection with local elastic distortions required for a particle or defect to move in the condensed state. This line of argument can be traced back to the pioneering work of Zener~\cite{1949_PR_76_1169} in the context of metallic crystalline materials and was taken further by Lawson~\cite{1957_JPCS_3_250, 1960_JCP_32_131} by directly relating the activation volume $\Delta V^{\#}$ to $\Delta S_0$. Keyes~\cite{1958_JCP_29_467} argued for a corresponding relation between $\Delta V^{\#}$ and $\Delta H_0$. These relations have provided a rationalization of the EEC effect between $\Delta H_0$ and $\Delta S_0$ that has been validated empirically in many polymeric and other condensed materials.~\cite{1962_JCP_37_2785, 1982_Polymer_23_473, 1993_JAP_74_1597, 1998_JNCS_235_237_737, 2005_JPCB_109_16567} Exactly solvable toy models of dynamics on hierarchical potential energy landscapes,~\cite{1991_JPCM_3_3941, 1987_JPAMG_20_5627, 1991_JPAMG_24_2807} as naturally found in disordered condensed materials, also give rise to the EEC. The large entropy of activation in condensed materials is found in these models to arise from the rapid proliferation of transition state paths over high barriers in such hierarchical spaces, an intuitive physical picture supporting the formal theoretical results of Yelon and coworkers.~\cite{1992_PRB_46_12244, 1990_PRL_65_618} A more physically tangible real-space approach to understanding relaxation in condensed fluids starts from a consideration of the nature of collective motion in the fluids required to enable atomic displacement. Based on this point of view, Barrer~\cite{1943_TFS_39_48, 1943_TFS_39_59} has offered an interesting heuristic `zone theory' of the EEC phenomenon in the high $T$ Arrhenius regime of liquids, thus having features in common with the heuristic arguments of AG in cooled liquids,~\cite{1965_JCP_43_139} as discussed in Section~\ref{Sec_AG}. Therefore, a physical approach of this kind focusing on the precise nature of collective motion in dense fluids~\cite{2013_JCP_138_12A541} might offer a deep understanding of the link between $\Delta H_0$ and $\Delta S_0$, as found in the relaxation, diffusion, and chemical reaction processes of so many condensed materials.~\cite{2001_CR_101_673, 2008_JPCB_112_15980}

\begin{figure*}[htb!]
	\centering
	\includegraphics[angle=0,width=0.975\textwidth]{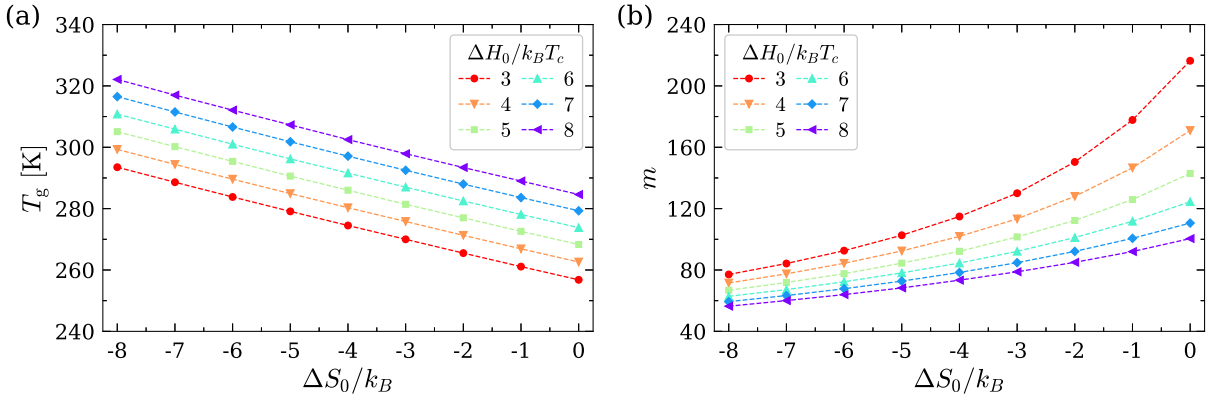}
	\caption{\label{Fig_HS}Influence of the high $T$ activation free energy parameters on polymer glass formation predicted by the GET. (a) $T_{\mathrm{g}}$ and (b) $m$ as a function of the activation entropy $\Delta S_0 / k_B$ for a range of $\Delta H_0$. The calculations utilize the parameter set specified in the caption of Figure~\ref{Fig_Sc}.}
\end{figure*}

\begin{figure*}[htb!]
	\centering
	\includegraphics[angle=0,width=0.975\textwidth]{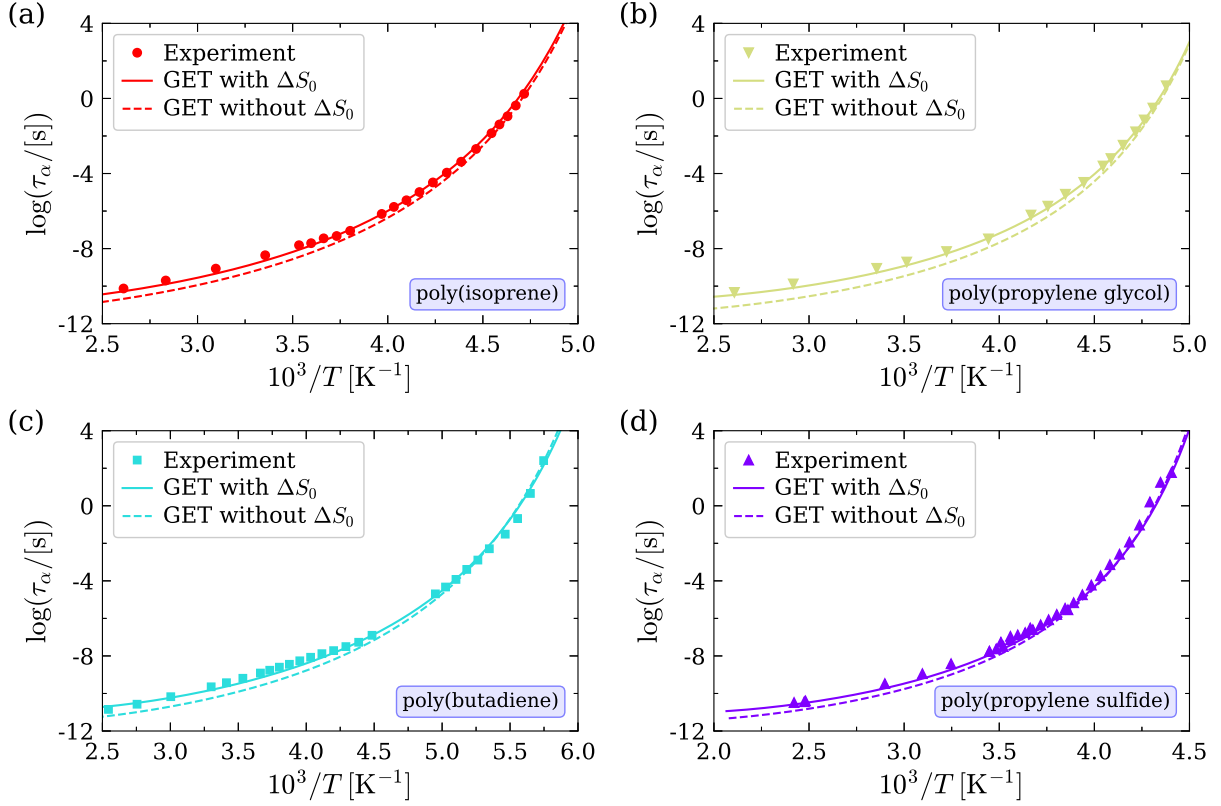}
	\caption{\label{Fig_FitExp}Description of experimental results for the $T$ dependence of $\tau_{\alpha}$ for several polymers based on the GET. (a) Poly(isoprene). (b) Poly(propylene glycol). (c) Poly(butadiene). (d) Poly(propylene sulfide). Experimental data were provided to us by the authors of ref~\citenum{2015_Mac_48_3005}. Solid and dashed lines represent our best fits based on the GET with and without $\Delta S_0$, respectively. The calculations based on the GET are performed for a melt of chains with the structure of PP at zero pressure, where $V_{\text{cell}} = 2.5^3 \text{\AA}^3$ and $N_c = 8000$. The molecular parameters $\epsilon$ and $E_b$ and energetic parameters $\Delta H_0$ and $\Delta S_0$ for the polymers are given in Table~\ref{Table_Fit}, along with the basic properties of polymers.}
\end{figure*}

\begin{table}
	\caption{\label{Table_Fit}Basic properties of the polymers considered in Figure~\ref{Fig_FitExp}. The listed properties include the molar mass $M$, glass transition temperature $T_{\mathrm{g}}$, and fragility index $m$, along with the model parameters, $\epsilon$, $E_b$, $\Delta H_0$, and $\Delta S_0$, used for the fits to the experimental results based on the GET.}
	\begin{tabular}{lccccccc}
		\hline
		polymer & $M\ [\mathrm{g}/\mathrm{mol}]$ & $T_{\mathrm{g}}\ [\mathrm{K}]$ & $m$ & $\epsilon/k_B\ [\mathrm{K}]$ & $E_b/k_B\ [\mathrm{K}]$ & $\Delta H_0/k_BT_c$ & $\Delta S_0/k_B$ \\
		\hline
		PI & 15700 & 207 & 84 & $189.8$ & $357.8$ & $5.0$ & $-1.65$ \\
		PPG & 18000 & 202 & 105 & $163.9$ & $395.5$ & $4.0$ & $-2.8$ \\
		PB & 87000 & 174 & 90 & $165.2$ & $282.4$ & $4.5$ & $-2.2$ \\
		PPS & 44000 & 229 & 116 & $185.5$ & $442.0$ & $4.6$ & $-1.8$ \\
		\hline
	\end{tabular}
\end{table}

Although the GET does not provide a method for estimating $\Delta H_0$ and $\Delta S_0$, we may examine the influence of these basic parameters on polymer glass formation within this TST-based model. As noted above, we have utilized simplifying approximations in the GET to avoid estimating these energetic parameters, i.e., $\Delta H_0 = 6k_B T_c$ and $\Delta S_0 / k_B = 0$, with the second simplifying assumption being `inherited' from AG.~\cite{1965_JCP_43_139} Here, we tentatively treat both $\Delta H_0$ and $\Delta S_0$ as adjustable parameters to better understand their potential relevance to polymer glass formation. Figure~\ref{Fig_HS} shows both $T_{\mathrm{g}}$ and $m$ as a function of $\Delta S_0$ for a range of $\Delta H_0$. Our calculations also include the predictions based on the assumptions, $\Delta H_0 / k_B T_c = 6$ and $\Delta S_0/k_B = 0$, of the original GET.~\cite{2008_ACP_137_125} We see that $T_{\mathrm{g}}$ increases linearly with $|\Delta S_0|$, while $m$ drops considerably as $|\Delta S_0|$ elevates for the range of $\Delta S_0$ considered here. This result clearly demonstrates that precise estimations of both $\Delta H_0$ and $\Delta S_0$ are crucial for transforming the GET into a quantitative predictive molecular theory. Next, we present our preliminary results on comparisons between the GET and experiment, in particular when the activation entropy is present.

Figure~\ref{Fig_FitExp} shows our GET description of experimental results for the $T$ dependence of $\tau_{\alpha}$ for several polymers, including poly(isoprene) (PI), poly(propylene glycol) (PPG), poly(butadiene) (PB), and poly(propylene sulfide) (PPS). These representative polymers are selected to demonstrate the significance of including $\Delta S_0$ in the GET. All the experimental data were provided to us by the authors of ref~\citenum{2015_Mac_48_3005}, and the molar mass for each polymer corresponds to the highest value reported there. For simplicity, our calculations based on the GET are performed for a melt of chains with the structure of PP at zero pressure, where $V_{\text{cell}} = 2.5^3 \text{\AA}^3$ and $N_c = 8000$, for all the polymers considered. In an effort to describe the experimental data, we first adopt the original GET, where $\Delta H_0 = 6k_BT_c$ and $\Delta S_0 = 0$, to fit the experimental values of $T_{\mathrm{g}}$ and $m$ for each polymer by adjusting $\epsilon$ and $E_b$ simultaneously. The fitted results are shown as dashed lines in Figure~\ref{Fig_FitExp}, and the best values for $\epsilon$ and $E_b$ are summarized in Table~\ref{Table_Fit}. We see that the GET, even without $\Delta S_0$, leads to a reasonable quantitative description of the experimental results in the $T$ regime near $T_{\mathrm{g}}$, but deviations are noticeable at high $T$. Based on the obtained molecular parameters $\epsilon$ and $E_b$, we then allowed both $\Delta H_0$ and $\Delta S_0$ to be adjusted. Utilizing the values of $\Delta H_0$ and $\Delta S_0$ in Table~\ref{Table_Fit}, Figure~\ref{Fig_FitExp} exhibits our calculations from the GET with $\Delta S_0$ as solid lines, indicating that the inclusion of $\Delta S_0$ enables a quantitative and accurate description of the experimental data for the $T$ dependence of $\log \tau_{\alpha}$ over the entire $T$ regime of glass formation accessible to experiment.

In the future work, we need to compare the GET predictions to measurements for more polymers to better understand both $\Delta H_0$ and $\Delta S_0$, as well as the interrelation of these two energetic parameters. Jeong and Douglas~\cite{2015_JCP_143_144905} have explored the determination of $\Delta H_0$ and $\Delta S_0$ for alkane melts based on MD simulations, and we have also investigated these parameters in coarse-grained polymer melts with variable pressure,~\cite{2016_MacroLett_5_1375, 2017_Mac_50_2585, 2020_Mac_53_6828} cohesive interaction strength,~\cite{2016_Mac_49_8355, 2020_Mac_53_9678} and chain stiffness.~\cite{2020_Mac_53_4796} The computational machinery certainly exists to make progress in understanding the activation parameters based on simulations, and we hope that such analyses will lead to corresponding advances in analytic theories of these fundamental energetic parameters. We think that a comprehensive understanding of these energetic parameters will provide the foundation on which a real theory of GF liquids will ultimately be based. For the present, we must admit that TST of liquid dynamics has some serious shortcomings that are an impediment to the GET and other models of glass formation based on the foundation of thermally activated transport.

\section{\label{Sec_String}String Model of Glass-Forming Liquids}

\subsection{Historical Background}

As in the case of the relation between the fluid entropy and the dynamics in GF liquids, the idea that the growth of the structural relaxation time should be associated with increased collective motion that accompanies the reduction of configurational entropy long predates the AG model.~\cite{1965_JCP_43_139} Most of the arguments along this line in the older literature of GF liquids are rather qualitative, but we mention a historically interesting suggestion by Mott at a Solvay conference on condensed materials in 1951,~\cite{1951_Mott} in which the activation energy of materials in a relatively disordered state was suggested to grow extensively with the mass of some dynamic cluster in the material. Later, this suggestion was subjected to experimental tests by Nachtreib and Handler~\cite{1955_JCP_23_1187} to interpret their self-diffusion measurements of white phosphorous over a large $T$ range, and the measurements seemed to confirm the heuristic picture of the $T$-dependent activation energy of Mott for diffusion in this material. Although these historical contributions are not often discussed now, we also acknowledge the historical contributions by Orowan~\cite{1967_GJI_14_191} to the understanding of the nature of molecular motions underlying flow processes in viscous fluids and solid crystalline materials. See the paper of Goldstein~\cite{1969_JCP_51_3728} for a recounting of this now largely inaccessible work. Moreover, the influential `zone theory' of Barrer~\cite{1943_TFS_39_48, 1943_TFS_39_59, 1996_JMS_109_1} likewise emphasized the necessity of collective atomic motion involving many particles in connection with molecular diffusion. These are the shoulders upon which later models of the dynamics of GF liquids rest.

With this kind of historical background in view, it seems natural that AG~\cite{1965_JCP_43_139} would propose that the activation energy in cooled liquids should grow in proportion to the number of particles in some sort of hypothetical CRR, thereby creating an enduring conceptual construct in modeling the dynamics of liquids. As discussed in Section~\ref{Sec_AG}, AG~\cite{1965_JCP_43_139} further proposed that the number of particles in the vaguely defined CRR should scale inversely with the configurational entropy $S_c$. The greatest strength of this assumption is apparently that the VFT relation~\cite{1921_PZ_22_645, 1925_JACS_8_339, 1926_ZAAC_156_245} can be rationalized when this assumption is combined with rough estimates of $S_c$ based on an approximation suggested earlier by Bestul and Chang.~\cite{1964_JCP_40_3731} The AG model is not so much a theory, but rather a set of hypotheses introduced to conform to prevailing theoretical ideas and correlations between properties seen experimentally. This is probably a quite good description of more recent theories of glass formation, so we do not think that the AG model should be judged harshly given its rather empirical origin. This `semi-empirical' model has been immensely successful for creating a qualitative picture of the origin of the strong variation of the relaxation and diffusion properties with $T$ and polymer concentration, and it has largely determined the language used to describe GF liquids, regardless of whether or not one believes in the assumptions of the AG model. The RFOT theory~\cite{1987_PRB_36_8552, 1989_PRA_40_1045, 2004_JCP_121_7347, 2007_ARPC_58_235} is based upon similar ideas to rationalize the slowing down of the dynamics of GF liquids in terms of an `entropic droplet model'. 

It is worth noting that DiMarzio never accepted the AG model, explaining why he never combined the AG model~\cite{1965_JCP_43_139} with the lattice model of GD developed for calculating $S_c$.~\cite{1958_JCP_28_373} This situation is apparent in his later work,~\cite{1997_JRNIST_102_135} where DiMarzio introduced an alternative to the AG model by asserting a rather different relation between the thermodynamics and the structural relaxation time in cooled liquids. We also mention a formal critique of the assumptions underlying the AG model by Dyre and coworkers.~\cite{2009_JNCS_355_624} As a counterweight to these criticisms, we mention some theoretical discussions that formalize the idea of a growing activation energy as $T$ is varied in the interfacial dynamics of crystalline materials~\cite{1970_SS_21_45} and in GF liquids,~\cite{2014_JCP_141_141102} showing that there are significant merits to the assumptions of the AG model.~\cite{1965_JCP_43_139}

Dudowicz et al.~\cite{1999_JCP_111_7116} have shown that the fluid configurational entropy scales inversely with the average length of dynamic polymers calculated from a model of living polymerization, a finding that has taken great significance that dynamic polymeric structures involving stringlike structures have been observed in many simulations of GF liquids~\cite{1998_PRL_80_2338, 2003_JCP_119_5920, 2011_PRL_106_115702, 2013_SoftMatter_9_241, 2013_JCP_138_12A541, 2014_NatCommun_5_4163, 2014_JCP_140_204509, 2015_PNAS_112_2966, 2015_JCP_142_234907} to date, in which the collective motion has been examined in great detail and the thermodynamic, geometric, and dynamic properties of the dynamic stringlike structures closely conform to equilibrium polymerization. Historically, the findings of Dudowicz et al.~\cite{1999_JCP_111_7116} served as a great impetus for the development of the string model of glass formation, in which the dynamic clusters in cooled liquids were identified as being equilibrium polymers and realizations of the hypothetical CRR of AG.~\cite{1965_JCP_43_139} Douglas et al.~\cite{2006_JCP_125_144907} have comparatively discussed the variation of the configurational entropy of fluids exhibiting equilibrium polymerization and the configurational entropy calculated from the GET, and strong similarities in the $T$ variation of the configurational entropies from the two models are indicated, such as analogs of the onset temperature $T_A$, the crossover temperature $T_c$, and the glass transition temperature $T_{\mathrm{g}}$, below which the configurational entropy saturates to a finite value in the polymerization model. Douglas et al.~\cite{2008_JCP_128_224901} have also shown that the `cooperativity' of the equilibrium polymerization transition, a description of the extent to which the transition resembles a phase transition, plays the role of fragility in the correspondence between the polymerization model and the GET.

\subsection{String Model Essentials}

As just discussed above and elaborated on in ref~\citenum{2008_ACP_137_125}, some of the assumptions made by AG~\cite{1965_JCP_43_139} now have some support from simulations and the analytic theory of fluids exhibiting self-assembly, providing the starting point for developing the string model of glass formation. A subsequent work~\cite{2014_JCP_140_204509} has found that the stringlike collective motion conforms quantitatively to the equilibrium polymerization model over the entire $T$ range below $T_A$ for which simulations can currently be performed for GF liquids. As we shall see below, the only unspecified parameters in these comparisons between the simulation data and the string model are the activation enthalpy $\Delta H_0$ and activation entropy $\Delta S_0$ of TST, which are determined in the high $T$ Arrhenius regime, i.e., $\Delta H_0$ from the slope of the variation of $\ln \tau_{\alpha}$ with $1/T$ and $\Delta S_0$ from the prefactor in the Arrhenius relation (eq~\ref{Eq_Arrhenius}). Our lack of a theoretical understanding of the dynamics of fluids in the Arrhenius regime is the greatest hurdle in predicting the dynamics of GF liquids from the string model.

\begin{figure*}[htb!]
	\centering
	\includegraphics[angle=0,width=0.975\textwidth]{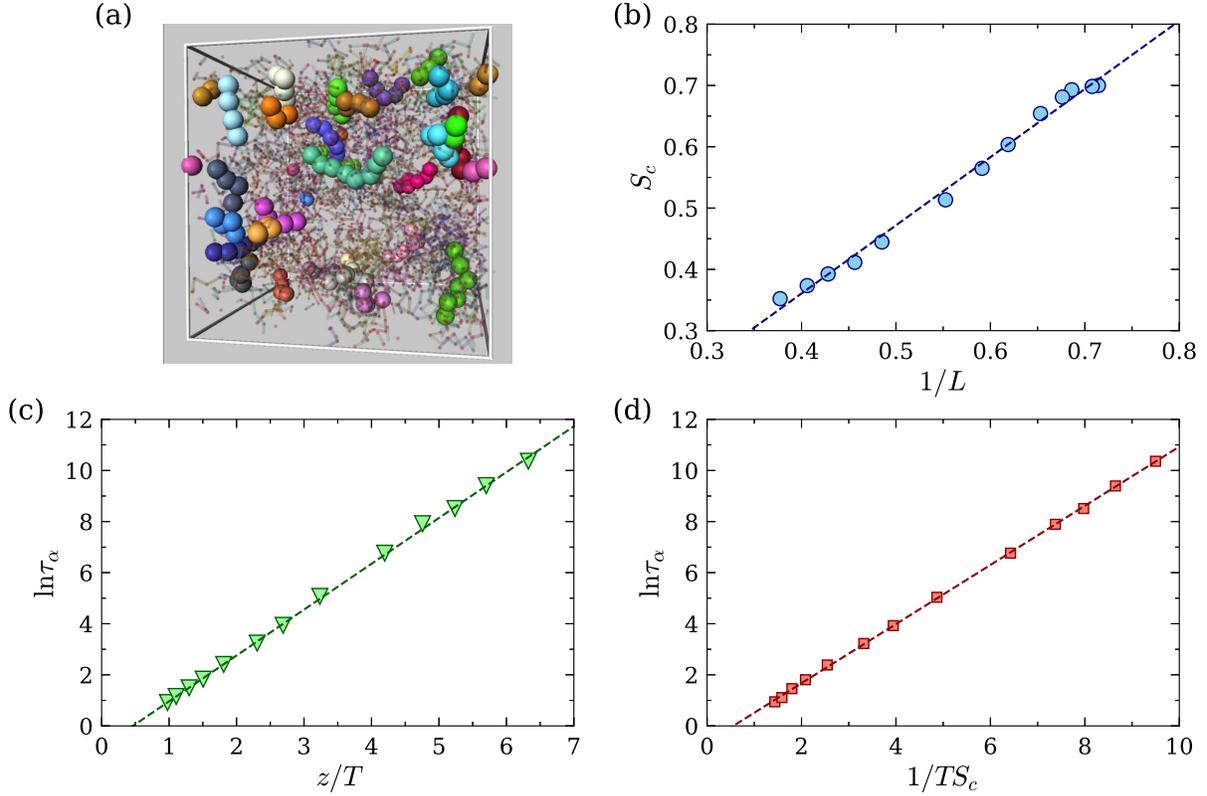}
	\caption{\label{Fig_CRR}Test of the average string length $L$ as a quantitative realization of the hypothetical cooperatively rearranging regions (CRR) of AG~\cite{1965_JCP_43_139} based on simulation results of a coarse-grained polymer melt. (a) Snapshot of stringlike clusters of polymer segments undergoing cooperative exchange motion over the lifetime over which these clusters persist. Each string is shown by large spheres in a different color. Only strings of length larger than $4$ are shown to aid the visualization, and the other segments are displayed as translucent think cylinders. (b) Configurational entropy $S_c$ versus $1 / L$. (c) $\ln \tau_{\alpha}$ versus $z/T$, where the size of the CRR is $z = L / L_A$ with $L_A$ being the value of $L$ at $T_A$. (d) $\ln \tau_{\alpha}$ versus $1/TS_c$. Lines in panels (b--d) are a guide to the eye. Here and in Figures~\ref{Fig_StringTest} and~\ref{Fig_String}, all quantities from the coarse-grained simulations are expressed in standard reduced Lennard-Jones units. See ref~\citenum{2020_Mac_53_6828} for details about reduced units and their mapping to laboratory units. Panels (b--d) were adapted with permission from ref~\citenum{2013_JCP_138_12A541}.}
\end{figure*}

As noted briefly above, the development of the string model~\cite{2014_JCP_140_204509} was motivated by extensive simulations of both polymeric and other GF liquids~\cite{1998_PRL_80_2338, 2003_JCP_119_5920, 2011_PRL_106_115702, 2013_SoftMatter_9_241, 2013_JCP_138_12A541, 2014_NatCommun_5_4163, 2014_JCP_140_204509, 2015_PNAS_112_2966, 2015_JCP_142_234907} indicating that the most mobile particles in GF liquids form stringlike structures defined in terms of their cooperative exchange motion over the lifetime in which the strings exist. The technical details for quantifying the stringlike cooperative motion have been repeatedly described in previous works and we refer the reader to ref~\citenum{2003_JCP_119_5920} for a discussion of this procedure. Such cooperative motion is broadly consistent with the philosophical assumptions of the AG model.~\cite{1965_JCP_43_139} Figure~\ref{Fig_CRR}a shows a snapshot of stringlike clusters of polymer segments undergoing cooperative exchange motion over the lifetime over which these clusters persist in a simulated polymer melt.~\cite{2013_SoftMatter_9_241} We note that there is little correlation between these stringlike motions and chain connectivity, so these stringlike motions are not `reptative', i.e., the segments in these motions are not correlated with the chain backbone. Such collective motion occurs in a rather similar fashion even in metallic GF liquids where there are no chemical bonds. Starr et al.~\cite{2013_JCP_138_12A541} have investigated what measure of dynamic heterogeneity may appropriately quantify the size scales of the CRR envisioned by AG based on simulations of a coarse-grained polymer melt, where $S_c$ has been calculated. It has been shown that $S_c$ is inversely proportional to the average string length $L$ to a good approximation, as illustrated in Figure~\ref{Fig_CRR}b. Moreover, taking the size of the CRR as $z = L / L_A$ with $L_A$ being the value of $L$ at $T_A$, the linear relation between $\ln \tau_{\alpha}$ and $z / T$ is confirmed by the analysis in Figure~\ref{Fig_CRR}b. Figure~\ref{Fig_CRR}d also verifies the linear relation between $\ln \tau_{\alpha}$ and $1 / TS_c$. Further, $L / L_A$ has been shown to track the normalized activation free energy, $\Delta G / \Delta G_0$.~\cite{2014_JCP_140_204509} Therefore, these results convincingly demonstrate that the dynamic strings provide a \textit{quantitative realization} of the hypothetical CRR of AG.~\cite{1965_JCP_43_139} 

In the string model of glass formation, applicable in the $T$ range below $T_A$ where relaxation is non-Arrhenius, the activation free energy $\Delta G$ for structural relaxation is proportional to the average string length $L$ normalized by its value at the onset temperature $T_A$. As a formal extension of TST,~\cite{1941_CR_28_301, Book_Eyring} the string model of glass formation defines the $T$-dependent activation free energy as $\Delta G(T) = \Delta G_0 (L/L_A)$, leading to the following expression for $\tau_{\alpha}$, 
\begin{equation}
	\label{Eq_String}
	\tau_{\alpha} = \tau_0 \exp\left( \frac{\Delta G_0}{k_BT} \frac{L}{L_A} \right).
\end{equation}
The parameter $\tau_0$ can be eliminated from a knowledge of $\tau_{\alpha}$ at $T_A$,~\cite{2015_PNAS_112_2966} and $\Delta H_0$ may be determined from the Arrhenius equation (eq~\ref{Eq_Arrhenius}) in the high $T$ regime where standard TST is assumed to be applicable as a descriptive framework for liquid dynamics, resulting in the following equation with $\Delta S_0$ being the only fitting parameter,
\begin{equation}
	\tau_{\alpha} = \tau_{\alpha} (T_A)\exp\left(\frac{\Delta H_0 - T\Delta S_0}{k_BT}\frac{L}{L_A} - \frac{\Delta H_0 - T_A\Delta S_0}{k_BT_A}\right).
\end{equation}
Notably, simulation studies indicate that collective motion does not completely vanish in the Arrhenius regime,~\cite{2013_JCP_138_12A541} i.e., $L_A$ is not equal to unity. Moreover, the string model asserts that $\Delta S_0$ cannot be neglected, but the spirit of the AG model is certainly preserved in this model, despite this and other technical differences.~\cite{2014_JCP_140_204509} Therefore, the essence of the string model of glass formation is that the activation free energy in the high $T$ Arrhenius regime is `renormalized' by the $T$-dependent factor $L/L_A$ quantifying the change in the extent of collective motion in the non-Arrhenius regime below $T_A$.

\begin{figure*}[htb!]
	\centering
	\includegraphics[angle=0,width=0.975\textwidth]{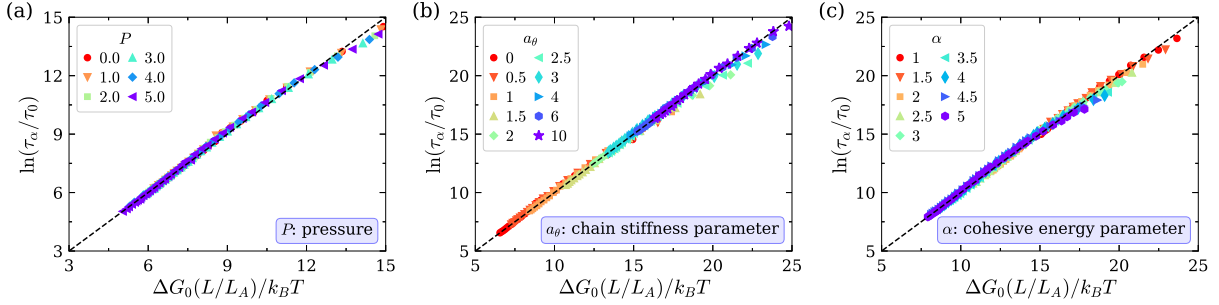}
	\caption{\label{Fig_StringTest}String model description of the relationship between $\tau_{\alpha}$ and $L$. (a--c) Results for a simulated coarse-grained polymer melt with variable pressure, chain stiffness, and cohesive interaction strength, respectively. Lines indicate $\ln (\tau_{\alpha}/\tau_0) = \Delta G_0(L/L_A)/k_BT$. Adapted with permission from refs~\citenum{2020_Mac_53_6828},~\citenum{2020_Mac_53_4796}, and~\citenum{2020_Mac_53_9678}.}
\end{figure*}

Equation~\ref{Eq_String} of the string model has been quantitatively confirmed in simulations of a number of polymeric and other GF liquids, including knotted ring and star polymer melts,~\cite{2019_JCP_150_101103, 2018_JCP_149_161101, 2020_JCP_152_054904} thin films on solid substrates,~\cite{2014_NatCommun_5_4163, 2015_JCP_142_234907} polymer nanocomposites having a range of concentrations and polymer-surface interaction strengths,~\cite{2015_PNAS_112_2966} polymer melts with variable pressure,~\cite{2016_MacroLett_5_1375, 2017_Mac_50_2585, 2020_Mac_53_6828} chain stiffness,~\cite{2020_Mac_53_4796} and cohesive interaction strength,~\cite{2016_Mac_49_8355, 2020_Mac_53_9678} polymer nanofibers,~\cite{2017_SoftMatter_13_1190} metallic liquids,~\cite{2015_JCP_142_164506} and superionic $\mathrm{UO}_2$.~\cite{2019_JCP_150_174506} For the purposes of illustration, we show in Figure~\ref{Fig_StringTest} that the string model describes well the simulation results in a coarse-grained polymer melt, where the pressure, chain stiffness, or cohesive interaction strength is systematically varied. The string model is further justified by a theoretical analysis by Freed,~\cite{2014_JCP_141_141102} where TST is extended to account for stringlike cooperative barrier crossing events in GF liquids. However, much work needs to be done to extend the string model to a wide range of other materials. We again emphasize that even standard TST in the Arrhenius regime remains in a relatively rudimentary theoretical state.

In passing, we mention that relaxation in GF liquids occurs as a multi-stage hierarchical process.~\cite{2018_JCP_148_104508} While we evidently focus on the long-time structural relaxation process involving both large-scale diffusive molecular motion and momentum diffusion, there is a `fast' relaxation process dominated by the inertial motion of the molecules whose amplitude grows upon heating. It is worth pointing out that the fast dynamics can be considered in a unified way with the long-time structural relaxation within the string model, as indicated in a recent work.~\cite{2018_JCP_148_104508}

\subsection{\label{Sec_LT}Temperature Dependence of String Length}

The string model~\cite{2014_JCP_140_204509} not only identifies the CRR with the stringlike cooperative motion, but it also provides an explanation for the origin and geometrical nature of these strings based on the framework of treating them as `initiated equilibrium polymers'.~\cite{1999_JCP_111_7116} This has been discussed in detail in ref~\citenum{2014_JCP_140_204509}. Here, we briefly review the most interesting predictions from the living polymerization model of strings.

In the model of initiated equilibrium polymerization,~\cite{1999_JCP_111_7116} the strings are dynamic or `equilibrium' polymers that form and disintegrate in equilibrium. These dynamic polymers have a $T$-dependent average length, or polymerization index, $L \equiv \langle L \rangle$. The fraction of linked mobile particles $\Phi$ serves as the order parameter for the self-assembly process of strings, which is related to $L$ via the relation,
\begin{equation}
	\label{Eq_L1}
	L = \frac{1}{1 - \Phi + \delta r/2},
\end{equation}
where $\delta r$ is the ratio of the initiator to the monomer volume fraction $\phi_0$. $\Phi$ is limited to the range between $\delta r$ and $1$. Taking the onset temperature $T_A$ as the reference, along with the approximation of $\delta r \approx \Phi_A$ with $\Phi_A$ being the value of $\Phi$ at $T_A$, eq~\ref{Eq_L1} can be written as~\cite{2014_JCP_140_204509} 
\begin{equation}
	\label{Eq_L2}
	L = \frac{L_A(1 - \Phi_A/2)}{1 - \Phi + \Phi_A/2}.
\end{equation}

In the living polymerization model, $\Phi$ is the extent of polymerization defined by the fraction of monomers forming polymeric structures, and, in simulations, this quantity can be interpreted to be the fraction of the highly mobile particles participating in the strings. The variation of $\Phi$ with $T$ is well described by the prediction from the polymerization model,~\cite{1999_JCP_111_7116}
\begin{equation}
	\label{Eq_Phi0}
	\Phi = 1 - \phi(T)/\phi_0,
\end{equation}
where the explicit expression for $\phi(T)$ is given in refs~\citenum{2014_JCP_140_204509} and~\citenum{1999_JCP_111_7116}. $\Phi$ is directly related to $S_c$ in the theory of equilibrium polymerization, a relation that has been validated in simulations of a coarse-grained GF polymer melt,~\cite{2014_JCP_140_204509} where $\Phi$ and $S_c$ are independently calculated. $\Phi$ is notably much easier to calculate than $S_c$ in simulations. As we have discussed earlier in Figure~\ref{Fig_CRR}, a simulation study based on the same polymer melt has also confirmed an inverse scaling between the average string length $L$ and $S_c$ to a very good approximation,~\cite{2013_JCP_138_12A541} a basic tenet of the AG theory~\cite{1965_JCP_43_139} and the thermodynamic relationship in the equilibrium polymerization model.~\cite{2006_JCP_125_144907}

While the full expression for $L$ predicted by the string model~\cite{2014_JCP_140_204509} is complicated mathematically, this expression can be simplified by using the high $T$ expansion of $\Phi$,~\cite{1999_JCP_111_7116}
\begin{equation}
	\label{Eq_LHighT}
	L = L_A \left(1 -\frac{\Phi_A}{2}\right) \left\{1 + \frac{\Phi_A}{2}\left[1 + \phi_0\exp\left(-\frac{\Delta G_p}{k_BT}\right)\right]\right\},
\end{equation}
where the free energy $\Delta G_p = \Delta H_p - T\Delta S_p$ describes the thermodynamics of string polymerization rather than the activation free energy of the fluid. Equation~\ref{Eq_LHighT} is expected to be valid in the $T$ range above $T_c$ but below $T_A$, which we have confirmed in simulations of coarse-grained polymer melts with variable cohesive energy and pressure.~\cite{2016_Mac_49_8355, 2017_Mac_50_2585} It is an important feature of equilibrium polymerization that the excess degree of polymerization in the high $T$ regime, $L_A - 1$, is directly related to the plateau of $L$ in the low $T$ regime, where both the degree of polymerization and the fluid configurational entropy saturate to finite values associated with the fully polymerized state. This feature of the polymerization model has highly non-trivial implications for the string model of glass formation, since it links collective motion in the high $T$ Arrhenius regime to the scale of collective motion in the equilibrium glass state, where relaxation again becomes Arrhenius in the string model of the dynamics of GF liquids under equilibrium conditions. Of course, equilibrium conditions at low $T$ are normally difficult to achieve in practice, so this prediction is somewhat of academic interest, just as the question of whether or not $S_c$ actually vanishes at a finite $T$.

\begin{figure*}[htb!]
	\centering
	\includegraphics[angle=0,width=0.975\textwidth]{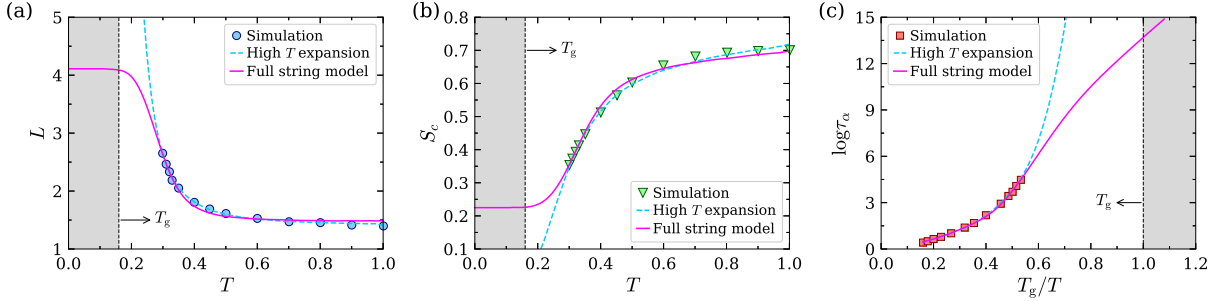}
	\caption{\label{Fig_String}$T$ dependence of the thermodynamic and dynamic properties of glass formation predicted by the string model, along with comparisons with simulation results of a coarse-grained polymer melt. (a) $L$ and (b) $S_c$ as a function of $T$. (c) Angell plot of $\log \tau_{\alpha}$. Solid and dashed lines correspond to the full form of the string model and its approximant based on the high $T$ expansion, respectively. The filled regions indicate the $T$ regime below $T_{\mathrm{g}}$ estimated from the full string model, which is determined by the condition, $\tau_{\alpha} = 100$ s, corresponding to $10^{14}$ in standard reduced Lennard-Jones units. Adapted with permission from ref~\citenum{2014_JCP_140_204509}.}
\end{figure*}

Figure~\ref{Fig_String}a examines the validity of the polymerization model in describing the variation of $L$ with $T$ determined from simulations of a coarse-grained polymer melt.~\cite{2014_JCP_140_204509} As can be seen, the string model, either in its full form (given in ref~\citenum{2014_JCP_140_204509}) or its simplified form based on the high $T$ expansion (eq~\ref{Eq_LHighT}), provides an excellent description of the available simulation data, which are restricted to a $T$ range between $T_A$ and $T_c$. However, large differences arise at low $T$ between the full and simplified string models, which has important implications for glass formation. While $L$ from the high $T$ expansion increases rapidly upon cooling at low $T$, the full form for $L$ instead exhibits a plateau at low $T$. Since $L$ inversely scales with $S_c$ in the string model, we see from Figure~\ref{Fig_String}b that the string model makes similar predictions for $S_c$. Note that the plateau of $S_c$ at low $T$ in the full string model is reminiscent of the prediction from the GET for fully flexible polymer melts, as discussed in Section~\ref{Sec_ScGET}. We note that a recent study on a metallic GF liquid, where the estimation of $S_c$ is less difficult than for polymers because the large vibrational contributions associated with molecular bonding to the fluid entropy are absent, has revealed the presence of an extended low $T$ plateau in $S_c$.~\cite{2020_PRB_101_014113}

Combing the functional form for $L$ provided by the living polymerization model with eq~\ref{Eq_String}, the string model goes beyond the relatively high $T$ regime of glass formation accessible to present simulations and make predictions for $\tau_{\alpha}$ over the entire $T$ regime of glass formation. This is illustrated in Figure~\ref{Fig_String}c, where we show the Angell plot of $\log \tau_{\alpha}$ determined from simulations, along with the descriptions based on the string model. It is evident that the full string model indicate a return to Arrhenius relaxation in the glass state, albeit with a higher effective activation free energy. We emphasize that the string model~\cite{2014_JCP_140_204509} implies no divergence in $\tau_{\alpha}$ at any finite $T$. This is in accord with the GET predictions for flexible polymer melts~\cite{2016_JCP_145_234509} and the observations on a number of liquids, such as water, silica, and metallic GF liquids,~\cite{2010_JCP_133_014508} which show a `fragile-to-strong' transition upon sufficient cooling, corresponding to a return to Arrhenius relaxation. The interesting thing about GF liquids exhibiting a fragile-to-strong transition is that the Arrhenius relaxation seems to occur in a $T$ regime where the liquid is apparently fully in equilibrium. Note that the position of the transition to a low $T$ Arrhenius behavior as well as the fragility of glass formation and the low $T$ plateau in the activation energy all depend on $\Phi_A$ in the string model, a quantity determined in the Arrhenius regime.

\subsection{String Model and Phenomenological Models of Glass Formation}

While the string model~\cite{2014_JCP_140_204509} aims to describe the dynamics of GF liquids in terms of parameters and quantities with well-defined meanings, there are popular models of glass formation that are highly phenomenological in nature. Here, we utilize the string model to understand some of these phenomenological models, which has been discussed in refs~\citenum{2014_JCP_140_204509} and~\citenum{2020_Mac_53_9678}. In particular, we show that the string model enables the rationalization of the empirical fitting function for $\tau_{\alpha}$ introduced by R{\"o}ssler and coworkers~\cite{2012_PRE_86_041507, 2013_JCP_139_084504, 2015_Mac_48_3005} and the Mauro-Yue-Ellison-Gupta-Allan (MYEGA) model.~\cite{2009_PNAS_106_19780}

As discussed in Section~\ref{Sec_LT}, the order parameter $\Phi$ for the assembly of mobile particles into strings can be derived from the equilibrium polymerization theory. However, the exact mathematical description of $\Phi$ is complicated. Douglas and coworkers~\cite{2006_JCP_125_144907, 2008_JCP_128_224901} found that a simple two-state model can be used to describe $\Phi$ in the cooperative polymerization model,
\begin{equation}
	\label{Eq_Phi1}
	\Phi = \frac{1}{1 + \exp[(\Delta H_p - T \Delta S_p )/k_BT]}.
\end{equation}
This two-state expression arises in many physical contexts and has often been further approximated by expanding around the free energy of association about the polymerization transition point $T_p$ corresponding to the condition of $\Phi = 1/2$, which gives rise to the following approximation,~\cite{2008_JCP_128_224901}
\begin{equation}
	\label{Eq_Phi2}
	\Phi \approx \frac{1}{1 + \exp[-\Delta H_p(T - T_p )/k_BT_p^2]} \equiv \frac{1}{1 + \exp[(T - T_p)/\mathcal{D}_0]},
\end{equation}
where $\mathcal{D}_0 \equiv k_B T_p^2/|\Delta H_p|$ reflects the width of the clustering transition. Reference~\citenum{2006_JCP_125_144907} provides more details regarding the above argument. The two-state model predicts that $L/L_A$ should equal $2$ at $T_p$ at which the $T$ dependences of both $\Phi$ and $s_c$ exhibit an inflection point, in accord with the observation of Ngai and coworkers~\cite{1999_JCP_111_3639, 1999_JPCB_103_4045} that the apparent activation energy near $T_c$ is normally about twice the high $T$ activation energy. In the Stockmayer fluid of dipolar particles~\cite{2005_PRE_71_031502} and tabletop measurements of driven magnetic particles thermalized by vertical shaking,~\cite{2005_PRE_72_031301} the average string length $L$ has been observed to be generally near $2$. In the GET, the crossover temperature $T_c$ of glass formation is identified by the temperature at which the $T$ dependence of $Ts_c$ has an inflection point, so we may expect a close connection between $T_p$ and $T_c$.

If the above approximations for $\Phi$ are combined with the string model, an approximation for the activation free energy emerges as,
\begin{equation}
	\label{Eq_Rossler}
	\Delta G_0 (L/L_A) \approx \Delta G_0 \{1 + \exp[-(T-T_p)/\mathcal{D}_0]\},
\end{equation}
which is the same form as the expression for the activation free energy proposed by R{\"o}ssler and coworkers.~\cite{2015_Mac_48_3005} In this two-state model expression, the activation free energy is somewhat artificially broken into a `local' contribution $\Delta G_0$ and a `cooperative' contribution representing the stringlike collective motion. This type of two-state model has a long history in the modeling of GF liquids.~\cite{2000_JNCS_274_131, 2012_SciRep_2_713} The ECNLE model of Schweizer and coworkers~\cite{2014_JCP_140_194506, 2014_JCP_140_194507, 2015_Mac_48_1901} assumes a similar decomposition of the activation free energy in local and collective parts where the collective motions are conceived to arise from the energetic costs arising from the emergence of caging in liquids at low $T$ or high densities. Tanaka and coworkers~\cite{2018_PNAS_115_9444} have recently invoked a mean-field two-state model of liquid dynamics that seems similar in spirit for the activation energy for structural relaxation in water. Note that eq~\ref{Eq_Rossler} is a rough mean-field model that neglects the distribution of barrier heights associated with the structural organization of the dynamic heterogeneity, an aspect that is emphasized in the string model.~\cite{2014_JCP_140_204509} 

We may also formally recover the recent fashionable MYEGA expression~\cite{2009_PNAS_106_19780} for the $T$ dependence of $\tau_{\alpha}$ by approximating $L$ by its high $T$ approximation in its well-known equilibrium polymerization model estimate,~\cite{1990_JPCM_2_6869, 2005_PRE_71_031502, 2003_JCP_119_12645, 2007_JCP_126_194903}
\begin{equation}
	L \approx L_0 \exp[\Delta H_p / (2 k_B T)],
\end{equation}
where $L_0$ is a constant determined by molecular parameters of the polymerization model. The Arrhenius variation of $L$ holds exactly for purely uncooperative or `isodesmic' equilibrium polymerization and provides a generic approximation for $L$ for equilibrium polymerization processes near and above the thermodynamic transition temperature $T_p$. However, this simple relation for $L$ no longer holds generally at low $T$ in the case in which the polymerization process becomes cooperative.~\cite{2014_JCP_140_204509} Correspondingly, we may thus expect the MYEGA equation,~\cite{2009_PNAS_106_19780} as well as the popular VFT relation,~\cite{1921_PZ_22_645, 1925_JACS_8_339, 1926_ZAAC_156_245} to only hold over a limited $T$ range. Therefore, we can also understand the success of these phenomenological fitting functions for $\tau_{\alpha}$ based on the string model.

\subsection{Opportunities and Challenges}

Although stringlike collective motion is a conspicuous feature of the dynamics of simulated GF liquids broadly in the regime where the dynamics becomes non-Arrhenius and has also been observed in GF colloidal fluids,~\cite{1999_PRE_60_5725, 2011_PRL_107_208303} the grain boundaries of colloidal and granular systems,~\cite{2011_PNAS_108_11323, 2010_PRE_81_041301} and colloidal fluids approaching their melting,~\cite{2014_PNAS_111_15356, 2017_PRL_118_088003} there is currently no obvious experimental method for identifying the strings in molecular fluids. There is clearly a need for getting information about the strings in materials of practical interest.

In our recent works,~\cite{2020_Mac_53_6828, 2020_Mac_53_7239} we hypothesized that the activation volume $\Delta V^{\#}$ might be related to the average string length $L$, based on a number of studies~\cite{1959_JPS_34_349, 1962_JCP_37_2785, 1975_JPS_13_1737, 1987_JPC_91_4169, 1994_TA_238_41, 2005_JPCB_109_16567, 2009_JCP_131_194511, 2011_JNCS_357_351, 2011_PRE_83_061508, 2012_JPCM_24_065105, 2015_EPJE_38_91} showing that the growth of $\Delta V^{\#}$ upon cooling occurs in parallel with the apparent activation energy of GF liquids below $T_A$. If such a relation could be established between $\Delta V^{\#}$ and $L$, then it would provide an accessible method for determining $L$, the central quantity in the string model of the dynamics of GF liquids. Since measurements of the length scales characterizing the extent of collective motion are often quite challenging to make and involve advanced instrumentation that is often not available in an industrial research setting, while $\Delta V^{\#}$ can be straightforwardly measured experimentally, a direct relation between $\Delta V^{\#}$ and $L$ is particularly appealing from an experimental viewpoint. Unfortunately, our study based on simulations and the GET indicates that there is no direct relation between $\Delta V^{\#}$ and the extent of collective motion quantified by $L$.~\cite{2020_Mac_53_6828, 2020_Mac_53_7239} 

Although the strings can be directly observed in measurements of colloidal suspensions, we continue the search for an accessible experimental metric for $L$ based on correlation studies. For instance, previous works~\cite{2013_SoftMatter_9_1254, 2013_SoftMatter_9_1266} on energy and mobility fluctuations in the glassy interfacial dynamics of nanoparticles indicate that the color of these fluctuations correlates strongly with $L$. The basic idea here is that mobility fluctuations associated with mobile particles, such as strings, should lead to observable effects on properties that measure fluctuations in the mobility, such as electrical conductivity.~\cite{2010_SoftMatter_6_5944, 2015_JCP_142_084704, 2017_JCP_147_194508} In the future work, we need to check these relationships between noise color associated with mobility fluctuations and $L$ in cooled liquids below $T_A$ and see if we can develop an accessible method for estimating $L$ in molecular and polymeric liquids. There is also the prospect of observing stringlike collective motion on the surface of crystalline and GF materials in real space based on ultrahigh resolution imaging.~\cite{1996_PRB_53_13547, 2003_NatMater_2_783} The imaging studies, even in the case of the strings in the colloidal systems,~\cite{2012_RPP_75_066501, 2017_MacroLett_6_27} needs improvement in order to compare quantitatively with the string model. Another promising possibility seems to be offered by single molecule fluorescence microscopy.~\cite{2016_SoftMatter_12_7299, 2017_Polymer_116_452} This technique allows for investigations of the rotational motion of single fluorescence probes with different sizes doped in a GF polymer, leading to the identification of a characteristic length scale that might be related to $L$, but much work remains to be done to establish such a relationship quantitatively. We look forward to participating in these developments in the future.

\section{Other Entropy-Based Models of Liquid Dynamics}

The strong correlation between the fluid entropy and the dynamics of GF liquids has led to alternative models to the AG model~\cite{1965_JCP_43_139} and its antecedents, the GET and the string model discussed above. For completeness, we provide a brief overview of these alternative models, with an emphasis on the hints provided by these model into how the GET might be developed in the future. 

\subsection{\label{Sec_RFOT}Random First-Order Transition Theory}

The RFOT theory~\cite{1987_PRB_36_8552, 1989_PRA_40_1045, 2004_JCP_121_7347, 2007_ARPC_58_235} is close in spirit to the AG model~\cite{1965_JCP_43_139} in the sense that it is predicated on the idea of dynamic domains in cooled liquids which grow upon cooling, accompanied by a corresponding drop in the configurational entropy, $S_c$. This mean-field model, initially formulated in the context of spin glass materials,~\cite{1987_PRB_36_8552} has also elements similar to the GD model~\cite{1958_JCP_28_373} of polymer glass formation in the sense that it predicts $S_c$ to vanish at a Kauzmann temperature, $T_K$.

The RFOT theory~\cite{1987_PRB_36_8552, 1989_PRA_40_1045, 2004_JCP_121_7347, 2007_ARPC_58_235} envisions that the liquid is divided into metastable regions with a characteristic correlation length $\xi$, describing the average `entropic droplet' or `mosaic' size. A scaling relation between $\xi$ and $S_c$ is suggested by a consideration of the balance between the surface and bulk free energies of these regions. This theory also assumes that the activation energy for relaxation scales with $\xi$ with a power. It should thus be appreciated that the general qualitative findings of this interesting mean-field model have many points of contact with the AG~\cite{1965_JCP_43_139} and GET models.~\cite{2008_ACP_137_125} A particularly interesting aspect of the RFOT theory is that it focuses on the geometrical size of what may be identified with the CRR of the AG model, rather than the number of particles within these hypothetical clusters.

As in the AG model,~\cite{1965_JCP_43_139} a basic problem about the implementation of the RFOT model~\cite{1987_PRB_36_8552, 1989_PRA_40_1045, 2004_JCP_121_7347, 2007_ARPC_58_235} for polymeric fluids is that $S_c$ can only be roughly estimated based on similar approximations introduced by Bestul and Chang~\cite{1964_JCP_40_3731} and AG,~\cite{1965_JCP_43_139} where the uncertainties are especially large in polymer fluids because of the relatively large vibrational entropy associated with molecular bonding in such fluids.~\cite{2013_JCP_138_12A541} We note that an interesting methodology for estimating $S_c$ for non-molecular GF liquids based on a RFOT framework has been introduced,~\cite{2014_PNAS_111_11668} so this situation could conceivably change in the future if this methodology can be extended to treat molecular GF liquids. Steven and Wolynes~\cite{2005_JPCB_109_15093} have performed a test of the RFOT model based on estimates of $S_c$ obtained from specific heat data for a range of small-molecule liquids where this type of $S_c$ estimate is expected to be a reasonable rough approximation. We view the estimation of $S_c$ from specific heat measurements to be inherently unreliable for quantitative analysis, however, and we strictly avoid any discussion of this method for estimating $S_c$. Although $S_c$ can be estimated for molecular fluids from simulations,~\cite{2013_JCP_138_12A541} including polymers, the method is limited in the $T$ range in which equilibrium simulations can be performed. The RFOT theory~\cite{1987_PRB_36_8552, 1989_PRA_40_1045, 2004_JCP_121_7347, 2007_ARPC_58_235} offers no method for calculating $S_c$ and other parameters of the model. The GET~\cite{2008_ACP_137_125} is founded on a statistical mechanical model of polymer fluids, which enables calculations of all the thermodynamic properties in terms of molecular parameters, including $S_c$. The LCT for polymer thermodynamics,~\cite{1998_ACP_103_335, 2014_JCP_141_044909} which is the thermodynamic component of the GET,~\cite{2008_ACP_137_125} has been validated by many studies.~\cite{2005_APS_183_63} By combining the LCT with the AG model,~\cite{1965_JCP_43_139} the GET~\cite{2008_ACP_137_125} is a highly predictive theory with no free parameters beyond the molecular and thermodynamic parameters governing the thermodynamic state of polymer fluids. Of course, the extension of the GET to include the entropy of activation, a quantity neglected by AG,~\cite{1965_JCP_43_139} makes the GET no longer fully predictive. We consider below how this limitation of the GET might be overcome. An interesting difference between the AG and RFOT models is that the dimensionless activation free energy is deduced to scale in the RFOT theory~\cite{2003_JCP_119_9088, 2008_JPCB_112_301} as $\Delta G / k_B T \sim 1 / S_c$, which is consistent with the phenomenological expression first noted by Bestul and Chang rather than the AG expression, $\Delta G / k_B T \sim 1 / (TS_c)$.~\cite{2003_JCP_119_9088} This RFOT expression is then consistent with the GET when the enthalpy $\Delta H_0$ of activation is set to zero, in which case the free energy of activation is taken to be dominated by the entropy of activation, $\Delta S_0 > 0$, an assumption suitable for the description of systems where particle interactions are dominated by repulsive interactions, as in the case of hard-sphere fluids at constant density. Note that the RFOT model emphasizes the configurational entropy per unit mass rather than the entropy density, which has ramifications for considering the glass formation of materials at constant pressure.

It might be possible to combine the LCT with the activation free energy scaling arguments underlying the RFOT approach~\cite{1987_PRB_36_8552, 1989_PRA_40_1045, 2004_JCP_121_7347, 2007_ARPC_58_235} to glass formation, but the model seems to involve unspecified parameters that we currently do not know how to specify and various assumptions that require careful assessment, as we have tried to perform in the GET. We plan to explore this alternative approach to the dynamics of GF liquids in the future.

At this stage, we mention an interesting exploration of the compatibility of the string model of glass formation and the RFOT scaling arguments for the dynamics of GF liquids by Starr et al.~\cite{2013_JCP_138_12A541} This work adopts the hypothesis that the `string' clusters exhibiting cooperative exchange motion, identified successfully as having the main attributes of the hypothetical CRR of AG,~\cite{1965_JCP_43_139} can also be identified with the hypothetical `entropic droplets' of the RFOT model as the logical compatibility of the AG and RFOT models would formally suggest. Verification of such a relation would be important because the RFOT theory offers no explicit algorithm for what might constitute an `entropic droplet' and is equally vague as the AG model regarding the physical nature of the CRR. It has often been assumed in theoretical discussions of the AG and RFOT models that the CRR are more or less spherical,~\cite{1965_JCP_43_139, 1982_JNCS_53_325, 2009_PNAS_106_11506, 2000_PNAS_97_2990} in line with the term droplet, but the basis of this assumption is rather unclear. Klein and coworkers~\cite{1986_PRL_57_2845, 1998_PRE_57_5707, 2000_PRL_85_1270, 1983_PRB_28_445} have considered the role of fluctuations on the formation of dynamic clusters in model GF liquids, indicating that while compact clump-like structures are found in mean-field theory descriptions of GF materials, the clusters become progressively more ramified, i.e., fractal, when fluctuation effects associated with the short-range attractive interactions of real liquids are incorporated into the theory, a finding having profound implications for both glass formation and crystal nucleation at high undercooling.~\cite{1990_JCP_93_711, 2006_PRL_97_105701} Resolving the physical nature of the `entropic droplets' is then a central question in developing the RFOT theory into a quantitative model of the dynamics of GF liquids and we discuss this matter below.

The RFOT scaling expression for the activation free energy reads $\Delta G \sim \xi^{\psi}$, where $\psi$ is a surface tension exponent whose value depends on the assumed geometry of the entropic droplets. See ref~\citenum{2013_JCP_138_12A541} for an extended discussion of $\psi$. As noted above, identifying $\xi$ with the radius of gyration of the strings $R_{\mathrm{g}, \mathrm{string}}$ indicates that $\Delta G = A R_{\mathrm{g}, \mathrm{string}}^{\psi}$ with $A$ and $\psi$ being adjustable constants.~\cite{2013_JCP_138_12A541} This preliminary test of the RFOT model indicates compatibility between the AG and RFOT models in the qualitative sense that the free energy barrier should grow with the size of regions exhibiting cooperative motion. However, the exponent $\psi$ is not compatible with compact liquid droplet estimates of this scaling exponent, suggesting that these field excitations should be modeled as being fractal structures, as in the case of the string model. We may also view this scaling of the activation free energy with as being compatible with the string model based on the observed scaling relation,~\cite{2013_JCP_138_12A541, 2015_JCP_142_164506} $R_{\mathrm{g}, \mathrm{string}} \sim L^{\nu}$, where $\psi = 1 / \nu$ in this interpretation of the $\Delta G$ scaling with $1 / \nu$ being the fractal dimension of the self-avoiding polymeric strings. This interpretation of the free energy scaling with the average string size would lead to $\psi = 5/3$ if we assume the Flory estimate of $\nu$.~\cite{Book_Freed} This estimate of $\psi$ roughly accords with the value of $\psi \approx 1.3$ estimated from simulations of polymer melts by Starr et al.~\cite{2013_JCP_138_12A541} 

Reference~\citenum{2013_JCP_138_12A541} has discussed the theoretical RFOT estimates of $\psi$ in comparison to computational estimates, along with the implications of the observed scaling of $\Delta G$ with $R_{\mathrm{g}, \mathrm{string}}$, so we do not elaborate on this matter further here. Our main point here is that the RFOT approach appears to be promising, provided that a more realistic physical description of the `entropic droplets' is considered. Despite the apparent success of the RFOT scaling, there are still issues that must be faced. In particular, this scaling approach still involves a number of practical issues regarding predictions of relaxation times for particular materials. We currently do not understand the meaning of the parameters $A$ and $\psi$ in the RFOT model so that quantitative predictions of relaxation times remain elusive. In our view, there seems to be no demonstrated advantage of the RFOT model~\cite{1987_PRB_36_8552, 1989_PRA_40_1045, 2004_JCP_121_7347, 2007_ARPC_58_235} over the AG model.~\cite{1965_JCP_43_139} This exercise is valuable, however, in the sense that it provides a new perspective on the AG model and indicates that further attention should be given to the geometrical structure of the strings in developing the entropy theory of glass formation into a more general and validated theoretical framework. This effort remains a work in progress.

\subsection{\label{Sec_Excess}Rosenfeld and Excess-Entropy Scaling}

There is another interesting approach to the dynamics of liquids based on an attempt to interrelate the rate of diffusion and relaxation to the fluid entropy. Based on a conception going back to Boltzmann that both the thermodynamics and dynamics of liquids is dominated by hard-core intermolecular interactions,~\cite{1971_JCP_54_5237} Rosenfeld~\cite{1977_PRA_15_2545, 1999_JPCM_11_5415, 2018_JCP_149_210901} developed a scaling model of the dynamics of liquids emphasizing a formal mapping of liquids onto hard spheres based on a consideration of the entropy in the liquid state, $S_{\text{liq}}$, relative to the entropy in the ideal gas state, $S_{\mathrm{id}}$, rather than the crystal or glass state as in the case of AG.~\cite{1965_JCP_43_139} See ref~\citenum{2018_JCP_149_210901} for a detailed discussion of Rosendeld's works. This `excess entropy', \textit{relative to the ideal gas}, denoted as $\widehat{S}_{\mathrm{ex}} = S_{\text{liq}} - S_{\mathrm{id}}$, should not be confused with other definitions of the `excess entropy' $S_{\mathrm{exc}}$ in the context of the AG model. In particular, Rosenfeld argued that transport properties of liquids such as the diffusion coefficient $D$ should scale with $\widehat{S}_{\mathrm{ex}}$ as,
\begin{equation}
	\label{Eq_Rosenfled}
	D \sim \exp(B \widehat{S}_{\mathrm{ex}}),
\end{equation}
where $B$ is an adjustable positive constant.~\cite{2018_JCP_149_210901} This particular exponential variation of dynamic properties is termed `Rosenfeld scaling' and the description of $D$ and $\tau_{\alpha}$ by more general functional forms is termed `excess-entropy scaling'.~\cite{2018_JCP_149_210901} These two distinct terms indicate that Rosenfeld scaling has its limitations, as we shall discuss below. There is also a popular extension of the Rosenfeld scaling model introduced by Dzugutov~\cite{1996_Nature_381_137} based on an approximation of $\widehat{S}_{\mathrm{ex}}$ by its two-body estimate from the radial distribution function $g(r)$,
\begin{equation}
	\label{Eq_S2}
	S_2 = -2 \pi \rho \int_0^{\infty} \{ g(r) \ln g(r) - [g(r) - 1] \} r^2 dr,
\end{equation}
where $r$ is the spatial distance. For readers familiar with the AG model,~\cite{1965_JCP_43_139} Rosenfeld scaling appears to be rather odd, but it must be remembered that the sign of $\widehat{S}_{\mathrm{ex}}$ is opposite to $S_c$. Note also that $\widehat{S}_{\mathrm{ex}}$ is in the numerator of eq~\ref{Eq_Rosenfled}, while $S_c$ is in the denominator in the corresponding AG relation for $D$ and $\tau_{\alpha}$. It is then easy to appreciate why confusion is sometimes encountered regarding how the fluid entropy should be related to liquid dynamics. Interestingly, simulation results have indicated that $\widehat{S}_{\mathrm{ex}}$ actually scales inversely with the product of the numerically estimated $S_c$ and $T$ in model GF liquids,~\cite{2006_JCP_125_076102} as required by self-consistency between Rosenfeld scaling and the AG model, a relation that seems rather mysterious to us. Particularly clear expositions of the excess-entropy scaling approach to liquid dynamics have been made by Chakraborty and coworkers~\cite{2006_JCP_124_014507} and Sastry and coworkers.~\cite{2014_PRL_113_225701} An extensive and excellent review on Rosenfeld scaling has recently been given by Dyre.~\cite{2018_JCP_149_210901} Speedy et al.~\cite{1988_MP_66_577} have discussed the limitations of the hard-core approximation in describing the dynamics of liquids. 

We can calculate $\widehat{S}_{\mathrm{ex}}$ from the LCT~\cite{1998_ACP_103_335, 2014_JCP_141_044909} to implement this type of entropy theory. However, we did not do so because numerous simulations have shown that while this model seems to hold rather well in the high $T$ regime, large deviations from Rosenfeld scaling are normally observed in GF liquids at low $T$.~\cite{2018_JCP_149_210901} In many fluids, the simple exponential form of $D$ and $\tau_{\alpha}$ in terms of $\widehat{S}_{\mathrm{ex}}$ is superseded by other functional forms that can be applied with impressive accuracy over a wide range of temperatures and densities, accounting for the intense interest in this approach to understanding and correlating the dynamic properties of liquids in terms of thermodynamic properties of liquids. Unfortunately, the particular functional form relating transport properties to $\widehat{S}_{\mathrm{ex}}$ has been found to be highly dependent on molecular structure in molecular fluids~\cite{2018_JCP_149_210901, 2012_CES_79_153} so that this approach to liquid dynamics offers no direct general relationship between transport properties and $\widehat{S}_{\mathrm{ex}}$. Thus, the applicability of this approach is largely limited to atomic and small-molecule liquids. Chopra et al.~\cite{2010_JPCB_114_16487} have discussed the origin of the failure of density-temperature scaling in molecular fluids and related this change of scaling to changes in the density-temperature scaling of transport properties, a phenomenon that is deeply related to the cooperative dynamics of GF liquids, as we have recently found elsewhere. We thus do not see how one could develop a quantitative theory of the dynamics of real polymeric GF liquids based on this theoretical framework since many polymer materials are only thermally stable against degradation below $T_A$, i.e., non-Arrhenius collective dynamics is inherent to this class of materials. Nonetheless, recent works suggest that the excess-entropy scaling approach has some important conceptual ideas to offer for the future development of the GET,~\cite{2008_ACP_137_125} which we briefly discuss below.

Given the apparently general observation that the excess entropy scaling is restricted to elevated temperatures and the emphasis of this model on repulsive interactions, it is natural to compare this model to the standard theory of activated transport where $D$ formally scales as, $D \sim \exp(\Delta S_0/ k_B)$, for molecular diffusion in systems in which attractive interactions are entirely neglected. This might then provide an approach to calculating $\Delta S_0$, one of the most elusive quantities in TST.~\cite{1946_JCP_14_591, 2020_Mac_53_6828} We hope to pursue this approach to understanding and numerically estimating $\Delta S_0$ in the future.

Finally, recent studies appear to indicate that $\widehat{S}_{\mathrm{ex}}$ might enable the estimation of the characteristic temperatures of glass formation. For instance, refs~\citenum{2017_JCP_147_024504} and~\citenum{2018_JCP_148_034504} suggest that the onset temperature $T_A$ may be calculated as the temperature at which $\Delta \widehat{S}_{\mathrm{ex}}$, defined as $\widehat{S}_{\mathrm{ex}}$ minus it two-body estimate $S_2$, changes sign. It has also been suggested that the crossover temperature $T_c$ might be inferred from the $T$ variation of $\widehat{S}_{\mathrm{ex}}$.~\cite{2017_JCP_147_024504, 2017_PRL_119_265502} It would be an invaluable advance to the GET to have an alternative thermodynamic-based approach for calculating $T_A$, and we plan to test this proposal against independent estimates of $T_A$ for our polymer model in the future. Over time, we hope to see the various `pieces' of the entropy theory of glass formation to grow together to form a highly predictive and validated framework from theoretical, computational, and experimental perspectives.
 
\section{Summary and Outlook}

The entropy perspective on the dynamics of glass-forming liquids and the general conception that collective motion accompanies the drop of configurational entropy has been developing steadily over the last century, and over this period, empirical correlations between thermodynamic and dynamic properties have gradually been superseded by quantitative relations organized around increasingly sophisticated theoretical frameworks for quantifying the dynamics and thermodynamics of condensed materials and for quantifying collective motion and its physical consequences on relaxation and diffusion. The generalized entropy theory and the string model are just the most recent manifestations of this conceptual development, along with the necessary experimental and computational works that give credence to these models through validation studies.

Although much progress has been made, there are still many `gaps' in the entropy approach to the dynamics of glass-forming liquids that will require significant effort in the future to obtain a more predictive model of the dynamics of glass-forming liquids. For example, our understanding of the activation free energy parameters has not advanced very much beyond the descriptions introduced by Eyring and coworkers.~\cite{Book_Eyring, 1940_JACS_62_3113, 1941_CR_28_301} A significant effort also needs to be made to develop a fundamental understanding of the activation parameters in the Arrhenius regime where the complicated effects of dynamic heterogeneity are not an important issue. It is very helpful that relaxation times are generally relatively short in the high temperature Arrhenius regime, which is a favorable aspect for simulation studies that could provide much information on which a sound theoretical treatment of the activation energy parameters could be made. There are also many experimental data now available for this purpose, so we are hopeful that much progress in this direction will be made in the near future. Simulations have shown that molecular topology can greatly influence the dynamics of polymeric glass-forming materials,~\cite{2018_JCP_149_161101, 2020_JCP_152_054904} so computational and experimental investigations of how molecular topology influence the high temperature activation energetic parameters and the overall dynamics of glass-forming liquids promise to be a future of development. We also anticipate that studies of nanocomposites, thin supported films, polymers with solvent additives, and polymer blends all offer promising areas for examining the activation free energy parameters and their large influence on the dynamics of glass-forming liquids. In short, we envision this type of study to form the foundation for a fundamental and predictive theory of glass-forming liquids.

There are also many issues to resolve in the entropy theory relating to the question of why dynamic heterogeneity in glass-forming liquids, such as the strings, generally adopts structural forms and obeys the thermodynamics consistent with existing models of self-assembly.~\cite{1999_JCP_111_7116, 2006_JCP_125_144907, 2008_JCP_128_224901, 2014_JCP_140_204509} Answering these questions will require a theoretical understanding of a different order. Equilibrium polymerization has long been understood to arise in connection with the description of fluctuation effects in materials undergoing second-order phase transitions,~\cite{1970_JPC_3_256, 1972_JPC_5_1399, 1972_JPC_5_1417, 1980_PRL_45_1748, 1981_PRL_47_457, 1995_PRE_51_1791, 1969_Symancik} and we expect that a deeper understanding of glass formation within appropriate field theoretical frameworks will naturally evolve in time. We also expect that the theoretical concepts and methods derived from the alternative entropy theories of glass formation, namely, the random first-order transition theory~\cite{1987_PRB_36_8552, 1989_PRA_40_1045, 2004_JCP_121_7347, 2007_ARPC_58_235} and the excess-entropy scaling approach,~\cite{1977_PRA_15_2545, 1999_JPCM_11_5415, 2018_JCP_149_210901} will play a large role in developing the generalized entropy theory further. We await these theoretical developments and quantitative computational investigations to validate these new concepts. For the present, the generalized entropy theory and the string model emphasized in the present work are promising working models. It seems inevitable that many of the currently disparate models of glass formation will come together to form a unified theory of glass formation. Hopefully, we will not have to wait another century for these developments to occur.

\newpage

\section*{Biographies}

\begin{figure*}[htb!]
	\centering
	\includegraphics[angle=0,width=0.3\textwidth]{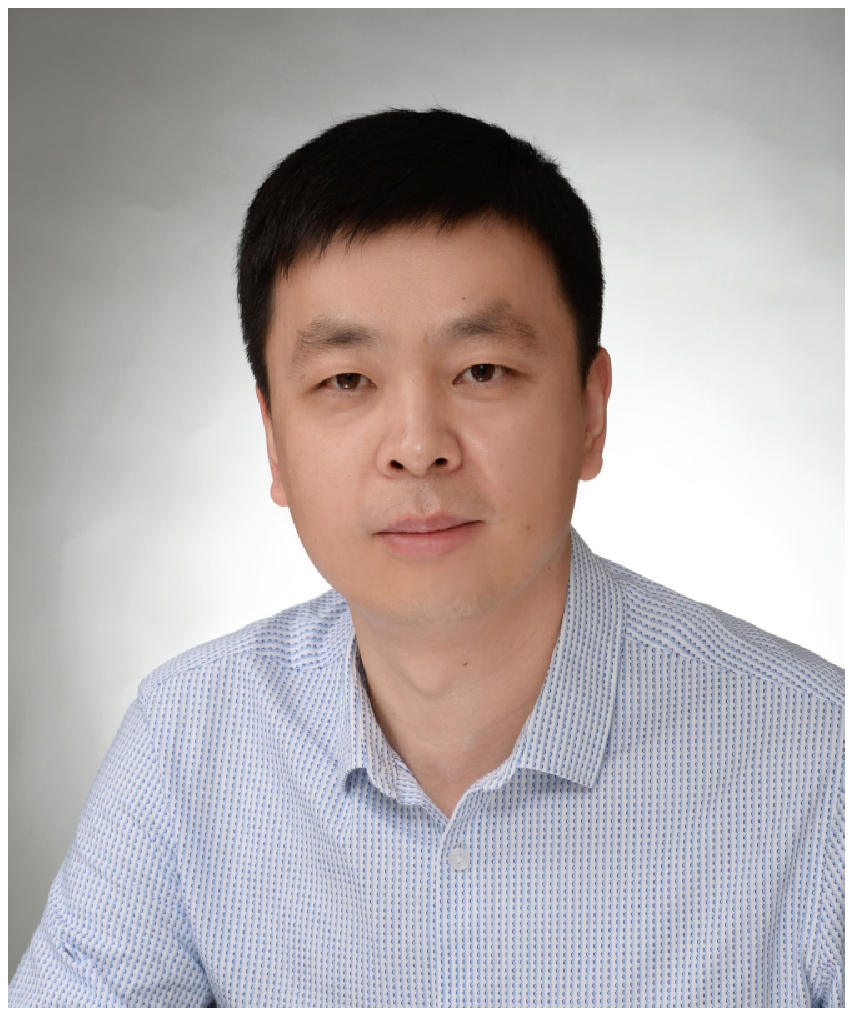}
\end{figure*}

Wen-Sheng Xu is a professor in the State Key Laboratory of Polymer Physics and Chemistry at Changchun Institute of Applied Chemistry, Chinese Academy of Sciences in China, where he has been a faculty member since 2019. He received a B.E. degree in Materials Science and Engineering at Tianjin University in 2007 and a Ph.D. degree in Chemistry and Physics of Polymers at Changchun Institute of Applied Chemistry, Chinese Academy of Sciences in 2012 under the guidance of Professors Li-Jia An and Zhao-Yan Sun. From 2013 to 2018, he was a postdoc first working with Professor Karl F. Freed at the University of Chicago and then with Dr. Yangyang Wang at Oak Ridge National Laboratory. His research is focused on the glass formation and rheology of polymeric materials.

\begin{figure*}[htb!]
	\centering
	\includegraphics[angle=0,width=0.3\textwidth]{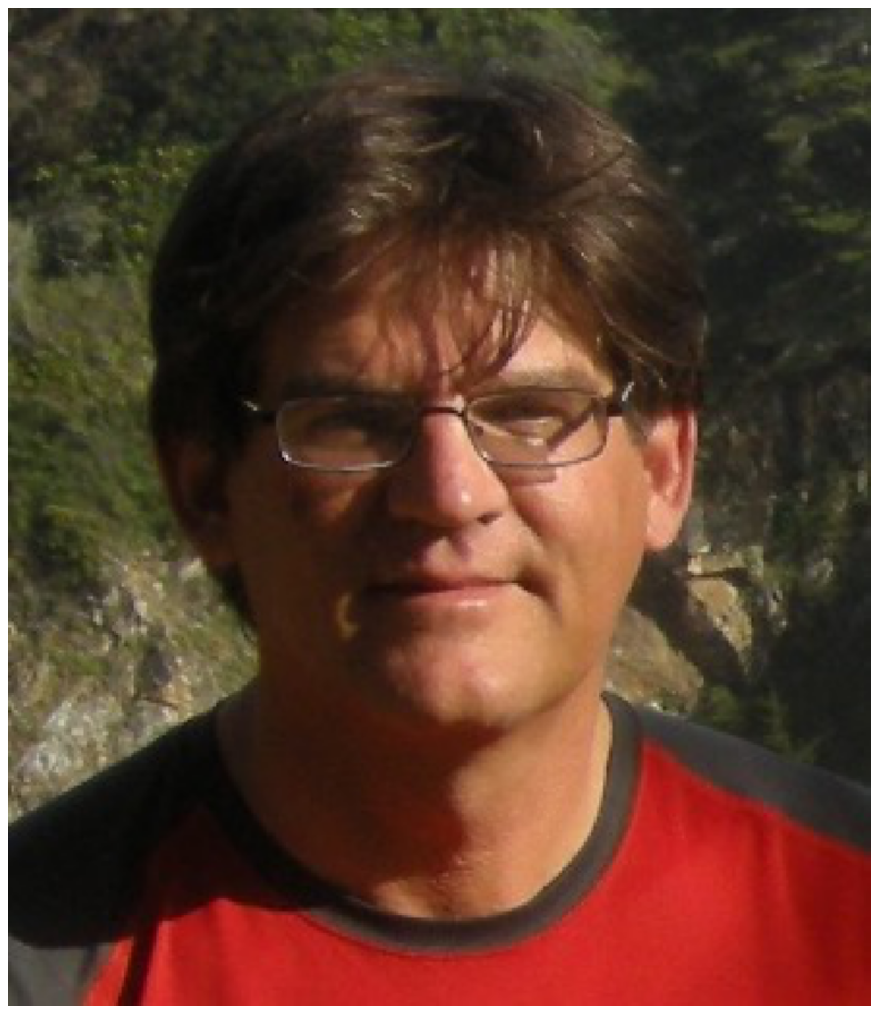}
\end{figure*}
 
Jack F. Douglas is a NIST Fellow in the Materials Science and Engineering Division of the National Institute of Standards and Technology (NIST) at the facility in Gaithersburg, Maryland. He obtained a B.S. degree in Chemistry and an M.S. degree in Mathematics from Virginia Commonwealth University and then a Ph.D. degree in Chemistry at the University of Chicago. After receiving his Ph.D. degree, he was a NATO Fellow at the Cavendish Laboratory, Cambridge and a NRC postdoctoral fellow at NIST before becoming a research scientist at NIST. His field of research includes the equilibrium and dynamic properties of polymer solutions and melts, fractional calculus and path integration, phase separation and critical phenomena, relaxation processes in glass-forming liquids, polycrystalline materials and nanoparticles, the elastic properties of gels, and the thermodynamics and dynamics of self-assembly processes.

\begin{figure*}[htb!]
	\centering
	\includegraphics[angle=0,width=0.3\textwidth]{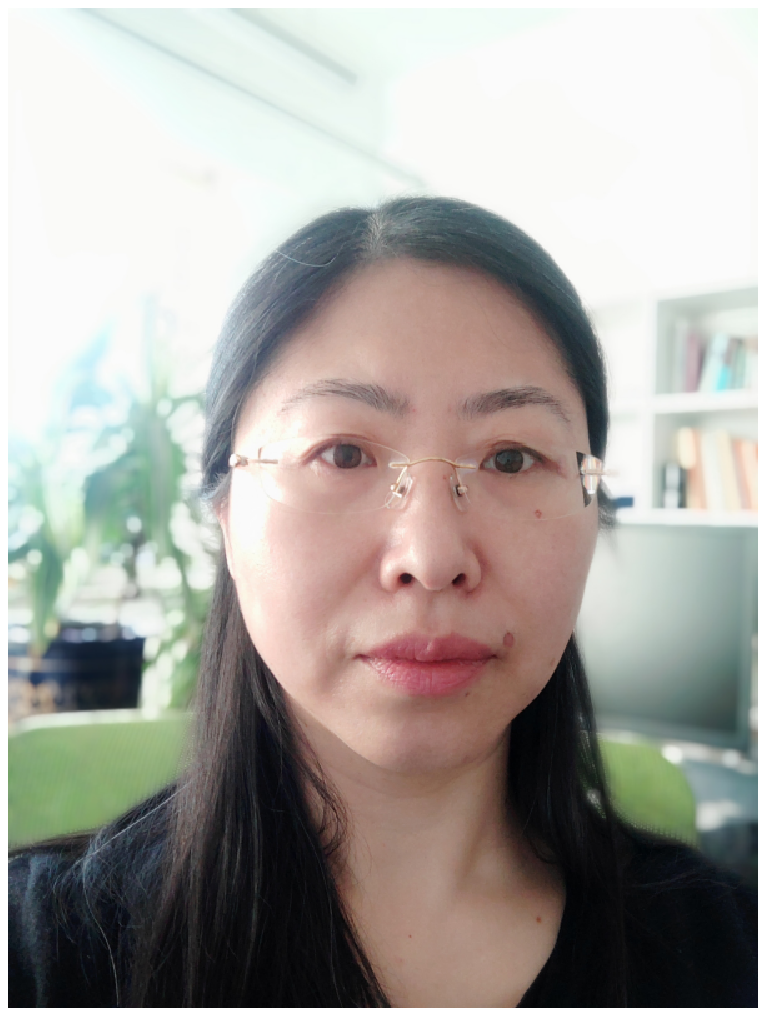}
\end{figure*}

Zhao-Yan Sun is a professor at Changchun Institute of Applied Chemistry, Chinese Academy of Sciences in China and a deputy director of the State Key Laboratory of Polymer Physics and Chemistry. She received a Ph.D. degree in Polymer Chemistry and Physics at Jilin University in 2001. She was a postdoctoral fellow in the Department of Chemistry at University of Dortmund from 2001 to 2002. Since 2003, she has been a faculty member at the present institution, initially as an assistant professor and then an associate professor, before becoming a full professor in 2010. She was awarded the Young Chemistry Award of the Chinese Chemical Society in 2005. Her research group focuses on the structure and dynamics of polymers and nanocomposites and the development of computer simulation methods.

\begin{acknowledgement}
W.-S.X. acknowledges the support from the National Natural Science Foundation of China (No. 21973089). Z.-Y.S acknowledges the support from the National Natural Science Foundation of China (Nos. 21833008 and 21790344), the National Key R\&D Program of China (No. 2018YFB0703701), the Jilin Provincial science and technology development program (No. 20190101021JH), and the Key Research Program of Frontier Sciences, CAS (QYZDY-SSW-SLH027). J.F.D. thanks his long-term collaborators, Karl F. Freed, Jacek Dudowicz, Francis W. Starr, Hao Zhang, and Wen-Sheng Xu, and many postdocs associated with the research groups at the National Institute of Standards and Technology, the University of Chicago, the University of Alberta, Wesleyan University, and Changchun Institute of Applied Chemistry, Chinese Academy of Sciences for the many collaborative efforts over many years that made his contributions to the entropy theory of glass formation possible. J.F.D also acknowledges the National Institute of Standards and Technology for its long-term support of this research effort. Z.-Y.S thanks Professor Li-Jia An for helpful discussions on glass formation over the years. W.-S.X. thanks Professor Karl F. Freed and Dr. Jack F. Douglas for numerous discussions concerning polymer glass formation over the years. W.-S.X. also thanks Professors Li-Jia An and Zhao-Yan Sun for their support and encouragement at every stage of his academic career. We are grateful to the authors of ref~\citenum{2015_Mac_48_3005} for sharing their experimental data with us and to Dr. Beatriz A. Pazmi{\~{n}}o Betancourt for help with the visualization of strings in simulations.
\end{acknowledgement}


\bibliography{refs}

\end{document}